\documentclass[a4paper,11pt]{article}
\usepackage{amsmath, amssymb, amsfonts, setspace, url, verbatim,xcolor, cite, graphics, graphicx, , slashed, mathtools}
\usepackage[T1]{fontenc}
\usepackage[utf8]{inputenc}
\usepackage{lmodern}
\usepackage{microtype}
\definecolor{vub}{RGB}{0,52,154}

\makeatletter 
\@mparswitchfalse%
\makeatother
\reversemarginpar

\usepackage[
colorlinks=true, 
linkcolor=vub,   
citecolor=vub,   
filecolor=magenta, 
urlcolor=cyan,
citebordercolor=red,
filebordercolor=red,
linkbordercolor=red,
urlbordercolor=red,
linktocpage=true
]
{hyperref}
\usepackage[a4paper,top=2cm,left=2.5cm,right=2.5cm,bottom=2cm]{geometry}
\usepackage{longtable}
\usepackage{subfigure}
\usepackage{multirow}
\usepackage{dsfont}
\usepackage{mathabx}
\usepackage{appendix}
\usepackage[shortlabels]{enumitem}
\definecolor{vubo}{RGB}{255,102,0}

\usepackage[textsize=scriptsize,linecolor=vub,backgroundcolor=vub!10,bordercolor=vub]{todonotes}

\usetikzlibrary{fit}

\def\rmd{\mathrm{d}}

\newcommand{\be}{\begin{equation}}
\newcommand{\ee}{\end{equation}}
\newcommand{\dd}{\mathrm{d}}
\newcommand{\cH}{\mathcal{H}}

\newcommand{\lambdaone}{\ell}
\newcommand{\Snew}{S'}

\tikzstyle{box_up} = [text width = 1.5cm, rounded corners=1pt, align=center, anchor=north,minimum height=2.5cm]
\tikzstyle{box_down} = [text width = 1.5cm, rounded corners=1pt, align=center, anchor=south,minimum height=2.5cm]
\tikzstyle{labels_up} = [text width = 1.75cm, rounded corners=1pt, align=center, anchor=south,font=\footnotesize]  
\tikzstyle{labels_down} = [text width = 1.75cm, rounded corners=1pt, align=center, anchor=north,font=\footnotesize]  
\tikzstyle{B} = [fill=blue!0,style=draw, rounded corners=1, line width=1pt,draw=black]  
\tikzstyle{F} = [fill=gray!15,style=draw, rounded corners=1, line width=1pt,draw=black]  
\tikzstyle{redarrows} = [thick,red,->,shorten >= 2pt, shorten <= 2pt]
\tikzstyle{bluearrows} = [thick,blue,->,shorten >= 5pt, shorten <= 5pt]

\usepackage{datetime}

\newcommand*\circled[1]{\footnotesize\tikz[baseline=(char.base)]{%
\node[shape=circle,fill=vub!15,draw,thick,inner sep=2pt] (char) {\bf #1};}}
            \usepackage{enumitem}

\begin{document}

\hfill IFT-UAM/CSIC-24-117

\vspace{1em} 

\begin{center}
{\huge
\bf The surprising structure \\  

\vspace{0em}
of non-relativistic 11-dimensional \\ 

\vspace{0.5em} supergravity}

\vspace{1.5em}
{\large
 Eric A. Bergshoeff{}$^{\,1}$, Chris D. A. Blair{}$^{\,2}$, Johannes Lahnsteiner{}$^{\,3}$, and Jan Rosseel{}$^{\,4}$}
\end{center} 
\begin{center}
\begin{flushleft}
\itshape \small
 ${}^1$Van Swinderen Institute, University of Groningen,
Nijenborgh 3, 9747 AG Groningen, The Netherlands \\ \vspace{0em}
 ${}^2$Instituto de Física Teórica UAM/CSIC, 
Universidad Autónoma de Madrid, Cantoblanco, Madrid 28049, Spain\\ \vspace{0em}
 ${}^3$Nordita, KTH Royal Institute of Technology and Stockholm University, Hannes Alfv\'{e}ns v\"{a}g 12, SE-106 91 Stockholm, Sweden \\ \vspace{0em}
 ${}^4$Division of Theoretical Physics, Rudjer Bo\v{s}kovi\'c Institute, Bijeni\v{c}ka 54, 10000 Zagreb, Croatia
 \end{flushleft}
\end{center}
{  {\tt e.a.bergshoeff@rug.nl}, {\tt c.blair@csic.es}, {\tt johannes.lahnsteiner@su.se}, {\tt Jan.Rosseel@irb.hr}
}

\vspace{2em}

\begin{abstract}
We study a non-relativistic limit of 11-dimensional supergravity. 
This limit leads to a theory with an underlying membrane Newton-Cartan geometry.
Consistency of the non-relativistic limit requires the imposition of constraints, requiring that certain bosonic and fermionic torsions and curvatures vanish.
We investigate the implications of two versions of these constraints.
In one version, we keep only 16 supersymmetry transformations, leading to a simple (purely bosonic) constraint structure but an unusual realisation of the supersymmetry algebra which does not close into diffeomorphisms.
In the other, we keep all 32 supersymmetry transformations.
This requires a complicated sequence of bosonic and fermionic constraints, eventually involving three derivatives of bosonic fields.
We argue, with a linearised calculation, that this sequence of constraints terminates.
Furthermore, we show that there exists a family of supersymmetric solutions satisfying these constraints, containing the non-relativistic limit of the M2 supergravity solution recently obtained by Lambert and Smith as a background relevant for non-relativistic holography.

\end{abstract}

\clearpage
\tableofcontents

\clearpage

\section{Introduction}

The eleven-dimensional supergravity \cite{Cremmer:1978km} is famously the unique maximal supergravity in the maximal number of dimensions where local supersymmetry can be formulated.
As with all statements about uniqueness, there are implicit assumptions.
In this paper, we will remove that of Lorentz invariance, and construct a new \emph{non-relativistic 11-dimensional supergravity}.

We do this by taking a certain non-relativistic limit of eleven-dimensional supergravity.
This is not the usual point particle non-relativistic limit but a membrane non-relativistic limit.
It invokes a $3+8$ split of the local tangent space, breaking the local $\mathrm{SO}(1,10)$ Lorentz transformations to local $\mathrm{SO}(1,2)$ and $\mathrm{SO}(8)$ transformations as well as off-diagonal \emph{boosts}.
The formulation and consistency of the limit relies on an interplay between the metric and the 11-dimensional three-form (to which a membrane couples electrically).

This limit is one example of a variety of string and brane non-relativistic limits which can be taken in string/M-theory theory and supergravity leading to non-relativistic (or non-Lorentzian) theories -- for reviews see \cite{Oling:2022fft,Bergshoeff:2022iyb}. These limits are interesting for several reasons.
Ambitiously, they may allow us to understand the formulation of quantum gravity in its non-relativistic corner, providing a zoo of decoupled corners of string and M-theory, related by dualities \cite{Gomis:2000bd,Danielsson:2000gi, Blair:2023noj}.
They are also of interest as a mechanism with which to construct interesting limits and new versions of the AdS/CFT correspondence 
\cite{Gomis:2005pg,Harmark:2017rpg,Harmark:2018cdl,Lambert:2024uue,Fontanella:2024rvn,Lambert:2024yjk}.

The best understood theory arising from this family of limits is non-relativistic string theory \cite{Gomis:2000bd,Danielsson:2000gi}.
Here the string worldsheet remains relativistic, but the target space geometry does not exhibit 10-dimensional Lorentz invariance. In curved backgrounds, the target space is described by String Newton-Cartan (SNC) geometry \cite{Andringa:2012uz,Bergshoeff:2018yvt}, which involves a 2+8 split of the local tangent space, into `longitudinal' and `transverse' directions.
The beta functionals of the SNC string have been shown to reproduce the equations of motion of the non-relativistic gravity theory that results from taking the limit of the bosonic NSNS part of the ten-dimensional supergravity \cite{Gomis:2019zyu, Bergshoeff:2019pij, Bergshoeff:2021bmc}.

A half-maximal supersymmetric extension of the SNC supergravity limit was constructed in \cite{Bergshoeff:2021tfn}, and supersymmetric brane solutions of this theory studied in \cite{Bergshoeff:2022pzk}.
This provides a concrete realisation of non-relativistic supergravity in ten dimensions.
This non-relativistic supergravity has some novel features compared to its relativistic counterpart.
It features emergent local symmetries: a bosonic scaling (dilatation) symmetry and a fermionic Stuckelberg shift (or superconformal) symmetry.
Consistency of the limit (including invariance of the action and closure of the supersymmetry algebra) requires imposing constraints on the bosonic geometry: these are that certain components of intrinsic torsion tensors as well as field strength components must vanish.

The goal of this paper is to obtain the extension to maximal supersymmetry, meaning $\mathcal{N}=2$ in ten dimensions or $\mathcal{N}=1$ in eleven dimensions.
We choose to work with the latter option, which given the distinguished unique nature of the corresponding supergravity, is both interesting and convenient (as the eleven-dimensional description unifies otherwise distinct sectors of the ten-dimensional theory).

Indeed, as a first step, it is known at the purely bosonic level how to formulate the eleven-dimensional counterpart of the SNC geometry.
This is provided by Membrane Newton-Cartan (MNC) geometry, which follows from taking a membrane non-relativistic limit of the 11-dimensional metric and three-form.
This limit was carried out for the bosonic part of 11-dimensional supergravity in \cite{Blair:2021waq}.
In particular, dimensional reduction produces the bosonic part of SNC supergravity with both the NSNS and RR sectors included.

The expectation is then that the supersymmetric extension will follow from a similar procedure as that used for $\mathcal{N}=1$ supergravity in ten dimensions.
One performs an invertible redefinition of \emph{all} eleven-dimensional fields (metric, three-form and gravitino) in terms of a parameter $c$ whose limit $c \rightarrow \infty$ defines the non-relativistic limit.
We insert these redefinitions into the action, the supersymmetry transformations, the equations of motion, etc., and demand that the limit $c \rightarrow \infty$ is well-defined.
As in the ten-dimensional $\mathcal{N}=1$ case \cite{Bergshoeff:2021tfn} consistency will turn out to require that certain bosonic intrinsic torsion tensors vanish, along with certain components of the non-relativistic four-form field strength.

The major technical complication that arises is that we must vary these constraints under all symmetries of the theory, and for consistency set to zero whatever new geometric objects arise under this variation.
In particular, and in contrast to the ten-dimensional $\mathcal{N}=1$ case, supersymmetry relates the initial bosonic constraints to new fermionic constraints, which in turn vary into components of curvature tensors of the MNC geometry.
Understanding whether this sequence of constraints terminates without completely trivialising the theory (so that only flat space would be a solution) turns out to be a challenging task to establish.
We argue in this paper that this is indeed the case, and exhibit a family of interesting maximally supersymmetric solutions to the constraints we obtain.

\vspace{0.5em}\noindent The essential results of this paper can be summarised as follows:

\begin{enumerate}[label=\protect\circled{\arabic*}]
  \item We derive the supersymmetry transformations \eqref{finiteSUSY} and action \eqref{Lag} (to quadratic order in fermions) of the non-relativistic 11-dimensional MNC supergravity. The field content of the theory includes the MNC vielbeins, a three-form, an auxiliary bosonic field acting as a Lagrange multiplier, and two gravitinos which are chiral as spinors of $\mathrm{SO}(1,2) \times \mathrm{SO}(8)$. In addition to diffeomorphisms and supersymmetry, the action is invariant under local $\mathrm{SO}(1,2)$, $\mathrm{SO}(8)$ and boost transformations, and two emergent symmetries: a bosonic dilatation symmetry \eqref{dil} and a fermionic shift (or superconformal) symmetry \eqref{fermionicshift}.
  \item We obtain the consistency conditions which are required to have a well-defined limit with finite supersymmetry transformations and a supersymmetric action. We show how these consistency conditions can be met by imposing constraints. We identify two natural sets of constraints:
  \vspace{-0.5em}
  \begin{itemize}
	\item {\bf Maximal supersymmetry:} we keep all 32 supersymmetry transformations, at the cost of imposing a tower of bosonic and fermionic constraints, which require that numerous components of torsion and curvature tensors must vanish. This possibility is analysed in section \ref{maxSUSYsection}. We are in particular able to show that the tower of constraints terminates at the level of a linearised calculation, and we expect that our analysis suffices to obtain the structure of constraints in the full theory.
	\item {\bf Half-maximal supersymmetry:} we keep only 16 supersymmetry transformations, and only require a limited number of (purely bosonic torsional) constraints. This theory however does not correspond to a standard supergravity theory, in the sense that the commutator algebra of two supersymmetry transformations does not yield a diffeomorphism. This possibility is analysed in section \ref{halfSUSYsection}.
  \end{itemize} 
  \item We find (in section \ref{woundM2}) that there is a surprising tension between maximal supersymmetry and a non-relativistic supergravity description of a Newton potential $\Phi(x)$ obeying the Poisson equation $\Delta \Phi = 0$.
  The maximal supersymmetry constraints in fact imply the much stronger condition that $\Phi(x)$ must be at most quadratic in (transverse) coordinates. 
  On the other hand, when we impose the constraints needed in the half-maximal case, $\Phi(x)$ obeys the usual Poisson equation without any other constraints. 
  \item We demonstrate that despite the constraints, the theory with maximal supersymmetry contains a non-trivial infinite family of maximally supersymmetric solutions; see section \ref{LS}. This family includes the background obtained by Lambert-Smith in \cite{Lambert:2024uue} by taking an MNC limit of the relativistic M2 brane soluton, leading  to a solution of the bosonic part of MNC supergravity which they proposed to be holographically dual to an analogous non-relativistic limit of ABJM.
\end{enumerate}

\noindent To establish these main results, we organise the paper as follows.

\vspace{1em}
\noindent In section \ref{sec_limit}, we explain how to take the Membrane Newton-Cartan limit of the 11-dimensional supergravity. In section \ref{sec_limit_setup}, we introduce the field redefinitions involved, and demonstrate that to obtain a finite limit of the action, we need to introduce an auxiliary bosonic field which plays the role of a Lagrange multiplier in the non-relativistic supergravity.
In section \ref{sec_limit_susy}, we focus on the limit of the supersymmetry transformations themselves.
We first obtain the divergent part of these transformations, from which we learn that constraints must be imposed on the bosonic geometry, and potentially also on the supersymmetry parameters, in order to have a consistent limit. 
Here we also identify associated emergent fermionic shift symmetries.
Then, in section \ref{limit_action_eom}, we write down the action of the non-relativistic supergravity (to quadratic order in fermions) that results from the limit, and analyse the resulting equations of motion. We also briefly identify certain additional equations of motion that do not follow from this action, as a consequence of the emergent local symmetries.

In section \ref{geometry_section}, we take a step back and consider a first principles approach to the geometry underlying the non-relativistic supergravity.
We introduce bosonic connections and curvatures in section \ref{sec_bosconns} and relate these to geometric structures that naively appear in the limit of the action and equations of motion.
In section \ref{maxSUSYconns}, we specialise this geometric structure to the case relevant for maximal supersymmetry, where we impose a particular set of bosonic constraints.
In this setting, we also introduce fermionic connections (for the emergent shift symmetry) and curvatures.
In section \ref{halfSUSYconns}, we briefly indicate the specialisation of the bosonic geometric structure to the case where we impose constraints compatible only with half-maximal supersymmetry.

In section \ref{maxSUSYsection}, we discuss the theory with maximal supersymmetry in more detail.
We summarise the supersymmetry transformations in section \ref{fullsupersummary}, outline the complicated tower of constraints that appears in section \ref{constraints_max} (relegating the more technical derivation of this tower to appendix \ref{tower_derivation}), and discuss the closure of the supersymmetry algebra in section \ref{fullclosure}.

In section \ref{halfSUSYsection}, we outline the differences that appear if we only keep half the supersymmetry transformation parameters.
This leads to the supersymmetry transformations summarised in section \ref{halfsusysummary}, a much simpler set of constraints as shown in \ref{halftower}, as well as an unusual version of the supersymmetry algebra displayed in section \ref{halfclosure}.

In section \ref{solutions}, we investigate possible supersymmetric solutions of the non-relativistic supergravity.
In section \ref{KSeqns} we write down the Killing spinor equations.
As a first exercise, we show in \ref{flatBG} that a flat background admits in fact an infinite number of Killing spinor solutions.
In section \ref{woundM2}, we analyse a solution which describes a Newton potential.
In section \ref{LS}, we obtain an infinite family of supersymmetric solutions containing amongst them the M2 limit of \cite{Lambert:2024uue}.

We conclude with a discussion in section \ref{discussion}.
The appendices contain additional technical material.

\section{The non-relativistic limit of 11-dimensional supergravity}
\label{sec_limit}

\subsection{Setting up the limit}
\label{sec_limit_setup}

\paragraph{The relativistic theory}

The basic fields of the 11-dimensional supergravity are the vielbein $E^{\hat a}{}_\mu$, where $\hat a=0,\dots,10$ is a local tangent space index, the three-form $C_{\mu\nu\rho}$ and the gravitino $\Psi_\mu$.
We follow the conventions of the textbook by Freedman and Van Proeyen \cite{Freedman:2012zz}, with minor field redefinitions.\footnote{For the gravitino, three-form and supersymmetry transformation parameter we have $\psi_\mu^{FVP} = \sqrt{2} {\Psi}_\mu$, $A_{\mu\nu\rho}^{FVP} = -\tfrac{1}{\sqrt{2}} C_{\mu\nu\rho}$ and $\epsilon^{FVP} = \sqrt{2} \epsilon$.}
The Lagrangian (excluding four-fermion terms) is:
\be
\begin{split}
\mathcal{L} & = E\left( R  - \tfrac{1}{2 \cdot 4!} F_{\mu\nu\rho\sigma} F^{\mu\nu\rho\sigma}- 2 \bar\Psi_\mu \gamma^{\mu\nu\rho} D_\nu \Psi_\rho
+ \tfrac{1}{48} F_{\rho\sigma\lambda\tau}( \bar \Psi_\mu \gamma^{\mu\nu\rho\sigma\lambda\tau} \Psi_\nu + 12 \bar \Psi^\rho \gamma^{\sigma\lambda} \Psi^\tau )
 \right)
\\ & \quad +\tfrac{1}{144^2} \epsilon^{\mu_1 \dots \mu_{11}} F_{\mu_1\dots\mu_4} F_{\mu_5\dots\mu_8} C_{\mu_9 \mu_{10} \mu_{11}}
 \,,
\end{split}
\label{RLag}
\ee
where $E \equiv \det E^{\hat a}{}_{\mu}$.
We use the epsilon symbol $\epsilon^{\mu_1 \dots \mu_{11}}$ with $\epsilon^{01\dots 10} = -1$. The supersymmetry transformations of the fields are:
\be
\begin{split}
\delta_\epsilon E^{\hat a}{}_{\mu}& = \bar \epsilon \gamma^{\hat a} \Psi_\mu \,,\quad \delta_\epsilon C_{\mu\nu\rho}  =  3  \bar\epsilon \gamma_{\hat a \hat b} E^{\hat a}{}_{[\mu} E^{\hat b}{}_{\nu }\Psi_{\rho]} \,, \\
\delta_\epsilon \Psi_\mu & = D_\mu ( \hat\omega) \epsilon - \tfrac12 \tfrac{1}{144} ( \gamma^{\nu\rho\sigma\lambda}{}_\mu - 8 \gamma^{\rho\sigma\lambda} \delta^\nu_\mu ) \hat F_{\nu\rho\sigma\lambda} \epsilon \,,
\end{split}
\label{susytransforig}
\ee
where the supercovariant derivative and field strength are defined by
\be
\begin{split}
\hat \omega_{\mu}{}^{\hat a \hat b} &= \omega_\mu{}^{\hat a \hat b} - \tfrac12 (  \bar\Psi_\mu \gamma^{\hat b} \Psi^{\hat a} -  \bar\Psi^{\hat a} \gamma_\mu \Psi^{\hat b} +  \bar\Psi^{\hat b} \gamma^{\hat a} \Psi_\mu )\,,
\\
\hat F_{\mu\nu\rho\sigma} & = F_{\mu\nu\rho\sigma} - 6  \bar\Psi_{[\mu} \gamma_{\nu\rho} \Psi_{\sigma]} \,,\quad F_{\mu\nu\rho\sigma}  = 4 \partial_{[\mu} C_{\nu\rho\sigma]} \,,
\end{split}
\ee
where we record that the ordinary spin connection obeys the vielbein postulate in the form
$
\partial_\mu E^{\hat a}{}_\nu + \omega_{\mu}{}^{\hat a}{}_{\hat b} E^{\hat b}{}_\nu - \Gamma_{\mu\nu}{}^\rho E^{\hat a}{}_\rho = 0
$.
The gamma matrices satisfy the algebra $\{ \gamma^{\hat a} , \gamma^{\hat b} \} = 2 \eta^{\hat a \hat b}$, and we use mostly plus signature.

\paragraph{The limit of the fields}

We now straightaway formulate the non-relativistic limit of the 11-dimensional supergravity.
We start with invertible field redefinitions of all fields, bosonic and fermionic, in which we introduce the parameter $c$ which we take to infinity to define the limit.
We first consider the bosonic fields, following \cite{Blair:2021waq} with some modifications, for which the field redefinition is:
\be
\begin{aligned}
E^{\hat a}{}_{\mu}& = ( c \tau^A{}_\mu \,, c^{-1/2} e^a{}_\mu ) \,, &  E^\mu{}_{\hat a} &= ( c^{-1} \tau^\mu{}_A \,, c^{+1/2} e^\mu{}_a ) \,,\\
C_{\mu\nu\rho} & = - c^3 \epsilon_{ABC} \tau^A{}_\mu \tau^B{}_\nu \tau^C{}_\rho + c_{\mu\nu\rho} \,.
\end{aligned}
\label{ExpBos}
\ee
where we let $\hat a = (A,a)$ where $A=0,1,2$ is a `longitudinal' flat index and $a=3,\dots,10$ is a `transverse' flat index.
This introduces the data needed to define a \emph{Membrane Newton-Cartan geometry}.
This geometry is equipped with a pair of vielbeins ($\tau^A{}_\mu, e^a{}_\mu)$ and their dual ``inverse vielbeins'' ($\tau^\mu{}_A, e^\mu{}_a)$.\footnote{Note that in the mathematical literature, what we call vielbeins are called solder forms. What we call ``inverse vielbeins'' are known there as ``vielbeins'' or frame fields.}
These obey the following duality and completeness relations:
\be
\tau^A{}_\mu e^\mu{}_a = 0 \,\quad
e^a{}_\mu \tau^\mu{}_A = 0 \,,\quad
e^\mu{}_a e^b{}_\mu = \delta_a^b \,,\quad
\tau^\mu{}_A \tau^B{}_\mu = \delta_A^B\,,\quad
\tau^\mu{}_A \tau^A{}_\nu + e^\mu{}_a e^a{}_\nu = \delta^\mu_\nu \,.
\label{NCcomplete} 
\ee
The determinant of the vielbein gives a measure factor for the non-relativistic theory:
\be
	E = c^{-1} \Omega \,,\quad
\Omega = - \tfrac{1}{3!8!} \epsilon^{\mu \nu\rho\sigma_1 \dots \sigma_8} \epsilon_{ABC} \epsilon_{a_1 \dots a_8} \tau^A{}_{\mu} \tau^B{}_{\nu} \tau^C{}_{\rho} e^{a_1}{}_{\sigma_1} \dots e^{a_8}{}_{\sigma_8} \,.
\ee
The field strength of the three-form $c_{\mu\nu\rho}$ is defined as usual as
\be
f_{\mu\nu\rho\sigma} = 4 \partial_{[\mu} c_{\nu\rho\sigma]} \,.
\ee
In the non-relativistic theory, we will use the inverse vielbeins $\tau^\mu{}_A$ and $e^\mu{}_a$ to write objects with flat indices.
Thus for example, $f_{abAB} = e^\mu{}_a e^\nu{}_b \tau^\rho{}_A \tau^\sigma{}_B f_{\mu\nu\rho\sigma}$.

As we will see shortly, there will in fact be an additional bosonic field present in the theory:
for the action to be finite in the limit, we will need to introduce an auxiliary field which in the limit serves as a Lagrange multiplier imposing a constraint on the transverse components $f_{abcd}$ of the field strength.

We now want to pair the above bosonic expansion with a prescription for the gravitino.
We mimic the ten-dimensional $\mathcal{N}=1$ approach \cite{Gomis:2004pw,Bergshoeff:2021tfn} and use the longitudinal gamma matrices, $\gamma^A = ( \gamma^0,\gamma^1,\gamma^2)$ to define projectors
\be
\Pi_\pm = \tfrac12 ( 1 \pm \gamma_{012} ) \,.
\ee
Then for the gravitino we posit the expansion
\be
\Psi_\mu = c^{-1} \psi_{+\mu} + c^{1/2} \psi_{-\mu} \,,
\label{ExpFer}
\ee
where $\psi_{\pm \mu}$ are two longitudinal chiral spinors with $\gamma_{012}\psi_{\pm\mu} = \pm\psi_{\pm\mu}$---or, equivalently, $\Pi_\pm \psi_{\pm \mu} = \psi_{\pm \mu}$.

\paragraph{The limit of Lorentz symmetry} 
By construction, the limit will break the local Lorentz symmetry of the original 11-dimensional theory.
This symmetry acts as
\be
\delta_\Lambda E^{\hat a}{}_\mu = - \Lambda^{\hat a}{}_{\hat b} E^{\hat b}{}_\mu \,,\quad 
\delta_\Lambda C_{\mu\nu\rho} = 0 \,,\quad
\delta_\Lambda \Psi_\mu = -\tfrac14 \Lambda^{\hat a \hat b} \gamma_{\hat a \hat b} \Psi_\mu \,.
\ee
If we make the redefinition $\Lambda_A{}^a = c^{-3/2} \lambda_A{}^a$ of the off-diagonal part of the Lorentz parameter, we find that the bosonic fields transform in the $c \rightarrow \infty$ limit as: 
\be
\delta e^a{}_\mu = - \Lambda^a{}_b e^b{}_\mu + \lambda_A{}^a \tau^A{}_\mu\,,\quad
\delta \tau^A{}_\mu = - \Lambda^A{}_B \tau^B{}_\mu \,,\quad
\delta c_{\mu\nu\rho} = - 3 \epsilon_{ABC} \lambda^A{}_ae^a{}_{[\mu} \tau^B{}_\nu \tau^C{}_{\rho]} \,.
\label{bosonic_transfs}
\ee
The surviving symmetries correspond to local longitudinal $\mathrm{SO}(1,2)$ Lorentz rotations, with parameters $\Lambda^A{}_B$, local transverse $\mathrm{SO}(8)$ rotations, with parameters $\Lambda^a{}_b$, and finally a Galilean boost symmetry with parameters $\lambda^A{}_a$.
The boost transformations \eqref{bosonic_transfs} act in a reducible yet indecomposable manner.
Note also that $\delta \tau^\mu{}_A =  -\lambda_A{}^a e^\mu{}_a$, $\delta e^\mu{}_a = 0$ such that the completeness and duality relations \eqref{NCcomplete} are invariant.

Now consider the fermions. If we have $\Psi_\mu = c^{a_+} \psi_{+\mu} + c^{a_-} \psi_{-\mu}$, for some numbers $a_\pm$, we find
\be
\delta \psi_{\pm \mu} = - \tfrac14 \Lambda^{ab} \gamma_{ab} \psi_{\pm \mu} - \tfrac14 \Lambda^{AB} \gamma_{AB} \psi_{\pm \mu} 
- \tfrac12 \lambda^{Aa} \gamma_A \gamma_a ( c^{-3/2} \psi_{\pm \mu} + c^{a_\mp - a_\pm - 3/2} \psi_{\mp \mu} ) \,.
\ee
In order to obtain a finite action of boosts on the fermions, we need $|a_+ - a_-| = 3/2$.
Our choice is $a_- = 1/2$ and $a_+ = -1$, such that under boosts in the $c \rightarrow \infty$ limit we have
\be
\delta \psi_{+\mu} = - \tfrac12 \lambda^{Aa} \gamma_A \gamma_a \psi_{-\mu} \,,\quad
\delta \psi_{-\mu} = 0 \,,
\ee 
again realising a reducible yet indecomposable structure. 

\paragraph{The limit of the action}

Now we discuss the result of substituting the field redefinition \eqref{ExpBos} and \eqref{ExpFer} into the 11-dimensional Lagrangian \eqref{RLag}.
The action can be expanded in powers of $c^3$:
\be
S = c^3 S_3 + c^0 S_0 + c^{-3} S_{-3} + \dots \,.
\ee
Note that this expansion only has a finite number of terms, but those beyond order $c^{-3}$ play no role in the story.
The divergent term is given explicitly by:
\be
S_3  = - \int d^{11} x \,  \Omega\,\tfrac{1}{4!}  \hat f^{(-)}{}^{abcd} \hat f^{(-)}{}_{abcd} \,,\\
\ee
where the anti-self-dual part of the transverse projection of the field strength $f_{\mu\nu\rho\sigma}$ appears in `supercovariant' form, with:
\begin{align}
\hat f_{abcd}  &= f_{abcd} - 6\,\bar \psi_{- [a} \gamma_{bc} \psi_{-d]} \,,\quad
\hat f^{(\pm)}{}_{abcd} = \tfrac{1}{4!} P^{(\pm)}{}_{abcd}{}^{efgh} \hat f_{efgh}  \,,
\label{deffabcd}
\end{align}
where $P^{(\pm)}{}^{abcd}{}_{efgh}  = \tfrac12 ( \delta^{abcd}_{efgh} \pm \epsilon_{abcd}{}^{efgh})$, with $\delta^{abcd}_{efgh} \equiv 4! \delta^{[a}_e \delta^{\phantom{[}b}_f \delta^{\phantom{[}c}_g \delta^{d]}_h$.
The convention here is to insert a factor of $1/n!$ by hand when contracting two rank $n$ antisymmetric tensors.

The selection of the anti-self-dual projection $\hat f^{(-)}$, as opposed to the self-dual projection $\hat f^{(+)}$, in the divergent part of the action is directly due to the choice to write the $c^3$ term in the expansion of the three-form in \eqref{ExpBos} with a minus sign.
Choosing a plus sign would cause $\hat f^{(+)}$ to appear instead.
Equivalently this would also follow if we change our conventions for orientation or the sign of the Chern-Simons term.
If we were to consider the worldvolume theory of a membrane in the eleven-dimensional background described by \eqref{ExpBos} (see for instance \cite{Garcia:2002fa, Gomis:2004pw,Kluson:2019uza,Roychowdhury:2022est}), this choice of sign can be viewed as selecting either the membrane or anti-membrane as the object which survives in the limit, which is natural to expect in a non-relativistic theory.

To eliminate this order $c^3$ term in the expansion of the action, we introduce an auxiliary field  $\lambdaone_{abcd}$ through a Hubbard-Stratonovich transformation\footnote{This refers to the replacement of the Lagrangian containing the term $c^3 \hat f^{(-)}{}^2$ quadratic in $\hat f^{(-)}$ with an equivalent Lagrangian in which $\hat f^{(-)}$ appears linearly, via the introduction of an auxiliary field: schematically $-c^3 \hat f^{(-)}{}^2 \longleftrightarrow -2 \hat{f}^{(-)} \lambdaone + c^{-3} \lambdaone^2$.
}
which leads to an equivalent action, $\Snew$ whose expansion in powers of $c^3$ is:
\begin{subequations}
\begin{align}
&\Snew_3  = 0 \,,\\
&\Snew_0  = S_0 - \int d^{11} x \,\Omega \,\tfrac{2}{4!}  \lambdaone_{abcd} \hat f^{(-) abcd} \label{Snew0} \,,\\
&\Snew_{-3} = S_{-3} + \int d^{11} x \,\Omega\, \tfrac{1}{4!}  \lambdaone_{abcd} \lambdaone^{abcd} \,.
\end{align}
\end{subequations}
By definition, $\lambdaone_{abcd}$ is anti-self-dual, $\tfrac{1}{4!} P^{(-)}{}_{abcd}{}^{efgh}\lambdaone_{efgh} =\lambdaone_{abcd}$.
Later on we will find it convenient to redefine this field: it will turn out that $\lambdaone_{abcd}$ does not transform supercovariantly, as would be expected of a bosonic matter field. 
The equation of motion of this auxiliary field sets
\be
\lambdaone_{abcd} = c^3 \hat f^{(-)}{}_{abcd}
  \,.
\label{defbarlambda}
\ee
The action $\Snew$ now has no divergent piece.
We are therefore able to take the limit $c \rightarrow \infty$, and identify $\Snew_0$ as the action for the non-relativistic supergravity.
In this action, the auxiliary field becomes a Lagrange multiplier, whose equation of motion sets\footnote{The field $\lambdaone_{abcd}$ corresponds to the (transverse part of the) Lagrange multiplier $\widetilde F_{\mu\nu\rho\sigma}$ used in \cite{Blair:2021waq}, which was introduced as the field strength of the \emph{subleading} part of the three-form.
The issue here is that it is $\widetilde{F}$ and not the corresponding potential which transforms naturally under the symmetries of the theory, and including subleading terms in $1/c^3$ means that the non-relativistic variables are not introduced in the form of an invertible field redefinition, unlike in \eqref{ExpBos}.
Therefore we do not follow this approach in the present paper.}
\be
\hat f^{(-)}{}_{abcd}  = 0 \,.
\ee
We will present the full expression for $S_0'$ in section \ref{limit_action_eom} below.

\subsection{The limit of the supersymmetry transformations}
\label{sec_limit_susy} 

\paragraph{General features}

We redefine the relativistic supersymmetric transformation parameter $\epsilon$ in the same fashion as the gravitino (see \eqref{ExpFer}), so that:
\be
\epsilon = c^{-1} \epsilon_+ + c^{1/2} \epsilon_- \,.
\label{ExpEps}
\ee
The supersymmetry transformations again admit an expansion of the form:
\be
\delta_\epsilon = c^3 \delta_3  + c^0 \delta_0 + c^{-3} \delta_{-3} + \dots \,.
\ee
Ideally, we would like to have $\delta_3 = 0$ and then identify $\delta_0 \equiv \delta_Q$ with the supersymmetry transformation of the non-relativistic supergravity.
In practice, we will find that our expansions are such that $\delta_3 = 0$ acting on the bosons, but $\delta_3 \neq 0$ acting on the fermions.
This has important consequences for the structure of the non-relativistic supergravity.

As the original action $S$ is invariant, we can expand $\delta_\epsilon S = 0$ to obtain
\be
  \delta_3 S_0 + \delta_0 S_3= 0 \,,\quad
 \delta_0 S_0 + \delta_{-3} S_3 + \delta_3 S_{-3} = 0 \,,
\ee
and so on.
When we switch to the equivalent action $\Snew$ with the additional auxiliary field $\lambdaone_{abcd}$, we have the conditions
\be
\delta_3 \Snew_0= 0 \,,\quad
\quad \delta_0 \Snew_0 + \delta_3\Snew_{-3} = 0 \,,
\label{deltaSprime} 
\ee
where now the supersymmetry variation includes that of $\lambdaone_{abcd}$, which is determined from \eqref{defbarlambda}.
Two important consequences can be drawn from these equations.

Firstly, for finite $c$, we see from the first equation of \eqref{deltaSprime} that whatever form the $\delta_3$ variations take, they give rise to a symmetry of the finite part of the action, $\Snew_0$.
This leads to an {\bf emergent fermionic shift symmetry} of the non-relativistic theory.

Secondly, we see that the variation of the finite part of the action is given by $\delta_0 \Snew_0 = - \delta_3 \Snew_{-3}$.
In the $c \rightarrow \infty$ limit, $\Snew_{-3}$ drops out of the theory.
Then $\delta_0 \Snew_0$ will only be zero if all terms that can arise from the divergent part $\delta_3$ of the supersymmetry variation vanish.
Evidently this is also needed simply to write down well-defined supersymmetry transformations of our fields.
Requiring $\delta_3$ to be zero will lead to {\bf constraints} that we must impose on our non-relativistic theory.

\paragraph{Transformation of the bosons}

We begin the detailed analysis by expanding the supersymmetry transformations \eqref{susytransforig} of the metric and three-form.
With the definitions \eqref{ExpBos}, \eqref{ExpFer} and \eqref{ExpEps} these lead to supersymmetry transformations of $\tau^A{}_\mu$, $e^a{}_\mu$  and $c_{\mu\nu\rho}$ for which there is no divergent piece, $\delta_3 = 0$.
In retrospect this justifies the choice of powers of $c$ appearing in \eqref{ExpFer} and \eqref{ExpEps}.
\begin{subequations}
\label{deltaQbos}%
The finite piece then gives:
\begin{align}
\delta_Q \tau^A{}_\mu & =
\bar\epsilon_- \gamma^A \psi_{-\mu}
 \,,\\
\delta_Q e^a{}_\mu & =  \bar\epsilon_+ \gamma^a \psi_{-\mu}
+ \bar\epsilon_- \gamma^a \psi_{+\mu} \,,
\\
\delta_Q c_{\mu\nu\rho} & =\notag
 6 \bar \epsilon_+   \epsilon_{ABC} \gamma^A \psi_{+[\mu} \tau^B{}_{\nu} \tau^C_{\rho]}+3 \bar\epsilon_- \gamma_{ab} \psi_{-[\mu} e^a{}_\nu e^b{}_{\rho]}
\\ & \quad
+6  \left(
\bar\epsilon_+ \gamma_{Aa} \psi_{-[\mu} \tau^A{}_\nu e^a{}_{\rho]} +
\bar\epsilon_- \gamma_{Aa} \psi_{+[\mu} \tau^A{}_\nu e^a{}_{\rho]}
\right) \,.
\end{align}
\end{subequations}
The supersymmetry transformation of the auxiliary field $\lambdaone_{abcd}$ is determined using \eqref{defbarlambda}, which owing to the factor of $c^3$ requires knowledge of the subleading variation $\delta_{-3}$.
In addition, the manipulations we will perform below on the supersymmetry variations of the fermions will modify the transformation of this field with terms involving subleading parts of the equations of motion of the fermions. 
For these reasons, we defer discussing the supersymmetry transformation of $\lambdaone_{abcd}$ until the end of this section.

\paragraph{Transformation of the fermions}

\begin{subequations}\label{divergferm}%
We now examine the supersymmetry transformation of the fermions.
In this case, there are terms at order $c^3$ acting on both $\psi_{\pm \mu}$, which can be determined to be:
\begin{align}
& \delta_3 \psi_{-\mu} = \tau^A{}_\mu \gamma_A \eta_{-}(\epsilon) \,,
\\ &\delta_3 \psi_{+\mu} =  \tau^A{}_\mu \rho_{+A} (\epsilon) - \tfrac12 e^a{}_\mu \gamma_a \eta_- ( \epsilon)
 \\ & \qquad\qquad \notag- \tfrac{1}{12} \tau^A{}_\mu  \gamma_A  \tfrac{1}{4!} \hat f^{(-)}{}_{abcd} \gamma^{abcd} \epsilon_+
  - \tfrac18 e^e{}_\mu \tfrac{1}{4!} \hat f^{(-)}{}_{abcd} \gamma^{abcd} \gamma_e \epsilon_-\,,
\end{align}
where we defined the following field dependent quantities:
\begin{align}
\label{defrhoeps}
\rho_{+A}(\epsilon) & = - \tfrac{1}{2} ( \eta_{AB} - \tfrac13 \gamma_A \gamma_B )\tfrac12 \hat T_{bc}{}^B \gamma^{bc} \epsilon_+
 \\ & \notag\qquad
 +\big(\tfrac12 \gamma^B \gamma^a \hat T_{a}{}_{\{ AB\} } + \tfrac{1}{4}  ( \eta_{AB} - \tfrac13 \gamma_A \gamma_B ) \tfrac{1}{3!} \hat f^{B}{}_{bcd} \gamma^{bcd} \Big) \epsilon_-\,,\\
\eta_-(\epsilon)&= (- \tfrac{1}{12} \tfrac{1}{4!} \hat f^{(+)}{}_{abcd} \gamma^{abcd}  - \tfrac{1}{6} \tfrac{1}{2} \gamma_A \gamma^{ab} \hat T_{ab}{}^A) \epsilon_-\,,
\label{defetaeps}
\end{align}
the former satisfying $\gamma^A \rho_{+A}(\epsilon) = 0$.
\end{subequations}
In addition to the self-dual part of $\hat f_{abcd}$ from equation \eqref{deffabcd}, we here have 
\be
 \hat f_{Aabc} \equiv f_{Aabc}
-3 \,\bar \psi_{-A} \gamma_{[ab} \psi_{-c]}
+ 6 \,\bar\psi_{+ [a} \gamma_{|A|b} \psi_{-c]}\,,
\label{deffAabc}
\ee
and certain components $T_{ab}{}^A$, $T_{a}{}^{\{AB\}}$ of
\be
T_{\mu\nu}{}^A = 2 \partial_{[\mu} \tau^A{}_{\nu]} \,,\quad \hat T_{\mu\nu}{}^A \equiv T_{\mu\nu}{}^A - \bar \psi_{-\mu} \gamma^A \psi_{-\nu}\,,
\label{defTorsion}
\ee
where the curly brackets in $T_{a\{AB\}}$ denotes the symmetric traceless part of the tensor, thus
\be
T_{a\{AB\}} = T_{a(AB)} - \tfrac13 \eta_{AB} \eta^{CD} T_{a(CD)}\,.
\label{defTSymmTraceless}
\ee
Recall that indices are flattened with the Membrane Newton-Cartan (inverse) vielbeins, and then raised and lowered with the flat longitudinal and transverse metrics.

\paragraph{Removing divergent terms using a trivial symmetry}

To remove the divergent terms involving $\hat f^{(-)}{}_{abcd}$, we can use a trivial `equation of motion symmetry', prior to taking the limit $c \rightarrow \infty$.
The variation of the action under a transformation of both the Lagrange multiplier and the $+$ projection of the gravitino is:
\be
\delta \Snew = \int d^{11} x \, \Omega\left( \tfrac{2}{4!}  \mathcal{E}(\lambdaone)^{abcd}\delta \lambdaone_{abcd}
 +\bar{\mathcal{E}}(\psi_+)^\mu\delta \psi_{+\mu}\right) \,,
\label{deltaSspecial}
\ee
where $\mathcal{E}(\lambdaone)^{abcd}$ is the equation of motion of the auxiliary field,
\be
\mathcal{E}(\lambdaone)^{abcd} = c^{-3} \lambdaone^{abcd} -  \hat f^{(-)}{}^{abcd}\,,
\ee
and $\mathcal{E}(\psi_+)$ is that of the gravitino, whose form we will not use explicitly in this section.
We have a trivial invariance of the action with parameter $Z_{+ abcd\mu}$ for the transformation:
\be
\begin{split}
\delta_{\text{trivial}} \lambdaone_{abcd} = \tfrac12 \bar{\mathcal{E}}(\psi_+)^\mu Z_{+ abcd \mu} \,,\quad
\delta_{\text{trivial}} \psi_+{}_\mu  =- \tfrac{1}{4!} \mathcal{E}(\lambdaone)^{abcd} Z_{+ abcd \mu} \,,
\end{split}
\ee
where $Z_{+ abcd\mu}$ is a spinor of the same longitudinal chirality as $\psi_{+\mu}$, and as a tensor is projected on the $abcd$ indices in the same manner as $\lambdaone_{abcd}$.
\begin{subequations}
\label{deltaextra}%
We now modify the supersymmetry transformations of $\psi_{+\mu}$ and the Lagrange multiplier simultaneously
by defining an ``extra'' modification of the supersymmetry variations with a field-dependent trivial symmetry such that the new transformations are
\begin{align}
\delta' & \equiv \delta_\epsilon + \delta_{\text{extra}}\,,
\label{def_deltaprime}
\\
\delta_{\text{extra}} \psi_{+\mu} & = - \tfrac{c^3}{12} \tau^A{}_\mu \gamma_A \tfrac{1}{4!} \mathcal{E}(\lambdaone)_{abcd} \gamma^{abcd} \epsilon_+ - \tfrac{c^3}{8} e^e{}_\mu \tfrac{1}{4!} \mathcal{E}(\lambdaone)_{abcd} \gamma^{abcd}  \gamma_e \epsilon_-\,,
\\
\delta_{\text{extra}} \lambdaone_{abcd} & = 
 \tfrac{c^3}{24} \tau^A{}_\mu \bar{\mathcal{E}}(\psi_+)^\mu\gamma_A \gamma_{abcd}\epsilon_+
+ \tfrac{c^3}{16} e^e{}_\mu \bar{\mathcal{E}}(\psi_+)^\mu\gamma_{abcd}\gamma_e \epsilon_-\,.
\end{align}
\end{subequations}
The variation of $\lambdaone_{abcd}$ can be checked to be automatically correctly projected, using the identities \eqref{gammaPi} and \eqref{proj_mixes}.
In terms of these improved supersymmetry transformations $\delta'$, the divergent terms take the form
\begin{subequations}
\label{delta3psifinal}
\label{deltaprimepsi3}%
\begin{align}
& \delta'_3 \psi_{+\mu} =  \tau^A{}_\mu \rho_{A+} (\epsilon) - \tfrac12 e^a{}_\mu \gamma_a \eta_- ( \epsilon) \,,
\\ & \delta'_3 \psi_{-\mu} =\tau^A{}_\mu \gamma_A \eta_{-}(\epsilon) \,,
\end{align}
\end{subequations}
while we have modified the finite transformation of $\psi_{+\mu}$ such that
\be
\label{deltaprimepsi0}
\delta_0'\psi_{+\mu} = \delta_0 \psi_{+\mu} - \tfrac{1}{12} \tau^A{}_\mu  \gamma_A  \tfrac{1}{4!} \lambdaone_{abcd} \gamma^{abcd} \epsilon_+
  - \tfrac18 e^e{}_\mu \tfrac{1}{4!} \lambdaone_{abcd} \gamma^{abcd} \gamma_e \epsilon_- \,.
\ee
We will discuss the modification of the transformation of $\lambdaone_{abcd}$ later.

\paragraph{Identifying emergent fermionic shift symmetries from remaining divergent terms}

From equation \eqref{delta3psifinal} we then identify emergent fermionic shift or Stuckelberg symmetries of the non-relativistic limit.
We know this exists because $\delta'_3 \Snew_0 = 0$ (up to trivial symmetries of the sort we have used above to remove other divergences).
The fermionic shift symmetries can be inferred from the structure of how the field dependent quantities $\eta_-(\epsilon)$ and $\rho_{+A}(\epsilon)$ appear in \eqref{delta3psifinal}, and are thus defined by:
\be
\begin{split}\label{fermionicshift}
\delta_{S} \psi_{+\mu} & =    \tau^A{}_\mu \rho_{+A}  - \tfrac12 e^a{}_\mu \gamma_a \eta_- \,,\\
\delta_{S} \psi_{-\mu} & =  \tau^A{}_\mu \gamma_A \eta_{-} \,,\\
\end{split}
\ee
for arbitrary local parameters $\eta_{-}$ and $\rho_{+A}$ with $\gamma^A \rho_{+A} = 0$.
Therefore the specific $\eta_-(\epsilon)$ and $\rho_{+A}(\epsilon)$ of \eqref{defrhoeps} and \eqref{defetaeps} are specific examples of such transformations. 

These symmetries can be viewed as fermionic counterparts of a bosonic emergent dilatation symmetry, which appeared already in the bosonic sector of the theory \cite{Blair:2021waq}.
This scales each field with a weight according to the power of $c$ that accompanies them in the field redefinition \eqref{ExpBos} or \eqref{ExpFer}, and so extends to the fermionic sector as well.
The explicit (infinitesimal) action of these emergent dilatations with local parameter $\alpha$ is: 
\begin{align}
    \delta_D \tau^A{}_\mu &= \alpha \,\tau^A{}_\mu\,, & \delta_D e^a{}_\mu &= -\tfrac12\alpha\, e^a{}_\mu\,, \notag\\
    \delta_D c_{\mu\nu\rho} &=0\,, & \delta_D\ell_{abcd} &= -\alpha\,\ell_{abcd}\,,\label{dil} \\
    \delta_D \psi_{-\mu} &= \tfrac12 \alpha\,\psi_{-\mu}\,, & \delta_D \psi_{+\mu} &= -\alpha\,\psi_{+\mu}\,.\notag
\end{align}
As the fermionic partner to a bosonic dilatation, we therefore also refer to the symmetries \eqref{fermionicshift} as superconformal transformations.

\paragraph{Identifying constraints}

The remaining issue with the divergent terms is that supersymmetry of the action $\Snew_0$ is not guaranteed, as $\delta'_0 \Snew_0 + \delta'_3 \Snew_{-3} = 0$.
We can fix this by finding constraints such that $\eta_-(\epsilon) = 0 = \rho_{+A}(\epsilon)$, where these field dependent quantities were defined in \eqref{defrhoeps} and \eqref{defetaeps}.
Different options for constraints that achieve this are possible:

\begin{enumerate}
\item \emph{Maximal supersymmetry.} 
We impose that all bosonic tensors appearing in \eqref{defrhoeps} and \eqref{defetaeps} must vanish:
\be
\hat T_{ab}{}^A = 0 \,,\quad
\hat T_{a}{}^{\{ AB \}} = 0 \,,\quad
\hat f_{abcd} = 0 \,,\quad
\hat f_{Aabc}  = 0 \,.
\label{geoconstraints}
\ee
Then we have supersymmetry transformations parametrised by $\epsilon_\pm$, i.e. 32 independent supersymmetries.
Accordingly, this leads to a theory with maximal supersymmetry. 

Strictly speaking, we did not have to impose $\hat{f}^{(-)}{}_{abcd} = 0$ to ensure that the divergent part of the supersymmetry transformation vanishes, however requiring a boost invariant set of constraints it is easily seen that this follows from the others. 

Consistency of the theory requires that the variation of the constraints \eqref{geoconstraints} under all symmetry transformations must again vanish. 
In particular, the supersymmetry variations of the constraints \eqref{geoconstraints} leads to a tower of constraints involving both fermionic and bosonic curvatures.
We will discuss how this works in detail in section \ref{maxSUSYsection}.

\item \emph{Half-maximal supersymmetry.} 
We impose here that only half of the supersymmetry parameters can be non-zero. 
In particular, we notice that \eqref{defrhoeps} and \eqref{defetaeps} nearly completely vanish when we set $\epsilon_-=0$. 
Heuristically, the condition $\epsilon_- = 0$ could be viewed as the BPS condition for the 11-dimensional membrane, which is suggested by interpreting the non-relativistic limit itself as being associated with a membrane.
Furthermore, as we will see later, it turns out that it is the transformation involving $\epsilon_-$ that is responsible for the complexity of the tower of constraints that appears in the preceding option.
This motivates imposing the following constraints:
\be
\epsilon_-=0\,,\quad \hat T_{ab}{}^A = 0\,,\quad \hat{f}_{abcd}=0 \,.
\label{halfsusy}
\ee
We have again included $\hat f_{abcd} = 0$ in the set of constraints, although only $\epsilon_- = 0$ and $\hat T_{ab}{}^A = 0$ are needed for the vanishing of the divergent terms.
Consistency of these constraints under supersymmetry variations leads solely to the additional condition that $\hat f_{abcd} = 0$, as we show in section \ref{halftower}.
This is a much simpler situation than the tower of constraints that appears in the option with maximal supersymmetry.

This leads to a theory with half-maximal supersymmetry. Note however that it turns out that this is not supersymmetry in the conventional sense, as the commutator algebra of two supersymmetry transformations does not give a diffeomorphism (see section \ref{halfclosure}).

\item \emph{Modify the limit.} Alternatively, we could imagine modifying the initial redefinition of the gravitinos in order to shift divergences to lower orders in $c$. We will not discuss this possibility in detail in this paper.

\end{enumerate}

\noindent We will discuss options 1 and 2 further in sections \ref{maxSUSYsection} and \ref{halfSUSYsection} respectively.
Next, we finish our derivation of the non-relativistic theory by finally writing down the explicit supersymmetry transformations that come out at order $c^0$ and then recording the result of the limit at the level of the action and equations of motion.

\paragraph{Finite part of the supersymmetry transformation of the fermions}

We now discuss the finite part of the supersymmetry transformations of the gravitinos $\psi_{\pm \mu}$.
Having used the equation of motion symmetry, this is given by a relatively complicated expression -- see appendix \ref{gorydetails}.
Combining this naive supersymmetry transformation with certain field dependent Stuckelberg shift transformations, we obtain an improved transformation taking the following form:
\begin{subequations}\label{finiteSUSY}%
\begin{align}
& \delta_Q \psi_{-\mu} = \notag (\partial_\mu + \tfrac14 \hat \omega_{\mu}{}^{ab} \gamma_{ab} + \tfrac14 \hat \omega_\mu{}^{AB}\gamma_{AB} - \tfrac12 \hat b_\mu ) \epsilon_-
\\ \notag
& \qquad \qquad +
\tfrac{1}{12} \tfrac{1}{6} e^a{}_\mu \hat f_{A bcd} ( \gamma_a{}^{bcd} - 6 \delta_a^b \gamma^{cd} ) \gamma^A \epsilon_-
\\ \notag &
\qquad\qquad
+ \tau_{A\mu} \left(
\tfrac12 \hat T_{a}{}^{\{AB\}} \gamma_B \gamma^a
+\tfrac14 ( \eta^{AB} - \tfrac13 \gamma^A \gamma^B ) \boldsymbol{\hat{f}}_B \right) \epsilon_+
\\ & \qquad\qquad
+ e^a{}_{\mu} \left(
\tfrac{1}{24} \gamma_a \boldsymbol{\hat{f}}^{(-)} - \tfrac{1}{12} \gamma_a {\boldsymbol{\hat T}^A} \gamma_A
+ \tfrac12 \hat T_{ab}{}^A \gamma^b \gamma_A - \tfrac{1}{8} \boldsymbol{\hat f}^{(+)} \gamma_a
\right) \epsilon_+
\,,\\ \notag
& \delta_Q \psi_{+\mu} = (\partial_\mu  + \tfrac{1}{4}\hat \omega_\mu{}^{ab} \gamma_{ab}
+ \tfrac{1}{4} \hat \omega_\mu{}^{AB} \gamma_{AB} 	+ \hat b_\mu )\epsilon_+
+ \tfrac12 \hat \omega_\mu{}^{Aa} \gamma_{Aa} \epsilon_-
\\ & \qquad \qquad \notag - \tfrac{1}{12}  \tau^A{}_\mu \gamma_A \boldsymbol{\lambda} \epsilon_+
 - \tfrac18 e^a{}_\mu {\boldsymbol{\lambda}}  \gamma_a \epsilon_-
\\ & \qquad\qquad
+\tfrac{1}{12} \tfrac{1}{6} e^a{}_\mu \hat f_{A bcd} ( \gamma_a{}^{bcd} - 6 \delta_a^b \gamma^{cd} ) \gamma^A \epsilon_+
 \,.
\end{align}
\end{subequations}
We find here  the following spin connections for longitudinal Lorentz rotations, transverse spatial rotations, boosts and dilatations:
\begin{subequations}\label{connections}%
\begin{align}
\hat \omega_\mu{}^{AB} & = \hat T_{\mu}{}^{[AB]} - \tfrac12 \tau^C{}_\mu \hat T^{AB}{}_C \,,\\
\hat \omega_\mu{}^{ab} & = \hat \Omega_\mu{}^{[ab]} - \tfrac12 e_{c\mu} \hat \Omega^{abc} + \tfrac14 \tau_{A\mu} \epsilon^{ABC} \hat{f}_{BC}{}^{ab}+\tfrac13 e^{[a}{}_\mu \hat T^{b] A}{}_A\,,\\
\hat \omega_\mu{}^{Aa} & = \tfrac12 \hat \Omega_\mu{}^{Aa} - \tfrac12 e_{\mu b} \hat \Omega^{Aab}
- \tfrac{1}{4} e^b{}{}_\mu {\hat f}^a{}_{b BC} \epsilon^{BCA}
- \tau^A{}_\mu\tfrac13  {\hat f}_{012}{}^a \,,\\
\hat b_\mu & = \tfrac13 e^a{}_\mu \hat T_a{}^A{}_A \,,
\end{align}
\end{subequations}
where $\hat \Omega_{\mu\nu}{}^a  = \Omega_{\mu\nu{}}{}^a - 2 \bar \psi_{+[\mu} \gamma^a \psi_{-\nu]}$ with $\Omega_{\mu\nu}{}^a = 2 \partial_{[\mu} e^a{}_{\nu]}$.
The field strengths $\hat f^{(\pm)}{}_{abcd}$ and $\hat f_{Aabc}$ were defined in \eqref{deffabcd} and \eqref{deffAabc}, respectively, and $\hat f_{ABab}$ and $\hat f_{ABCa}$ are defined in equations \eqref{defhathatfABab} and \eqref{defhathatfABCa}, respectively.
All hatted quantities are supercovariant: their supersymmetry variations do not involve derivatives of $\epsilon$.
The boldface notation denotes contraction with transverse gamma matrices:
\be
\boldsymbol{ \hat f} \equiv \tfrac{1}{4!} \hat f_{abcd} \gamma^{abcd}\,,\quad
\boldsymbol{ \hat f}_{A} \equiv \tfrac{1}{3!} \hat f_{Aabc} \gamma^{abc}\,,\quad
\boldsymbol\lambda \equiv \tfrac{1}{4!} \lambda_{abcd} \gamma^{abcd} \,,
\ee
where $\lambda_{abcd}$ is defined by
\be
\lambda_{abcd} \equiv
\lambdaone_{abcd} + \tfrac18\,\bar\psi_{+e}\big\{\gamma^{ef},\gamma_{abcd}\big\}\psi_{+f}
=
\lambdaone_{abcd} - \tfrac14 P^{(-)}{}_{abcd}{}^{efgh} \bar\psi_{+e} \gamma_{fg} \psi_{+h} \,.
\label{deflambda}
\ee

\paragraph{The supersymmetry transformation of the auxiliary field} 

It turns out that it is $\lambda_{abcd}$ which transforms naturally under the symmetries of the non-relativistic theory. 
The supersymmetry transformation of $\lambda_{abcd}$ follows from its definition \eqref{deflambda} using the algebraic equation of motion, and thus takes the form:
\be
\begin{split} 
\delta'_\epsilon \lambda_{abcd} & =
 c^3 \left( \delta'_\epsilon \hat f^{(-)}_{abcd}
+ \tfrac{1}{24} \tau^A{}_\mu \bar{\mathcal{E}}(\psi_+)^\mu \gamma_A \gamma_{abcd} \epsilon_+
+ \tfrac{1}{16} e^e{}_\mu \bar{\mathcal{E}}(\psi_+)^\mu \gamma_{abcd} \gamma_e \epsilon_-
\right)
\\ & \quad
- \tfrac12 P^{(-)}{}_{abcd}{}^{efgh} \bar\psi_{+e} \gamma_{fg} ( \delta'_\epsilon \psi_{+h})\,,
\end{split}
\label{deltalambdadesired}
\ee
where $\mathcal{E}(\psi_+)^\mu$ is the equation of motion of $\psi_{+\mu}$, and $\delta'_\epsilon$ denotes the supersymmetry transformations obtained after using the trivial symmetry ($\hat f^{(-)}{}_{abcd}$ only contains $\psi_{-a}$ whose transformations are unaffected by this redefinition).

Unpacking \eqref{deltalambdadesired} is reasonably complicated, and we provide more details of the technicalities involved in appendix \ref{app_lambda}.
Firstly it can be checked (using the finite equations of motion of the fermions which are obtained in section \ref{limit_action_eom} below) that the divergent part of the supersymmetry transformation is guaranteed to vanish once we impose the constraints \eqref{geoconstraints} or \eqref{halfsusy}.
To obtain the finite transformation, we need to know both the equations of motion and the supersymmetry transformations at subleading order.
It turns out to be sufficient for our purposes to determine these via a `linearised' calculation. 
This calculation firstly shows that $\lambda_{abcd}$ transforms supercovariantly, i.e., without any derivatives of the supersymmetry parameter, as befits a bosonic field, while $\lambdaone_{abcd}$ does not.
We then find that the finite transformation involves natural combinations of gravitino derivatives which can be viewed as arising from certain covariant gravitino curvatures, to be introduced below. 
At least in the theory with maximal supersymmetry, when the constraints \eqref{geoconstraints} are imposed, we can then invoke covariance to write down the full non-linear transformation.
This can be found in appendix \ref{app_lambda}, and we will review it when we analyse the realisation of maximal supersymmetry in section \ref{maxSUSYsection}.

\subsection{The limit of the action and equations of motion}
\label{limit_action_eom}

\newcommand{\Rl}{\mathring{R}}
\newcommand{\omegal}{\mathring{\omega}}
\newcommand{\Dl}{\mathring{D}}
\newcommand{\bl}{\mathring{b}}

\paragraph{The geometry determined by the limit} 

To formulate the action and equations of motion, we will make use of the bosonic parts of the supercovariant connections $\hat \omega$ arising in the supersymmetry transformations \eqref{finiteSUSY}, which we define as  
\begin{subequations}\label{connections_nohats_limit}%
\begin{align}
 \omegal_\mu{}^{AB} & =  T_{\mu}{}^{[AB]} - \tfrac12 \tau^C{}_\mu  T^{AB}{}_C \,,\\
 \omegal_\mu{}^{ab} & =  \Omega_\mu{}^{[ab]} - \tfrac12 e_{c\mu}  \Omega^{abc} + \tfrac14 \tau_{A\mu} \epsilon^{ABC} {f}_{BC}{}^{ab}+\tfrac13 e^{[a}{}_\mu  T^{b] A}{}_A\,,
 \label{connections_nohats_ab}\\
 \omegal_\mu{}^{Aa} & = \tfrac12  \Omega_\mu{}^{Aa} - \tfrac12 e_{\mu b}  \Omega^{Aab}
- \tfrac{1}{4} e^b{}{}_\mu { f}^a{}_{b BC} \epsilon^{BCA}
- \tau^A{}_\mu\tfrac13  { f}_{012}{}^a \,,\\
 \bl_\mu & = \tfrac13 e^a{}_\mu  T_a{}^A{}_A \,,
\end{align}
\end{subequations}
\begin{subequations} 
\label{curvatures_fromlimit}%
together with associated curvatures
\begin{align}
\Rl_{\mu\nu}{}^{AB} & = 2 \partial_{[\mu} \omegal_{\nu]}{}^{AB} + 2 \omegal_{[\mu}{}^{AC} \omegal_{\nu]C}{}^B \,,\\
\Rl_{\mu\nu}{}^{ab} &= 2\,\partial_{[\mu}\omegal_{\nu]}{}^{ab} + 2\,\omegal_{[\mu}{}^{ac}\omegal_{\nu]c}{}^b{} \,,\\
\Rl_{\mu\nu}{}^{Aa} &= 2\,\partial_{[\mu}\omegal_{\nu]}{}^{Aa} + 2\,\omegal_{[\mu}{}^{AB}\omegal_{\nu]B}{}^a + 2\,\omegal_{[\mu}{}^{ab}\omegal_{\nu]}{}^{A}{}_{b} + 3\,\bl_{[\mu}\omegal_{\nu]}{}^{Aa}\,,\\
\Rl_{\mu\nu} & = 2 \partial_{[\mu} \bl_{\nu]} \,.
\end{align} 
\end{subequations} 
In the next section, we will construct the geometry of the Membrane-Newton Cartan supergravity theory from the ground up, and discuss the properties of these connections and curvatures in more detail.
In particular we will see that when any of the objects appearing in the maximal supersymmetry constraints \eqref{geoconstraints} are non-zero, the true covariant derivatives and curvatures of the geometry will require modifications.

\paragraph{The action of the non-relativistic supergravity} 

We now present the full action of the non-relativistic theory, corresponding to $S'_0$ of \eqref{Snew0}.
This follows from a lengthy but straightforward calculation on inserting the field redefinitions \eqref{ExpBos} and \eqref{ExpFer} into the original Lagrangian \eqref{RLag}. 
The result is:
\begin{subequations}
\label{Lag}%
\begin{align}
S_{\text{non-rel}}  = \int \dd^{11} x \,\left(  \Omega[ \mathcal{L}_{R}  + \mathcal{L}_{\psi\psi}] + \mathcal{L}_{\text{CS}} \right)\,, 
\end{align}
where the bosonic contributions in the Lagrangian follow from:
\begin{align}
\mathcal{L}_R & = \Rl_{ab}{}^{ab} - T_{a\{ AB \} } T^{a\{ A B\} } - \tfrac{1}{12} f_{abcA} f^{abcA} 
- \tfrac{2}{4!} \lambda_{abcd} \hat f^{(-) abcd} 
\,,
\label{LR_limit} 
\\[5pt]
\mathcal{L}_{\text{CS}} & = \tfrac{1}{144^2} \epsilon^{\mu_1 \dots \mu_{11}} f_{\mu_1 \dots \mu_4} f_{\mu_5 \dots \mu_8} c_{\mu_9 \mu_{10} \mu_{11}} \,,
\end{align}
with curvature scalar $\Rl_{ab}{}^{ab} \equiv e^\mu{}_a e^\nu{}_b \Rl_{\mu\nu}{}^{ab}$, 
and (in addition to the fermions appearing in $\hat f^{(-)}{}_{abcd}$ of \eqref{LR_limit}) we have the following terms bilinear in the fermions:
\begin{align} 
\notag
 \mathcal{L}_{\psi\psi} & =
-\bar \psi_{-a} \gamma^{abc} 2  \Dl_{[b} \psi_{+c]} 
- \bar \psi_{-A} \gamma^{Aab} 2 \Dl_{[a} \psi_{-b]}
- 2 \bar \psi_{-a} \gamma^{Aab} 2 \Dl_{[b} \psi_{-A]} 
- \bar \psi_{+a} \gamma^{abc} 2 \Dl_{[b} \psi_{-c]} 
\\\notag & \quad
-2 T^{a\{AB\}} \bar \psi_{- A} \gamma_B \psi_{- a} 
\\\notag & \quad
+ T_{ab}{}^A \Big(
-  \tfrac12 \epsilon_A{}^{BC} \bar\psi_{- B}  \gamma^{ab} \psi_{- C} 
+  \tfrac{1}{2} \bar\psi_{+ c}  \gamma^{abcd} \gamma_A \psi_{+d} 
+  \bar  \psi_{+}^a  \gamma_A \psi_{+}^b 
\\\notag & \quad\qquad\qquad
+ \tfrac{1}{2}
 ( \bar \psi_{+A} \gamma^{abc} \psi_{-c} - \bar \psi_{-A} \gamma^{abc} \psi_{+c} )
- \epsilon_{ABC} ( \bar \psi_{+}{}^B\gamma^{Ca} \psi_{-}{}^b+ \bar \psi_{-}{}^B\gamma^{Ca} \psi_{+}{}^b)
\\\notag & \quad\qquad\qquad
- \tfrac12 ( \bar\psi_{+c} \gamma_{AB} \gamma^{abc} \psi_-^B - 2 \bar\psi_{+}^a \gamma^b \psi_{-A} 
+\bar\psi_{-c} \gamma_{AB} \gamma^{abc} \psi_+^B - 2 \bar\psi_{-}^{a} \gamma^b \psi_{+A} 
)\Big)
\\\notag & \quad
+ \tfrac{1}{48} f_{Aabc} \Big( 4 ( \bar \psi_{+d} \gamma^{Aabcde} \psi_{-e} + \bar \psi_{-d} \gamma^{Aabcde} \psi_{+e} ) 
+ 8 \bar\psi_{- d} \gamma^{abcd} \gamma^{AB} \psi_{- B}
\\\notag & \qquad\qquad\qquad\quad
+ 24 ( \bar \psi_{-}{}^A \gamma^{ab} \psi_{-}{}^c + \bar\psi_{+}{}^a \gamma^b\gamma^A \psi_-{}^c + \bar\psi_{-}{}^a \gamma^b\gamma^A \psi_+{}^c)
\Big)
\\\notag & \quad
+   \tfrac{1}{48} f_{abcd} \Big( 
 2 ( \bar \psi_{+ A} \gamma^{Aabcde} \psi_{-e} + \bar \psi_{- A} \gamma^{Aabcde} \psi_{+e} )
+ \bar\psi_{- A} \gamma^{abcd} \gamma^{AB} \psi_{- B} \Big)
\\ & 
\quad
+\tfrac12 f^{(+)}_{abcd} \bar \psi_{+}^a \gamma^{bc} \psi^d_+ 
-\tfrac12 f^{(-)}_{abcd} \bar \psi_{+}^a \gamma^{bc} \psi^d_+ 
\,.\label{Lpsipsi}  
\end{align}
\end{subequations}
The covariant derivatives of the fermions are defined by
\begin{subequations}
\label{Dpsi_limit}%
\begin{align}
\Dl_\mu \psi_{-\nu} &  = \left( \partial_\mu  + \tfrac14 \omegal_\mu{}^{AB} \gamma_{AB} + \tfrac14 \omegal_\mu{}^{ab} \gamma_{ab} - \tfrac12 \bl_\mu \right) \psi_{-\nu} \,,\\
\Dl_\mu \psi_{+\nu} &  = \left( \partial_\mu  + \tfrac14 \omegal_\mu{}^{AB} \gamma_{AB} + \tfrac14 \omegal_\mu{}^{ab} \gamma_{ab} +  \bl_\mu \right) \psi_{+\nu} + \tfrac12 \omegal_\mu{}^{Aa} \gamma_{Aa} \psi_{-\nu} \,.
\end{align}
\end{subequations}
Note that by $\Dl_a \psi_{\pm b}$ we mean $\Dl_a \psi_{\pm b} \equiv e^\mu{}_a e^\nu{}_b \Dl_{\mu} \psi_{\pm \nu}$, and so on.

The form of the bosonic part of the action follows from the expansion of the relativistic action followed by an integration by parts.
Everything appearing in the bilinear terms of \eqref{Lpsipsi} comes directly from the expansion of the relativistic Lagrangian, with the exception of the very final term, which arises on making the redefinition \eqref{deflambda} of the original auxiliary field $\lambdaone_{abcd}$.


\paragraph{Bosonic equations of motion following from the action} 

Let us first consider the bosonic equations of motion with vanishing fermions.
These were written down in \cite{Blair:2021waq} in terms of a certain affine connection which we do not make use of in this paper.
We will therefore rederive the equations of motion by explicitly varying \eqref{Lag}.
First of all, we remind the reader that $f^{(-)}{}_{abcd} = 0$ is the equation of motion of $\lambda_{abcd}$.

Next we consider the variation with respect to the vielbein variables. 
Ignoring fermionic contributions, this leads to:
\be
\begin{split}
\delta_{\tau,e} S = \int \dd^{11}x\, & \Omega \,\Big(
\delta e^\mu{}_a \left(
2 \Rl_{\mu b}{}^{a b} - e^{\mu}{}_{a} \mathcal{L}_{R}
- \tfrac12 f_{\mu bc A} f^{abc A} 
- \tfrac13 \lambda^{abcd} f_{\mu bcd} 
\right) 
\\ & \qquad
+ \delta \tau^\mu{}_A 
\left(
- \tau^A{}_\mu \mathcal{L}_R 
- \tfrac16 f_{abc \mu} f^{abcA} 
\right) 
\Big) 
\\ & 
 \qquad + \delta_{\tau,e} \omega_{\mu}{}^{ab} T_{ab}{}^A \tau^\mu{}_A  
- 2 \delta_{\tau,e} (T_{a\{ AB \}} ) T^{a\{A B\}} \Big)\,.
\end{split} 
\label{deltaStaue}
\ee
We comment that if the intrinsic torsion $T_{ab}{}^A$ was to vanish, then there would be no terms involving the variation of the spin connection $\omega_\mu{}^{ab}$, i.e. its variation would be a total derivative.
In fact both sets of constraints \eqref{geoconstraints} and \eqref{halfsusy} that we will consider impose that $T_{ab}{}^A$ is zero, so compatibility with supersymmetry implies that we could always ignore the spin connection variation in \eqref{deltaStaue}. Let us nevertheless continue without imposing any constraints (which must of course only be imposed after variation and not directly at the level of the action).
Projecting onto longitudinal trace, longitudinal traceless and transverse components, the equations of motion of $\tau^\mu{}_A$ can be written as:\footnote{Note that the traceless projection \eqref{tauEom2} is given in \cite{Blair:2021waq} with incorrect coefficients: equation (3.18) of that reference should read: 
$0  =   H^{\mu\nu}  \Dl_{\mu }T_{\nu \{AB\}}+ T^{aC}{}_C T_{a\{ AB\}}  -2 T_{a[C (A]} T^{a}{}_ {\{B) D\}}\eta^{CD} 
     +  \tfrac{1}{12} f_{\{A}{}^{ abc} f_{B \} abc} 
     + \tfrac14 f_{ab}{}^{CD} \epsilon_{CD\{A} T^{ab}{}_{B\} }
$.}
\begin{subequations}\label{tauEom}%
\begin{align}
\Rl_{ab}{}^{ab}  & =  T^{a\{AB\}} T_{a\{AB\}} 
+ \tfrac{1}{12} \epsilon_{ABC} T_{ab}{}^A f^{BCab} + \tfrac{1}{36} f_{Aabc} f^{Aabc}
\label{tauEom1} \,,\\
\label{tauEom2}
 0  &= e^{\mu a} \Dl_{\mu} T_{a\{AB\}}  + \tfrac{1}{12} f_{\{A}{}^{abc} f_{B\} abc} \,,\\
0 & = e^\nu{}_b  \Dl_\nu T^{ab}{}_A + T_{b\{A B\}} T^{ab}{}^B  -\tfrac16 f^{abcd} f_{Abcd} + \tfrac12 \epsilon_{ABC} f^{abcB} T_{bc}{}^C  \,.
\label{tauEom3}
\end{align}
\end{subequations}
\begin{subequations}\label{EEom}%
Similarly, the independent components of the equations of motion of $e^\mu{}_a$ are:
\begin{align}
0 = & \nonumber
\Rl_{Aca}{}^c +\tfrac14 f^{Babc} f_{bcAB} - \tfrac16 \lambda_{acde} f_A{}^{cde} 
+ \tfrac{1}{18} f_{012}{}^b T_{ab}{}_A - \tfrac{1}{12} \Omega_{AB}{}^b T_{ab}{}^B \\ 
& 	+ \tau^{\nu B} \Dl_\nu T_{a \{ AB \} } +  \Omega^B{}_c{}^c T_{a\{AB\}}  
- \Omega^B{}_{(ab)} T^b{}_{\{AB\}} - \tfrac14 \epsilon^{BCD} f_{CDab} T^b{}_{\{AB\}} 
\,,
\label{EEom1}
\\
0 \nonumber  = &2 \Rl_{(a|c|b)}{}^c - \delta_{ab} \mathcal{L}_R - \tfrac12 f_{acdB} f_b{}^{cdB} - \tfrac13 \lambda_{acde} f_b{}^{cde} 
- 2 T_a{}^{\{BC\}} T_{b\{BC\} } 
\\ & \qquad+ \Omega_{A}{}^c{}_{(a} T_{b)c}{}^A + \Omega_{A(a}{}^c T_{b)c}{}^A \,.
\label{EEom2}
\end{align} 
The covariant derivatives used here are defined explicitly as follows: 
\begin{align} 
\Dl_\mu T_{ab}{}^A & = \partial_\mu T_{ab}{}^A + 2 \omegal_{\mu}{}^c{}_{[a} T_{b]c}{}^A + \omegal_\mu{}^A{}_B T_{ab}{}^B - 2 \bl_\mu T_{ab}{}^A \,,\\
\nonumber
\Dl_\mu T_{a\{AB\}} & = \partial_\mu T_{a\{AB\}} - \omegal_\mu{}^b{}_a T_{b\{AB\}}- \omegal_\mu{}^C{}_{A} T_{a\{CB\}} - \omegal_\mu{}^C{}_{B} T_{a\{AC\}} 
\\ &\qquad + \omegal_{\mu \{A}{}^b T_{|ab|B\}} - \tfrac12 \bl_\mu T_{a\{AB\}}
\,.
\end{align} 
The bare $\Omega_{\mu\nu}{}^a$ terms appearing in \eqref{EEom1} and \eqref{EEom2} may appear strange, but reflect the fact that the curvature $\Rl_{\mu\nu}{}^{ab}$ and the covariant derivatives $\Dl$ are not fully covariant when $T_{ab}{}^A$ and $T_{a}{}^{\{AB\}}$ are non-zero -- we will discuss this later and show how to obtain fully covariant definitions which will incorporate these $\Omega_{\mu\nu}{}^a$ terms when we construct the supersymmetric Membrane Newton-Cartan geometry from first principles. 
Finally, note that the term $\lambda_{acde} f_b{}^{cde}$ in \eqref{EEom2} is automatically symmetric as $\lambda$ and $f$ are oppositely projected (using $f^{(-)}_{abcd} = 0$).
\end{subequations} 

Lastly, we have the equation of motion of the three-form: 
\be
\begin{split} 
&
\partial_\sigma \left(
\Omega \left[
4 \tau^{[\mu}{}_A e^{\nu}{}_b e^\rho{}_c e^{\sigma]}{}_d f^{Abcd} 
- 6 \epsilon^{ABC} e^{[\mu}{}_a e^\nu{}_b \tau^\rho{}_A \tau^{\sigma]}{}_B T^{ab}{}_C 
+2 e^{[\mu}{}_a e^\nu{}_b e^\rho{}_c e^{\sigma]}{}_d \lambda^{abcd} 
\right] 
\right) 
\\ & + \tfrac12 \tfrac{1}{4!4!} \epsilon^{\mu\nu\rho \sigma_1 \dots \sigma_8} f_{\sigma_1 \dots \sigma_4} f_{\sigma_5 \dots \sigma_8} 
=0 \,.
\end{split}
\label{EomChere}
\ee
Note that the total longitudinal projection of \eqref{EomChere} implies 
\be
\tfrac12 T_{ab}{}^A T^{ab}{}_A = - \tfrac12 \tfrac{1}{4!4!} \epsilon^{a_1 \dots a_8} f_{a_1 \dots a_4} f_{a_5 \dots a_8} 
= - \tfrac{1}{2} \tfrac{1}{4!} f^{(+)}{}_{abcd} f^{(+)abcd} \,,
\ee
using $f^{(-)}{}_{abcd} = 0$ in the second equality.
This means $T_{ab}{}^A = 0\Rightarrow f^{(+)}{}_{abcd} = 0$.

\paragraph{Fermionic equations of motion following from the action} 

\newcommand{\rl}{\mathring{r}}

Now we discuss the fermionic equations of motion. 
It's convenient to write the variation of the action \eqref{Lag} with respect to the gravitinos as 
\be
\delta S_{\text{non-rel}} = \int \dd^{11} x \,\Omega ( \delta \bar \psi_{+ \mu} \mathcal{E}(\psi_+)^\mu + \delta \bar \psi_{-\mu} \mathcal{E}(\psi_-)^\mu ) \,,
\label{deltaSgrav} 
\ee
up to total derivatives.
The transverse and longitudinal projections of the variation with respect to $\psi_{+\mu}$ are:
\begin{subequations}\label{eom_psiplus}%
\begin{align} 
\mathcal{E}(\psi_+)^a  & = - 2 \gamma^{abc} \rl_{- bc } 
+ T_{bc}{}^A \left( \tfrac12  \gamma^a \gamma^{bc} \gamma^B\gamma_A \psi_{-B}
+ \gamma_A ( \gamma^{abcd} \psi_{+d} + 2 \delta^{ab} \psi_+^c ) \right)
\notag \\ & \quad
-\tfrac{1}{6} f_{Abcd} \gamma^A (\gamma^{abcde} \psi_{-e} + 6 \delta^{ab} \gamma^c \psi_{-}{}^d ) \notag \\ & \quad 
- \tfrac{1}{24} f_{bcde} \gamma^A \gamma^{bcdea} \psi_{-A} 
+ ( f^{(+)a}{}_{bcd} - f^{(-)a}{}_{bcd} ) \gamma^{bc} \psi_{+d} \,,
\\
\mathcal{E}(\psi_+)^A  & =
 -\tfrac12 T_{ab}{}^B \gamma^A \gamma_B \gamma^c \gamma^{ab} \psi_{-c} + \tfrac{1}{24} f_{abcd} \gamma^A \gamma^{abcde} \psi_{-e} \,.
\end{align}
\end{subequations} 
The transverse and longitudinal projections of the variation with respect to $\psi_{-\mu}$ are:
\begin{subequations}\label{eom_psiminus}%
\begin{align} 
\mathcal{E}(\psi_-)^a  & = -2 \gamma^{abc} \rl_{+ab} 
\notag
\\ & \quad 
+ \tfrac12 T_{bc}{}^A ( 3 \eta_{AB} - \gamma_{AB})  \gamma^{abc} \psi_{+B} 
+ T^a{}_b{}^A \gamma_A \gamma^B \psi_{+B} 
\notag
\\ & \quad + 2 T_{b}{}^{\{AB\}} \gamma^a \gamma^b \gamma_A \psi_{-B}  
- \tfrac16 f_{Abcd} ( \gamma^A \gamma^{abcde} \psi_{+e} + \gamma^{AB} \gamma^{bcd} \psi_{-B} ) 
\notag
\\ & \quad - \tfrac12 f_{A}{}^a{}_{bc} ( \gamma^{bc} \psi_-^A - 2 \gamma^b \gamma^A \psi_{+}^c) 
- \tfrac{1}{24} f_{bcde} \gamma^A \gamma^{abcde} \psi_{+A} \,,
\\
\mathcal{E}(\psi_-)^A & =- 2 \gamma^{Aab} \rl_{-ab}  \notag + T_{ab}{}^B \left( \gamma^{AC} \gamma_B \gamma^{ab} \psi_{-C} 
+ \gamma_B \gamma^A \gamma^a \psi_+^b - ( \tfrac32 \delta^A_B +  \tfrac12 \gamma^A{}_B ) \gamma^{abc} \psi_{+c} \right)\\
& \quad- 2 T^{a\{AB\}}\gamma_B \gamma_a \gamma^b \psi_-^b 
 + \tfrac16 f_{Babc} ( \gamma^{AB} \gamma^{abcd} \psi_{-d} + 3 \eta^{AB} \gamma^{ab} \psi_{-}{}^c ) \notag
\\ & \quad + \tfrac{1}{24} f_{abcd} \left( \gamma^A \gamma^{abcde} \psi_{+e} + \gamma^{AB} \gamma^{abcd} \psi_{-B} \right) \,.
\end{align}
\end{subequations} 
The equations of motion are of course then $\mathcal{E}(\psi_+)^a = \mathcal{E}(\psi_+)^A = \mathcal{E}(\psi_-)^a = \mathcal{E}(\psi_-)^A = 0$.
Here we defined certain `gravitino curvatures':
\be
\rl_{-ab} \equiv 2 \Dl_{[a} \psi_{-b]} \,,\quad
\rl_{+ab} \equiv 2 \Dl_{[a} \psi_{+b]} + \tfrac13 \gamma_{[a} \gamma^A ( \Dl_{b]} \psi_{-A} - \Dl_{|A|} \psi_{-b]} ) - \tfrac14 \boldsymbol{\lambda} \gamma_{[a} \psi_{-b]}  \,.
\label{gravitinocurvatures_limit} 
\ee
which we will encounter again later.

The invariance of the action under the fermionic shift symmetries \eqref{fermionicshift} can be verified directly using \eqref{deltaSgrav} and the above expressions.
We trivially have $\bar \rho_{A+} \mathcal{E}(\psi_+)^A = 0$.
Invariance under the shifts with the parameter $\eta_-$ follows from the fact that $\gamma_A \mathcal{E}(\psi_-)^A = \tfrac12 \gamma_a \mathcal{E}(\psi_+)^a$.

\paragraph{Equations of motion not following from the action}

A common feature of the non-relativistic theories obtained by limits of (super)gravity is that the equations of motion following from the non-relativistic action do not constitute the full set of equations of motion of the theory.
In particular, the Poisson equation for a Newtonian gravitational potential is an example of such an equation.\footnote{When going beyond limits and using expansions (with additional fields and symmetries), one can construct an action that reproduces the Poisson equation for the Newton potential. This has been developed in the context of general relativity, see \cite{Hartong:2022lsy} for a recent review.} 

Based on what happens in the string Newton-Cartan theory \cite{Bergshoeff:2021bmc}, we identify the Newtonian potential with the longitudinal part of the three-form $c_{012}=\tau^\mu{}_0 \tau^\nu{}_1 \tau^\rho{}_2 \,c_{\mu\nu\rho}$. 
It is easy to check that there is no term involving a Laplacian of this component appearing in the equations of motion derived from the action. 
Instead, the Poisson equation can be obtained from the variation of the subleading action, and it transforms into the other equations of motion under boosts.
The Poisson equation for the 11-dimensional  supergravity was obtained in this way in \cite{Blair:2021waq}, and has a simple geometric expression in terms of the curvatures we have introduced:
\be
\Rl_{AB}{}^{AB} + \Rl_{Aa}{}^{Aa} + \tfrac{1}{4!} \lambda_{abcd} \lambda^{abcd}=0 \,,
\label{Poisson}
\ee
involving the traces of both the longitudinal Lorentz and boost curvatures.
Expanding the definitions of these curvatures one indeed finds a transverse Laplacian of $c_{012}$.

The fact that this equation does not appear as an equation of motion following from the variation of the action can be traced back to the presence of emergent local dilatations in of the non-relativistic action, which effectively removes one degree of freedom that would otherwise be present. 
\begin{subequations} 
\label{superfish}%
Similarly, the variation of the subleading action with respect to the fermionic shift symmetries implies there should likewise be additional fermionic equations of motion, which we can call super-Poisson equations.
In appendix \ref{RS_subleading}, we determine the form of the super-Poisson equations \eqref{superfish} by obtaining the variation with respect to the fermionic shift symmetry of the subleading gravitino kinetic terms, assuming the bosonic fields are constant. 
This implies that we should have super-Poisson equations of the form:
\begin{align} 
\gamma^a \rl_{+a} & + \dots = 0 \,,
\label{superfish1} \\
\gamma^a \rl_{-Aa}&  + \dots = 0 \,,
\label{superfish2} 
\end{align}
with the ellipses denoting possible covariant terms involving the geometric quantities which are set to zero in the maximal supersymmetry constraints \eqref{geoconstraints}, and
\end{subequations} 
\begin{subequations}
\label{gravitinocurvatures_limit2}%
where the following gravitino curvatures appear
\begin{align}
\rl_{+a} & = \gamma^A (\Dl_{A} \psi_{+a} - \Dl_{a} \psi_{+A} )- \tfrac14 \gamma_a \gamma^{BC} \Dl_B \psi_{-C} \,,\\
\rl_{-Aa} & = \Dl_A \psi_{-a} - \Dl_a \psi_{-A} - \tfrac{1}{12} \gamma_A {\boldsymbol{\lambda}} \psi_{+a} + \tfrac18 \boldsymbol{\lambda} \gamma_a \psi_{-A} - \tfrac13 \gamma_A \gamma^B (\Dl_B\psi_{-a} - \Dl_a \psi_{-B} )\,.
\label{rlAa}
\end{align} 
\end{subequations} 
The geometric origin of these curvatures will be elucidated below. 
The analysis of appendix \ref{RS_subleading} suffices to determine the form of the terms involving derivatives of the gravitinos in \eqref{gravitinocurvatures_limit2}, and we can then invoke covariance to replace $\partial_\mu$ with $\Dl_\mu$. The terms involving $\lambda_{abcd}$ in \eqref{rlAa} are added to obtain a supercovariant expression, given the supersymmetry transformation rules \eqref{finiteSUSY}.

\section{The geometry of the non-relativistic supergravity}
\label{geometry_section}%
Now we will study in more detail the geometry underlying the non-relativistic theory appearing in the limit described in the previous section.
We first outline a general approach to the bosonic connections and curvatures of the most general Membrane Newton-Cartan geometry from first principles.
Then, we specify these geometrical objects to the cases where we obtain a maximal or half-maximal supersymmetric theory in the limit, by imposing the constraints \eqref{geoconstraints} or \eqref{halfsusy}, respectively. In the former case, we also introduce fermionic curvatures which will appear when we consider the tower of constraints arising from repeated supersymmetry variations of the initial set \eqref{geoconstraints}.

\subsection{Bosonic symmetries, connections and curvatures} 
\label{sec_bosconns}

\paragraph{Bosonic symmetries}
First let's collect the bosonic symmetries (excluding diffeomorphisms) of the non-relativistic geometry.
Let $\Lambda^A{}_B$ and $\Lambda^a{}_b$ be parameters for SO$(1,2)$ and SO$(8)$ rotations and $\lambda_A{}^a$ parameters for boosts.
These are the symmetries inherited from the Lorentz symmetry of the original relativistic theory.
The non-relativistic supergravity also features an emergent local scaling symmetry, referred to as dilatations.
Let $\alpha$ denote the parameter of this symmetry.
The transformations of the bosons under these symmetries are as follows:
\begin{subequations}\label{bosonictransf}%
\begin{align}
\delta \tau^A{}_\mu  &= \alpha \,\tau^A{}_\mu- \Lambda^A{}_B \tau^B{}_\mu \,, & \delta \tau^\mu{}_A &= -\alpha \,\tau^\mu{}_A+\Lambda^B{}_A \tau^\mu{}_B  - \lambda_A{}^a e^\mu{}_a \,,\\
 \delta e^a{}_\mu  &= -\tfrac12 \alpha \,e^a{}_\mu- \Lambda^a{}_b e^b{}_\mu + \lambda_A{}^a \tau^A{}_\mu \,,&
\delta e^\mu{}_a  &= \tfrac12 \alpha \,e^\mu{}_a+\Lambda^b{}_a e^\mu{}_b\,,\\
\delta c_{\mu\nu\rho} &= 3 \partial_{[\mu} \theta_{\nu\rho]}- 3 \,\epsilon_{ABC} \lambda^A{}_a e^a{}_{[\mu} \tau^B{}_{\nu} \tau^C{}_{\rho]} \,, \\
\delta \lambda_{abcd} & = - \alpha\, \lambda_{abcd}+ 4 \Lambda^e{}_{[a} \lambda_{|e|bcd]} & \quad & \hspace{-8.2em} +  \tfrac{1}{4!} P^{(-)}_{abcd}{}^{efgh} 4 \lambda^A{}_{[e} f_{|A|fgh]} \,,
\end{align}
where we also have $\theta_{\mu\nu}$ appearing as the gauge symmetry parameter of the three-form.
The fermionic fields transform as:
\begin{align}
\label{bosonictransf_ferminus}
\delta \psi_{-\mu}& = + \tfrac12 \alpha \psi_{-\mu}- \tfrac14 \Lambda^A{}_B \gamma_A{}^B \psi_{-\mu} - \tfrac14 \Lambda^a{}_b \gamma_a{}^b \psi_{-\mu}\,,\\
\delta \psi_{+\mu}& = - \alpha \psi_{+\mu} - \tfrac14 \Lambda^A{}_B \gamma_A{}^B \psi_{+\mu} - \tfrac14 \Lambda^a{}_b \gamma_a{}^b \psi_{+\mu}- \tfrac12 \lambda_A{}^a \gamma^A{}_a \psi_{-\mu}  \,.
\label{bosonictransf_ferplus}
\end{align}
\end{subequations}
The dilatation symmetry will provide a useful book-keeping device for us.
We will say that some field $X$ has dilatation weight $w$ if it transforms as $\delta X = w \alpha X$ under dilatations.

\paragraph{Bosonic connections}
We now introduce spin connections $\big\{b_\mu,\omega_\mu{}^{AB},\omega_\mu{}^{ab},\omega_\mu{}^{Aa}\big\}$ for dilatations, $\mathrm{SO}(1,2)$ rotations, $\mathrm{SO}(8)$ rotations and boosts, respectively.
Here $\omega_\mu{}^{AB}, \omega_{\mu}{}^{ab}$ and $b_\mu$ have dilatation weight zero, while $\omega_\mu{}^{Aa}$ has dilatation weight $-\tfrac32$.
Given a bosonic field $X^{Aa}$ of dilatation weight $w$, and transforming under boosts as $\delta X^{Aa} = \Lambda_{Bb} Y^{Bb Aa}$, where $Y^{BbAa}$ has dilatation weight $w+3/2$, we define a derivative that is covariant with respect to SO$(1,2)$, SO$(8)$, dilatation and boost transformations as
\be
D_\mu X^{Aa} = \partial_\mu X^{Aa} + \omega_\mu{}^A{}_B X^{Ba} + \omega_\mu{}^a{}_b X^{Ab} - w b_\mu X^{Aa} - \omega_\mu{}{}_{Bb} Y^{BbAa} \,.
\label{covderiv}
\ee
Similarly on a pair of fermions $\chi_\pm$, where $\chi_-$ has weight $w$ and $\chi_+$ has weight $w-3/2$, such that under boosts $\delta \chi_+ = -\tfrac12 \Lambda_A{}^a \gamma^A{}_a \chi_-$, we can define analogous covariant derivatives by:
\begin{subequations}
\label{Dfermion}%
\begin{align}
D_\mu \chi_- &  = \left( \partial_\mu  + \tfrac14 \omega_\mu{}^{AB} \gamma_{AB} + \tfrac14 \omega_\mu{}^{ab} \gamma_{ab} - w b_\mu \right) \chi_- \,,\\
D_\mu \chi_+ &  = \left( \partial_\mu  + \tfrac14 \omega_\mu{}^{AB} \gamma_{AB} + \tfrac14 \omega_\mu{}^{ab} \gamma_{ab} - (w-3/2) b_\mu \right) \chi_+ + \tfrac12 \omega_\mu{}^{Aa} \gamma_{Aa} \chi_- \,.
\end{align}
\end{subequations}
\begin{subequations}\label{connectiontransf}%
These derivatives will be covariant -- in the sense that the transformation of the derivatives will not involve any derivatives of the symmetry parameters -- provided that the connections transform as follows: 
  \begin{align}
    \delta b_\mu &= \partial_\mu \alpha + \dots  \,,\quad
    \delta \omega_\mu{}^{AB} = \partial_\mu\Lambda^{AB} + \dots \,,\\ 
    \delta \omega_\mu{}^{ab} &= \partial_\mu\Lambda^{ab} + \dots \,,\quad 
    \delta \omega_\mu{}^{Aa} = \partial_\mu\Lambda^{Aa} + \dots \,,
  \end{align}
\end{subequations}
where the ellipses denotes further terms which will depend on the symmetry parameters (but not their derivatives), the connections and geometric tensors. 

\begin{subequations}\label{firstordercurvatures}%
To determine explicit expressions for these connections, we write down natural `torsions' (or `first-order curvatures') analogous to Cartan's first structure equations in Riemannian geometry:\footnote{Unlike the first Cartan structure equations of Riemannian geometry, the equations \eqref{firstordercurvatures} contain covariantisations with respect to dilatations, as well as an extra equation for the three-form field.}
  \begin{align}
    \mathcal{T}_{\mu\nu}{}^A &\equiv 2 D_{[\mu} \tau^A{}_{\nu]} =  2\,\partial_{[\mu}\tau^A{}_{\nu]} + 2\,\omega_{[\mu}{}^A{}_B\tau^B{}_{\nu]} - 2\,b_{[\mu}\tau^A{}_{\nu]}\,, \\
    \mathcal{T}_{\mu\nu}{}^a &\equiv 2 D_{[\mu} e^a{}_{\nu]} = 2\,\partial_{[\mu}e^a{}_{\nu]} + 2\,\omega_{[\mu}{}^a{}_b e^b{}_{\nu]} + \,b_{[\mu}e^a{}_{\nu]} -2\,\omega_{[\mu A}{}^a\tau^A{}_{\nu]}\,, \\
    \mathcal{F}_{\mu\nu\rho\sigma} &\equiv 4 D_{[\mu}c_{\nu\rho\sigma]} =  4\,\partial_{[\mu}c_{\nu\rho\sigma]} + 12\,\epsilon_{ABC}\omega_{[\mu}{}^A{}_a e^a{}_\nu \tau^B{}_\rho \tau^C{}_{\sigma]}\,.
  \end{align}
\end{subequations}
Contracting the curved indices with inverse frame fields, it is easily seen that the following components of \eqref{firstordercurvatures} are independent of the connections: 
\be
\mathcal{T}_{ab}{}^A = T_{ab}{}^A \,,\quad
\mathcal{T}_{a}{}^{\{AB\}} = T_a{}^{\{AB\}} \,,\quad
\mathcal{F}_{abcd} = f_{abcd} \,,\quad
\mathcal{F}_{Aabc} = f_{Aabc} \,,
\label{intrinsictorsions}
\ee
and so carry intrinsic information about the geometry in the form of intrinsic torsion and four-form field strength components.
We immediately notice that this list contains all the tensors which appeared in the divergent part \eqref{divergferm} of the fermionic supersymmetry transformations.

Technically, the tensors appearing in the constraints \eqref{geoconstraints} are the supercovariant completions of \eqref{intrinsictorsions}: for simplicity we are restricting here to a purely bosonic discussion with the understanding that it is really the supercovariant versions that we make use of in the supersymmetric theory. This can be accommodated by adding appropriate gravitino bilinears to \eqref{firstordercurvatures}, leading to the supercovariant versions of the connections appearing in \eqref{connections}.

Depending on the constraints we impose, some or all of \eqref{intrinsictorsions} will be set to zero.
To solve for the connections, we will further set all remaining components of the torsions \eqref{firstordercurvatures} to zero.
\begin{subequations}
\label{omegaBW}%
These give `conventional constraints', which are identically solved by connections of the following form: 
\begin{align}
 \omega_\mu{}^{AB} & = \omegal_\mu{}^{AB} - 2 \tau^{[A}{}_\mu B^{B]}\ \,,\\
 \omega_\mu{}^{ab} & =  \omegal_\mu{}^{ab} \,, \\
 \omega_\mu{}^{Aa} & = \omegal_\mu{}^{Aa} - \tfrac12 e^a{}_\mu B^A + \tau_B{}_\mu W^{\{AB\} a}\,,\\
 b_\mu & = \bl_\mu + \tau^A{}_\mu B_A  \,,
\end{align}
in terms of the connections \eqref{connections_nohats_limit} obtained directly from the limit, and two unknown quantities $B^A$ and $W^{\{AB\}a}$ which reflect the fact that the connection components $b_A$ and $\omega^{\{AB\} a}$ are in fact not determined by the conventional constraints: indeed the first-order expressions \eqref{firstordercurvatures} simply do not contain these components of the connections.
\label{connections_nohats}
\end{subequations}This can also be phrased as the statement that the equations \eqref{firstordercurvatures} are invariant under shifts of the connections of the form\footnote{It can be useful to think about the shift in the dilatation connection as a special conformal transformation in view of its analogy with conformal gravity.  This is not the case with respect to the ambiguity in the boost connection $W^{\{AB\}a}$, which has no analogue in the Riemannian case.}
\begin{align}\label{connectionshifts}
  &b_\mu\to b_\mu + \tau^A{}_\mu \beta_A\,, \qquad \qquad\omega_\mu{}^{AB}\to\omega_\mu{}^{AB} -2\,\tau^{[A}{}_\mu \beta^{B]}\,,  \notag\\
  &\omega_\mu{}^{Aa}\to\omega_\mu{}^{Aa} -\tfrac12\,\beta^A e^a{}_\mu  + \tau_{B\mu} w^{\{AB\}a}\,,
\end{align}
where $\beta_A$, $w^{\{AB\}a}$ are arbitrary parameters, which act directly as shifts of $B^A$ and $W^{\{AB\}a}$ appearing in \eqref{omegaBW}.

In the non-relativistic theory, any connection in the equivalence class parametrized by $\big\{B_A,W_{\{AB\}a}\big\}$ offers in principle a valid choice.
It is clear however that the limit gives rise to the particular choice with $B_A = 0$ and $W_{\{AB\}}{}^a = 0$.
As they can not arise in the limit, all physical quantities must turn out to be independent of the choice of $B_A$ and $W_{\{AB\}}{}^a$.  
We will use this observation frequently below.

The transformation of the connections under an arbitrary variation of the bosonic fields can be derived from the conventional constraints. 
To remain agnostic about which components of the intrinsic torsions \eqref{intrinsictorsions} are non-zero, let
\be
(\Delta \tau)_{\mu\nu}{}^A \equiv 2 D_{[\mu} \delta \tau^A{}_{\nu]} -\delta \mathcal{T}_{\mu\nu}{}^A \,,\quad
( \Delta e)_{\mu\nu}{}^a \equiv 2 D_{[\mu} \delta e^a{}_{\nu]} \,,\quad
(\Delta c)_{\mu \nu \rho \sigma} \equiv 4 D_{[\mu} \delta c_{\nu\rho\sigma]} - \delta \mathcal{F}_{\mu\nu\rho\sigma} \,.
\ee
Then, varying \eqref{firstordercurvatures} and solving for the variations of the connections, we find
\begin{subequations}
\label{deltaomega}%
\begin{align}
\label{deltaomega_b}
\delta b_\mu & = \tau^A{}_\mu \tau^\nu{}_A \delta b_\nu + \tfrac13 e^a{}_\mu (\Delta \tau)_{aA}{}^A \,,\\
\label{deltaomega_AB}
\delta \omega_\mu{}^{AB} & = \tau^C{}_\mu \left(
(\Delta \tau)_{C}{}^{[AB]} -\tfrac12 (\Delta \tau)^{AB}{}_C 
- 2 \delta^{[A}_C \tau^{|\nu |B]} \delta b_\nu\right) 
+ e^a{}_\mu (\Delta \tau)_{a}{}^{[AB]} 
 \,,\\\notag
 \delta \omega_\mu{}^{ab} & =
 \tau^C{}_\mu \left(
(\Delta e)_C{}^{[ab]} + \tfrac14 \epsilon_C{}^{AB} ( \Delta c)^{ab}{}_{AB}
 \right) 
 \\ \label{deltaomega_ab}
 & \quad
 + e^c{}_\mu \left(
 (\Delta e)_{c}{}^{[ab]} -\tfrac12 (\Delta e)^{ab}{}_c 
 + \tfrac13 \delta^{[a}_c (\Delta \tau)^{b]}{}_A{}^A 
 \right)  \,,\\
 \delta \omega_{\mu}{}^{Aa} & = \notag
 \tfrac12 \tau^B{}_\mu ( \Delta e)_{B}{}^{Aa} 
 +
 \tfrac{1}{18} \tau^A{}_\mu \epsilon^{BCD} ( \Delta c)_{BCD}{}^a
 - e_b{}_\mu \left( (\Delta e)^{A(ab)} + \tfrac14  \epsilon^{ABC} (\Delta c)^{ab}{}_{BC} \right)
 \\ \label{deltaomega_Aa}
 & \quad
 - \tfrac12 e^a{}_\mu \tau^{\nu A} \delta b_\nu 
 + \tau_{B \mu} \tau^{\nu\{A} \delta \omega_{\nu}{}^{B\}a}
  \,.
\end{align}
\end{subequations} 
The projections $ \tau^{\mu A} \delta b_\mu$ and $\tau_{B \mu} \tau^{\nu\{A} \delta \omega_{\nu}{}^{B\}a}$ are not determined by this procedure, as they involve the variation of the parts of the connections that cannot be determined. 

\paragraph{Bosonic curvatures}

\newcommand{\modR}{\Delta R}
\begin{subequations}\label{secondordercurvatures}%
Next we define curvatures for the above connections by:
\vspace{-0.5em}
\begin{align}
  R_{\mu\nu} &= 2\,\partial_{[\mu}b_{\nu]} + \modR (\mathcal{T}, \mathcal{F})_{\mu\nu} \,,\\
  R_{\mu\nu}{}^{AB} &= 2\,\partial_{[\mu}\omega_{\nu]}{}^{AB} + 2\,\omega_{[\mu}{}^{AC}\omega_{\nu]C}{}^B{}+ \modR (\mathcal{T}, \mathcal{F})_{\mu\nu}{}^{AB}  \,,\\
  R_{\mu\nu}{}^{ab} &= 2\,\partial_{[\mu}\omega_{\nu]}{}^{ab} + 2\,\omega_{[\mu}{}^{ac}\omega_{\nu]c}{}^b{}+ \modR (\mathcal{T}, \mathcal{F})_{\mu\nu}{}^{ab}  \,,\label{Rmunuab}\\
  R_{\mu\nu}{}^{Aa} &= 2\,\partial_{[\mu}\omega_{\nu]}{}^{Aa} + 2\,\omega_{[\mu}{}^{AB}\omega_{\nu]B}{}^a + 2\,\omega_{[\mu}{}^{ab}\omega_{\nu]}{}^{A}{}_{b} + 3\,b_{[\mu}\omega_{\nu]}{}^{Aa}
  + \modR (\mathcal{T}, \mathcal{F})_{\mu\nu}{}^{Aa}\,,
\end{align}
\end{subequations}
where the terms $\modR(\mathcal{T},\mathcal{F})$ denote terms which are linear in the possible non-zero components of the intrinsic torsion tensors \eqref{intrinsictorsions}.
These arise because the transformations of the connections determined by \eqref{deltaomega} involve `non-covariant' pieces involving these torsions.
In general, we will not need the explicit expressions for these terms.
However, to illustrate the general principle let us consider the definition of $R_{\mu\nu}{}^{ab}$. 
We first need to compute the general variation \eqref{deltaomega_ab} of the $\mathrm{SO}(8)$ spin connection under the  bosonic symmetries \eqref{bosonictransf}, assuming the only non-zero components of the torsions are those of \eqref{intrinsictorsions}.
The result of this computation is 
\be
\begin{split}
  \delta\omega_\mu{}^{ab} & = \partial_\mu \Lambda^{ab} + \omega_\mu{}^{a}{}_c \Lambda^{cb} + \omega_\mu{}^b{}_c \Lambda^{ac} 
   +\tau_{A\mu}\big(2\,\lambda_B{}^{[a}T^{b]\{AB\}} +\tfrac12\,\epsilon^{ABC}\lambda_B{}^cf_{Cc}{}^{ab}\big)
   \\&\quad + \tfrac12\, e_{c\mu}\big(2\,\lambda_A{}^{[a}T^{b]cA} - \lambda_{A}{}^cT^{abA}\big) -\tfrac13 \, e^{[a}{}_\mu T^{b]cC}\lambda_{Cc}\,.
\end{split}
\label{deltaomega_ab_mod}
\ee
The first three terms here are what would be conventionally expected.
The remainder show that there is an additional transformation under boosts involving the intrinsic torsions \eqref{intrinsictorsions}.
From this we infer that the curvature \eqref{Rmunuab} requires the following terms: 
\be
\begin{split}
 \modR(\mathcal{T},\mathcal{F})_{\mu\nu}{}^{ab} & = 4\,\tau_{A[\mu}\omega_{\nu]B}{}^{[a}T^{b]\{AB\}} +\tau_{A[\mu}\epsilon^{ABC}\omega_{\nu]B}{}^cf_{Cc}{}^{ab}
   \\ & \quad  +2\,e_{c[\mu}\omega_{\nu]A}{}^{[a}T^{b]cA} - e_{c[\mu}\omega_{\nu]A}{}^cT^{abA} - \frac23\,e^{[a}{}_{[\mu}T^{b]cC}\omega_{\nu] Cc}\,,
\end{split}
\label{modRab} 
\ee
\begin{subequations}
in order to transform covariantly.
Using the general solution \eqref{omegaBW} for the connections, we can then relate the full curvature $R_{\mu\nu}{}^{ab}$ to the curvature we used to write down the action and equations of motion arising from the limit.  
\label{RtoR}%
We have:
\begin{align} 
2  R_{(a|c|b)}{}^c & \notag = 2 \Rl_{(a|c|b)}{}^c + \Omega_{A(a}{}^c T_{b)c}{}^A + \Omega_{A}{}^c{}_{(a} T_{b)c}{}^A 
\\ & \quad + \tfrac12 \epsilon_{ABC} T_{(a}{}^{cA} f_{b)c}{}^{BC} 
- \tfrac{1}{12} \delta_{ab} \epsilon_{ABC} T^{cd}{}^A f_{cd}{}^{BC} \,,\\
R_{ab}{}^{ab} & = \Rl_{ab}{}^{ab} - \tfrac{1}{12} \epsilon_{ABC} T_{ab}{}^A f^{abBC} \,,\\
\notag R_{Aca}{}^c & = \Rl_{Aca}{}^c 
+ \tfrac14 f^{Babc} f_{bcAB} + \tfrac{1}{18} f_{012}{}^b T^a{}_{bA} - \tfrac{1}{12} \Omega_{AB}{}^c T^a{}_c{}^B
\\ & \quad - \Omega^B{}_{(ab)} T^b{}_{\{AB\}} + \Omega^B{}_{b}{}^b T_{a\{AB\}} 
-\tfrac14 \epsilon^{BCD} f_{abCD} T{}^b{}_{\{AB\}}
\notag
\\ & \quad 
+ \tfrac72 B^B T_{a\{AB\}} - \tfrac16 W_{\{AB\}}{}^b T^a{}_b{}^B\,.
\label{RtoRAa}
\end{align} 
\end{subequations} 
Comparing \eqref{RtoR} with the transverse vielbein equations of motion \eqref{EEom1} and \eqref{EEom2}, we see that all bare $\Omega_{\mu\nu}{}^a$ terms appearing there are indeed absorbed into the definitions of the full curvatures.
The undetermined connection components that appear in \eqref{RtoRAa} cancel out if we also define the covariant derivative appearing in \eqref{EEom1} in terms of the general connections \eqref{omegaBW}; then this equation of motion is compactly written as $R_{Aca}{}^c - \tfrac16 \lambda_{abcd} f_A{}^{bcd} + D^B T_{a\{AB\}} = 0$.

\subsection{The geometry with maximal supersymmetry}
\label{maxSUSYconns}

In this subsection, we develop in more detail the geometry of the non-relativistic supergravity in the case where we impose the constraints \eqref{geoconstraints} and keep maximal supersymmetry. 
Imposing these constraints and simultaneously solving \eqref{intrinsictorsions} for the connection components can be achieved by setting all components of \eqref{intrinsictorsions} to zero:
\begin{align}\label{conventionalconstraints}
  &\mathcal{T}_{\mu\nu}{}^A = 0\,, && \mathcal{T}_{\mu\nu}{}^a = 0\,, && \mathcal{F}_{\mu\nu\rho\sigma} = 0 
\end{align}
We can use \eqref{deltaomega} to work out the transformation of the connections under the bosonic symmetries \eqref{bosonictransf}. This results in:
\begin{subequations}
\label{deltaomega_full}%
\begin{align}
\delta b_\mu  & = \partial_\mu \alpha + \tau^A{}_\mu \tau^\nu{}_A ( \delta b_\nu - \partial_\nu \alpha) \,,\\
\delta \omega_\mu{}^{AB} & = \partial_\mu \Lambda^{AB} + \omega_\mu{}^A{}_C \Lambda^{CB} + \omega_\mu{}^B{}_C \Lambda^{AC}	- 2 \tau^{[A|}{}_\mu \tau^{\nu |B]}( \delta b_\nu - \partial_\nu \alpha) \,,\\
\delta \omega_\mu{}^{ab} & = \partial_\mu \Lambda^{ab} + \omega_\mu{}^a{}_c \Lambda^{cb} + \omega_\mu{}^b{}_c \Lambda^{ac} \,,\\
\delta \omega_\mu{}^{Aa} & = \partial_\mu\lambda^{Aa}+\omega_\mu{}^A{}_B\lambda^{Ba} + \omega_\mu{}^a{}_b\lambda^{Ab} + \tfrac32\,b_\mu\lambda^{Aa}
   -\Lambda^A{}_B\omega_\mu{}^{Ba} + \omega_\mu{}^{Ab}\Lambda_b{}^a - \tfrac32\,\alpha\omega_\mu{}^{Aa} \notag
\\ & \qquad + \tau_{B \mu} \tau^{\nu \{A} ( \delta \omega_\nu{}^{B\} a} - D_\nu \lambda^{B\} a})
- \tfrac12 e^a{}_\mu \tau^{\nu A} ( \delta b_\nu - \partial_\nu \alpha ) 
 \,.
\end{align} 
\end{subequations}
We observe that, as expected, the components of the connections that cannot be uniquely determined -- that is, those in which $B^A$ and $W^{\{AB\}a}$ appear in \eqref{omegaBW} -- have  an undetermined transformation.

Now consider the curvatures \eqref{secondordercurvatures}.
As all intrinsic torsion components are zero, the curvatures are in this case given by \eqref{secondordercurvatures} with the terms $\modR_{\mu\nu}(\mathcal{T}, \mathcal{F})$ identically vanishing. 
However, if we use a connection of the form \eqref{omegaBW} with non-zero $B^A$ and $W^{\{AB\}a}$ then the curvatures will also contain these in principle undetermined quantities which can not be present in any expression obtained from the original limit of the relativistic theory.
(Note that if we set $B^A = W^{\{AB\}a} = 0$, the general definition of the curvatures \eqref{secondordercurvatures} with \eqref{conventionalconstraints} imposed reduces to the expressions \eqref{curvatures_fromlimit} we used to formulate the action and equations of motion of the non-relativistic theory.)
Furthermore, due to the fact that the general connections contain undetermined components leading to undetermined transformations \eqref{deltaomega_full}, the transformations of the curvatures will in general be ambiguous. Another way to state this is that under the connection shift \eqref{connectionshifts}, the curvatures are generically not invariant.

This simply means that only curvature components independent of the undetermined connection components can appear in the theory.
It is straightforward to verify that the following projections have this property:
\begin{subequations}\label{tildedcurv_orig}
\begin{align}
  \breve{R}_{ab}  &\equiv  R_{ab} \,,&
  \breve{R}_{ab}{}^{AB}  &\equiv   R_{ab}{}^{AB} \,, &
  \breve{R}_{Aa}{}^{BC} &\equiv  R_{Aa}{}^{BC}+ 2\,\delta_A{}^{[B}R^{C]}{}_a  \,,\\
  \breve{R}_{\mu\nu}{}^{ab} &\equiv  R_{\mu\nu}{}^{ab} \,, & 
  \breve{R}_{ab}{}^{Cc} &\equiv  R_{ab}{}^{Cc}-R^C{}_{[a}\delta^c{}_{b]}\,, &
  \breve{R}_{Aa}{}^{Ab}  &\equiv  R_{Aa}{}^{Ab} + \tfrac18\,R_{AB}{}^{AB}\delta_a{}^b\,.
\end{align}
\end{subequations}
These will all transform covariantly, i.e. as determined by their index structure and dilatation weight inherited from those of the vielbeins and connections.\footnote{\label{fn:SCTgauge}The components that do not appear in \eqref{tildedcurv_orig} are the ones that are ambiguous. In analogy with conformal gravity, one can thus use them to write a gauge field for special conformal transformations
$ f_{\mu A} \equiv e^a{}_\mu R_{aA} + \tfrac14\,\tau^B{}_\mu R_{bBA}{}^b$. By definition, this combination transforms as $f_{\mu A}\to f_{\mu A} + D_\mu \beta_A$ under the shifts \eqref{connectionshifts}.}
This can be verified by an explicit calculation. 
Let us give the result for the boost and dilatation transformations: 
\begin{subequations} 
\begin{align}
\delta \breve{R}_{ab}  & = \alpha \breve{R}_{ab} 
\,,\quad
\delta \breve{R}_{\mu\nu}{}^{ab}  = 0\,,
\quad
\delta \breve{R}_{ab}{}^{AB}  =  \alpha \breve R_{ab}{}^{AB} 
\,, \\
\delta \breve{R}_{Aa}{}^{BC} & = - \tfrac12 \alpha \breve{R}_{Aa}{}^{BC} 
- \lambda_A{}^b \breve{R}_{ab}{}^{BC} - 2 \delta_A^{[B} \lambda^{C]b} \breve{R}_{ba} \,,\\
\delta \breve{R}_{ab}{}^{Ac} &=  -\tfrac12 \alpha \breve{R}_{ab}{}^{Ac}  
+ \lambda^{B c} \breve{R}_{ab}{}^A{}_B+ \lambda^{Ad} \breve{R}_{ab}{}^c{}_d 
+ \tfrac32 \lambda^{Ac} \breve{R}_{ab} - \lambda^{Ad} \breve{R}_{d[a} \delta^c_{b]}
\,, \\
\delta \breve{R}_{Aa}{}^{Ab}  &=  -2 \alpha \breve{R}_{Aa}{}^{Ab} 
+ \lambda^{Ac} ( - \breve{R}_{caA}{}^b + \breve{R}_{Aa}{}^b{}_c ) - \lambda_{A}{}^{b} \breve{R}_{Ba}{}^{AB} + \tfrac14 \delta_a^b \lambda_A{}^c \breve{R}_{Bc}{}^{AB}\,.
\end{align}
\end{subequations}
The number of independent unambiguous curvature components reduces further if we take into account the algebraic Bianchi identities which follow straightforwardly from the definitions \eqref{firstordercurvatures} and \eqref{secondordercurvatures} of the intrinsic torsions and the curvatures.
We derive these identities in appendix \ref{appBI}.
When we have imposed the constraints as in \eqref{conventionalconstraints}, the Bianchi identities in particular imply that
\be
0 = \breve{R}_{ab} = \breve{R}_{Aa}{}^{BC} = \breve{R}_{ab}{}^{AB} =  \breve{R}_{A[a}{}^A{}_{b]} =
\breve{R}_{[abc]d}=  \breve{R}_{A[abc]} \,,\quad \breve{R}_{abCc} = -2\,\breve{R}_{C[ab]c}\,.
\ee
Hence the only non-zero, independent well-defined curvature components present are simply $\breve{R}_{\mu\nu}{}^{ab}$ and $\breve{R}_{A(a}{}^{A}{}_{b)}$.

\paragraph{Fermionic curvatures}  
We now introduce fermionic curvatures which will appear in the variation of the bosonic constraints.
As well as transforming under the bosonic symmetries as in \eqref{bosonictransf_ferminus} and \eqref{bosonictransf_ferplus}, the gravitinos transform under supersymmetry and the fermionic shift symmetry \eqref{fermionicshift}. 
To covariantise derivatives of the gravitinos, we  
introduce `superconformal' gauge fields $\varphi_{-\mu}$ and $\varphi_{+A\mu}$, the latter satisfying $\gamma^A \varphi_{+A\mu} = 0$, which we want to transform as 
\be
\delta_S \varphi_{-\mu} = \partial_\mu \eta_- + \dots\,,\quad
\delta_S \varphi_{+A\mu} = \partial_\mu \rho_{A+} + \dots \,.
\ee
such that the following gravitino curvatures
\be
\label{fermioniccurvatures}
    r_{-\mu\nu} \equiv 2 \widehat D_{[\mu} \psi_{-\nu]} \,,\quad 
   r_{+\mu\nu} \equiv 2 \widehat D_{[\mu} \psi_{+\nu]}\,,
\ee
are covariant, with
\begin{subequations}
\label{hatDpsi}%
\begin{align}
\widehat D_\mu \psi_{-\nu} &  = D_\mu \psi_{-\nu}  - \,\tau^A{}_{\nu}\gamma_A\varphi_{-\mu}\,,\\
\widehat D_\mu \psi_{+\nu} &  = D_\mu \psi_{+\nu}  + \tfrac{1}{12}\,\tau^A{}_{\nu}\gamma_A\boldsymbol{\lambda}\psi_{+\mu}
 + \tfrac{1}{8}\,e^a{}_{\nu}\boldsymbol{\lambda}\gamma_a\psi_{-\mu} - \,\tau^A{}_{\nu}\varphi_{+A\mu} +\tfrac12 e^a{}_{\nu}\gamma_a\varphi_{-\mu}\,.
\end{align}
\end{subequations} 
The terms involving $\lambda$ in the second equation ensure supercovariance of $r_{\pm \mu\nu}$.
The bosonic covariant derivatives act according to the definition \eqref{Dfermion} with $w=1/2$, i.e.
\begin{subequations}
\label{Dpsi}%
\begin{align}
D_\mu \psi_{-\nu} &  = \left( \partial_\mu  + \tfrac14 \omega_\mu{}^{AB} \gamma_{AB} + \tfrac14 \omega_\mu{}^{ab} \gamma_{ab} - \tfrac12 b_\mu \right) \psi_{-\nu} \,,\\
D_\mu \psi_{+\nu} &  = \left( \partial_\mu  + \tfrac14 \omega_\mu{}^{AB} \gamma_{AB} + \tfrac14 \omega_\mu{}^{ab} \gamma_{ab} +  b_\mu \right) \psi_{+\nu} + \tfrac12 \omega_\mu{}^{Aa} \gamma_{Aa} \psi_{-\nu} \,.
\end{align}
\end{subequations}
The gravitino curvatures \eqref{gravitinocurvatures} are analogous to the bosonic torsions/first-order curvatures of \eqref{firstordercurvatures}.
In particular, certain projections of the fermionic curvatures \eqref{fermioniccurvatures} are independent of the superconformal connections, while setting other projections to zero allows us to determine the gauge fields $\varphi$ as dependent quantities.
\begin{subequations}\label{gravitinocurvatures}%
It can be verified that the following combinations
  \begin{align}
  \breve{r}_{-ab} &\equiv r_{-ab} 
  \,,&
  \breve{r}_{-Aa} &\equiv r_{-Aa} - \tfrac13 \gamma_A \gamma^B r_{-Ba} 
  \,,\\
  \breve{r}_{+ab} &\equiv r_{+ab} + \tfrac{1}{3} \gamma_{[a|} \gamma^B r_{-B|b]}\,,& 
  \breve{r}_{+a} &\equiv \gamma^A r_{+Aa} - \tfrac{1}{8} \gamma_{a}\gamma^{BC} r_{-BC} \,,
  \end{align}    
\end{subequations}
are independent of $\varphi_{-\mu}$ and $\varphi_{+A\mu}$.
Under the superconformal transformations \eqref{fermionicshift}, $\breve{r}_{-ab}$, $\breve{r}_{-Aa}$ and $\breve{r}_{+ab}$ turn out to be invariant while $\delta_S \breve{r}_{+a}=-\tfrac14\,\boldsymbol{\lambda}\gamma_a\eta_-$. 
In addition, these combinations transform under the bosonic symmetries according to their index structure under $\mathrm{SO}(1,2)\times\mathrm{SO}(8)$ and under dilatations and boosts as
\begin{subequations} 
\begin{align}
  \delta \breve{r}_{-ab} &= +\tfrac32\,\alpha\,\breve{r}_{-ab} \,,\\
  \delta \breve{r}_{-Aa} &= -\lambda^{Bb}\big(\eta_{AB} - \tfrac13\gamma_A\gamma_B\big)\breve{r}_{-ba}\,,\\
  \delta \breve{r}_{+ab} &=  \lambda^{Cc}\gamma_C\,\big(-\tfrac32\gamma_{[a}\breve{r}_{-bc]} + \tfrac23\gamma_{[a}\breve{r}_{-b]c}\big)\,,\\
  \delta \breve{r}_{+a} &= -\tfrac32\alpha\,\breve{r}_{+a} + \lambda^{Ab}\big(-\gamma_b\breve{r}_{-Aa} + \tfrac14\,\gamma_a\breve{r}_{-Ab} + \gamma_A\breve{r}_{+ab}\big) \,.
\end{align}
\end{subequations} 
We can find dependent expressions for the superconformal gauge fields by setting the other components of \eqref{fermioniccurvatures} to zero.
We have enough equations to solve for all components except for the completely gamma-traceless part of $\varphi_{+(AB)}$. This amounts to three undetermined coefficients -- analogous to the situation encountered with the bosonic connections and the undetermined components appearing in \eqref{omegaBW}.
The explicit expressions will not be particularly important; for completeness we provide the details in appendix \ref{app_varphi}.

The undetermined bosonic connection components appearing in the gravitino curvatures can be absorbed by a redefinition of the superconformal gauge fields 
\be
\varphi_{-\mu} \rightarrow \varphi_{-\mu} + \tfrac12 \gamma^A B_A \psi_{-\mu} 
\,,\quad
\varphi_{+A\mu} \rightarrow \varphi_{+A\mu} 
- \tfrac32 ( B_A - \tfrac13 \gamma_A \gamma^B B_B) \psi_{+\mu}
+ \tfrac12 W_{\{AB\} a} \gamma^B \gamma^a \psi_{-\mu} \,,
\ee
such that the full expressions $r_{\pm \mu\nu}$ are unambiguously defined. 
As a result, we note that the connection-free curvatures $\breve{r}_{\pm}$ of \eqref{gravitinocurvatures} can be directly identified with the curvatures $\rl_\pm$ of \eqref{gravitinocurvatures_limit} and \eqref{gravitinocurvatures_limit2} which appeared when we obtained the fermionic equations of motion after the limit.

Next, we should define fermionic curvatures for the superconformal gauge fields $\varphi_{-\mu}$ and $\varphi_{+A\mu}$. 
These curvatures also appear in the sequence of constraints.
Their definition is:
\begin{subequations}\label{superconformal}%
  \begin{align}
    S_{-\mu\nu} &= 2\,\partial_{[\mu}\varphi_{-\nu]} + \frac12\,\omega_{[\mu}{}^{ab}\gamma_{ab}\varphi_{-\nu]} + \frac12\,\omega_{[\mu}{}^{AB}\gamma_{AB}\varphi_{-\nu]} + b_{[\mu}\varphi_{-\nu]} \notag\\
    &\quad-e^a{}_{[\mu}\gamma^A\psi_{-\nu]}R_{aA} - \tfrac14\,\tau^A{}_{[\mu}\gamma^B\psi_{-\nu]}R_{aAB}{}^a \,,\\
    S_{+A\mu\nu} &= 2\,\partial_{[\mu}\varphi_{+A\nu]} + \frac12\,\omega_{[\mu}{}^{ab}\gamma_{ab}\varphi_{+A\nu]} + \frac12\,\omega_{[\mu}{}^{BC}\gamma_{BC}\varphi_{+A\nu]} +2\,\omega_{[\mu A}{}^B\varphi_{+\nu]B} \notag\\
    &\quad +4\,b_{[\mu}\varphi_{+A\nu]} -3\,\big(\eta_{AB} -\tfrac13\,\gamma_A\gamma_B\big)\omega_{[\mu}{}^{Bb}\gamma_b\varphi_{-\nu]} \notag\\
    &\quad + 3\,\big(\delta_A{}^B - \tfrac13\,\gamma_A\gamma^B\big)\big(e^b{}_{[\mu}\psi_{+\nu]}R_{bB} + \tfrac14\,\tau^C{}_{[\mu}\psi_{+\nu]}R_{cCB}{}^b\big) \,,
\end{align}
\end{subequations}
where the nonlinear terms involving bosonic curvatures $R$ are needed for (super)covariance.
The dilatation weight of $S_{-\mu\nu}$ is $-1/2$ and that of $S_{+A\mu\nu}$ is $-2$. Just as in the bosonic case, not all components of the superconformal curvatures are unambiguously defined. 
We discuss this in more detail in appendix \ref{appBI}, where we also obtain the Bianchi identities which further constrain the expressions which we can expect to find appearing in the non-relativistic theory, given that it must have an unambiguous origin from the limit. Remarkably this leaves only 
\begin{align}
\label{SpABa}
  S_{+[AB]a}
\end{align}
as an unconstrained component of the superconformal curvatures (subject to the algebraic identity \eqref{eq:thelastSCBianchi}). The rest are either ambiguously defined or can be expressed as derivatives of first order gravitino curvatures.
Note that the curvature components \eqref{SpABa} have dilatation weight $-5/2$, and must be boost invariant as there is no available object of the correct weight for them to transform into.

\subsection{The geometry with half-maximal supersymmetry} 
\label{halfSUSYconns}

We now briefly discuss the geometry of the non-relativistic half-supersymmetric case, where we impose the constraints \eqref{halfsusy}. 
Anticipating the fact that the variation of the constraints will require $f_{abcd} = 0$, we can impose the required constraints while simultaneously solving \eqref{intrinsictorsions} for the connection components by imposing:
\begin{align}
  \mathcal{T}_{\mu\nu}{}^A =  2\,e^a{}_{[\mu}\tau_{B\nu]} T_a{}^{\{AB\}}\,, \quad \mathcal{T}_{\mu\nu}{}^a = 0\,, \quad
  \mathcal{F}_{\mu\nu\rho\sigma} = 
   4\,\tau^A{}_{[\mu}e^a{}_{\nu} e^b{}_{\rho} e^c{}_{\sigma]} f_{Aabc}\,.
\end{align}
Seen as conventional constraints, these are still solved by the usual connection expressions \eqref{omegaBW}. 
As not all components of the torsions are set to zero, we have to take into account intrinsic torsion contributions to the connection transformations and curvatures, as we discussed above.
For example, specifying \eqref{deltaomega_ab_mod} to this case we find
\be
\begin{split}
  \delta\omega_\mu{}^{ab} & = \partial_\mu \Lambda^{ab} + \omega_\mu{}^{a}{}_c \Lambda^{cb} + \omega_\mu{}^b{}_c \Lambda^{ac} 
   +\tau_{A\mu}\big(2\,\lambda_B{}^{[a}T^{b]\{AB\}} +\tfrac12\,\epsilon^{ABC}\lambda_B{}^cf_{Cc}{}^{ab}\big)\,,
   \end{split}
\label{deltaomega_ab_mod_half}
\ee
with curvature defined by \eqref{Rmunuab} with: 
\be
\begin{split}
 \modR(\mathcal{T},\mathcal{F})_{\mu\nu}{}^{ab} & = 4\,\tau_{A[\mu}\omega_{\nu]B}{}^{[a}T^{b]\{AB\}} +\tau_{A[\mu}\epsilon^{ABC}\omega_{\nu]B}{}^cf_{Cc}{}^{ab}\,.
 \end{split}
\label{modRab_half} 
\ee
In principle, one should similarly go through all the other bosonic, fermionic, and superconformal curvatures and would find analogous modifications. We will not give these expressions here since the analysis of the tower of constraints will not require these details. 


\section{Non-relativistic supergravity with maximal supersymmetry}
\label{maxSUSYsection}

In this section, we analyse the theory that results from imposing the geometric constraints \eqref{geoconstraints}.
This keeps all 32 supersymmetry transformation parameters, $\epsilon_\pm \neq 0$.
Varying the constraints under supersymmetry leads to a hierarchy of further constraints, all of which must vanish for consistency of the theory.
This tower of constraints involves bosonic and fermionic connections and curvatures.

\subsection{Supersymmetry transformations}
\label{fullsupersummary}

Let's start by summarising the supersymmetry rules arising when we impose the geometric constraints \eqref{geoconstraints}.
These are:
\begin{subequations}
\begin{align}
\delta_Q \tau^A{}_\mu & =
\bar\epsilon_- \gamma^A \psi_{-\mu}
 \,,\\
\delta_Q e^a{}_\mu & =  \bar\epsilon_+ \gamma^a \psi_{-\mu}
+ \bar\epsilon_- \gamma^a \psi_{+\mu} \,,
\\ \nonumber
\delta_Q c_{\mu\nu\rho} & =
 6 \bar \epsilon_+   \epsilon_{ABC} \gamma^A \psi_{+[\mu} \tau^B{}_{\nu} \tau^C{}_{\rho]}
 +3 \bar\epsilon_- \gamma_{ab} \psi_{-[\mu} e^a{}_\nu e^b{}_{\rho]}
\\  & \quad
+6  \left(
\bar\epsilon_+ \gamma_{Aa} \psi_{-[\mu} \tau^A{}_\nu e^a{}_{\rho]} +
\bar\epsilon_- \gamma_{Aa} \psi_{+[\mu} \tau^A{}_\nu e^a{}_{\rho]}
\right) \,,
\\
\delta_Q \psi_{-\mu} & = D_\mu \epsilon_-  \,, \\ 
\delta_Q \psi_{+\mu} & = D_\mu \epsilon_+  - \tfrac{1}{12}  \tau^A{}_\mu \gamma_A \boldsymbol{\lambda} \epsilon_+
 - \tfrac18 e^a{}_\mu {\boldsymbol{\lambda}}  \gamma_a \epsilon_- \,,
\end{align}
where the covariant derivatives of the supersymmetry parameters are: 
\begin{align} 
D_\mu \epsilon_- & = 
 (\partial_\mu + \tfrac14 \hat \omega_{\mu}{}^{ab} \gamma_{ab} + \tfrac14 \hat \omega_\mu{}^{AB}\gamma_{AB} - \tfrac12 \hat b_\mu ) \epsilon_-\,,\\
 D_\mu \epsilon_+ & = 
 (\partial_\mu  + \tfrac{1}{4}\hat \omega_\mu{}^{ab} \gamma_{ab}
+ \tfrac{1}{4} \hat \omega_\mu{}^{AB} \gamma_{AB} 	+ \hat b_\mu )\epsilon_+
+ \tfrac12 \hat \omega_\mu{}^{Aa} \gamma_{Aa} \epsilon_-\,,
\end{align} 
\end{subequations}
where we use the supercovariant completions of the spin connections, based on the expressions \eqref{connections}.
The limit of course gives the specific bosonic connections \eqref{connections_nohats_limit} whereas the underlying geometry allows for the more general solution \eqref{omegaBW} containing undetermined components.
In the supersymmetry transformations, these undetermined components can be viewed as a fermionic Stuckelberg shift transformation \eqref{fermionicshift} with $\eta_- = - \tfrac12\,B^A\gamma_A\epsilon_-$ and $\rho_{+A} = \frac32\big(\eta_{AB}-\tfrac13\gamma_A\gamma_B\big)B^B\epsilon_-  + \tfrac12\,W_{\{AB\}}{}^a\gamma^B\gamma_a\epsilon_-$.

We also need the supersymmetry transformation of the Lagrange multiplier field $\lambda_{abcd}$, which follows from  \eqref{deltalambdadesired}.
This is quite complicated as it requires knowing the equations of motion of $\psi_{+\mu}$ at order $c^{-3}$.
However, we can determine the variation of $\lambda_{abcd}$ using what is essentially a linearised calculation, as outlined in appendix \ref{app_lambda}.
Then we extend the result to a covariant expression using the geometric quantities introduced above, obtaining:
\be
\begin{split} 
\delta_Q \lambda_{abcd}&  = 
\tfrac14 \bar \epsilon_- \gamma^e \gamma_{abcd} \breve{r}_{+e} 
+2 \bar \epsilon_+ \gamma_{[abc}{}^e \breve{r}_{+d]e} \,.
\end{split}
\label{deltaQlambda}
\ee 
Both terms are automatically projected correctly such that the right-hand side is anti-self-dual.
The transformation \eqref{deltaQlambda} can further be verified to be consistent with the symmetries of the theory and their closure (see below).


\subsection{Constraints for maximal supersymmetry} 
\label{constraints_max} 

As we have seen, the maximally supersymmetric theory requires the constraints \eqref{geoconstraints}.
The variation of these constraints under supersymmetry leads to additional constraints which must be imposed for consistency. 
Repeated variations leads to a tower of constraints. 

Schematically, the variation of the intrinsic torsions of \eqref{geoconstraints} leads to the appearance of the gravitino curvatures, \eqref{gravitinocurvatures}, which in turn vary into bosonic curvatures \eqref{tildedcurv_orig}. Carrying out these and further variations quickly becomes complicated. 
In appendix \ref{tower_derivation} we explain how to obtain the structure of the tower of constraints, eventually using linearised calculations to identify which curvature structures appear.
While we do not have a full proof at the non-linear level, what we observe strongly suggests that the full set of constraints to be imposed consists of the following.

\vspace{1em}

\begin{subequations}
\label{allconstraints}%
\noindent {\bf Bosonic constraints}
\begin{align}
\hat T_{ab}{}^A & = 0 \,,\quad \hat T_a{}^{\{AB\}} = 0 \,,\quad \hat f_{Aabc} = 0 \,,\quad \hat f_{abcd} = 0 \,,\\
\breve{R}_{ab}{}^{cd} &= 0 \,,\quad 
\breve{R}_{Aa}{}^{bc} = 0 \,,\quad 
D_a \breve{R}_{A(b}{}^A{}_{c)} = 0\,,\quad D_a \lambda_{bcde} = 0 \,,
\label{allconstraints_bos2}
\end{align} 

\noindent {\bf Fermionic constraints}
\begin{align}
\breve{r}_{-ab} = 0 \,,\quad \breve{r}_{-Aa} = 0 \,,\quad \breve{r}_{+ab} = 0 \,,\quad
D_a \breve{r}_{+b} = 0 \,.
\end{align} 
In addition, the constraint variation requires the super-Poisson equation $\gamma^a \breve{r}_{+a}= 0$.
We take the point of view that this equation is anyway to be imposed as an equation of motion, and do not regard it as a `new' condition obtained from the constraint variation. To some extent this is a matter of semantics.
\end{subequations}

\begin{figure}[ht]
\centering
\begin{tikzpicture}
\draw (0,0) node [labels_up] {Level 1\\ $\partial$, [2] };
\draw (0,0) node (B1) [B,box_up] { $\hat T_{ab}{}^A$ \\ $\hat f_{abcd}$};

\def\x{2.4}
\draw (\x,0) node [labels_up] { Level 2\\ $\partial$, [$\tfrac32$]};
\draw (\x,0) node (F1) [F,box_up] {$\breve{r}_{-ab}$};

\draw (2*\x,0) node [labels_up] {Level 3 \\ $\partial^2$, [1] };
\draw (2*\x,0) node (B2) [B,box_up] {
 \\ $\breve{R}_{ab}{}^{cd}$ \\ 
 };

\begin{scope}[xshift =-2.4cm]
\draw (1*\x,-6) node [labels_down] { Level 1 \\ $\partial$, [$\tfrac12$]};
\draw (1*\x,-6) node (B1down) [B,box_down] {$\hat T_a{}^{\{AB\}}$\\$\hat f_{Aabc}$ };

\draw (2*\x,-6) node (F1labelsdown) [labels_down] {Level 2 \\ $\partial$, [0] };
\draw (2*\x,-6) node (F1down) [F,box_down] {$\breve{r}_{+ab}$ \\ $\breve{r}_{-Aa}$};

\draw (3*\x,-6) node [labels_down] { Level 3 \\ $\partial^2$, [$-\tfrac12$] };
\draw (3*\x,-6) node (B2down) [B,box_down] {
$\breve{R}_{Aa}{}^{bc}$ \\ 
$D_a \lambda_{bcde}$};

\draw (4*\x,-6) node [labels_down] { Level 4 \\ $\partial^2$, [$-1$] };
\draw (4*\x,-6) node  (F2) [F,box_down] {$D_a \breve{r}_{+b}$};

\draw (5*\x,-6) node [labels_down] { Level 5 \\ $\partial^3$, [$-\tfrac32$]};
\draw (5*\x,-6) node (B3) [B,box_down] {\scriptsize $D_a\breve{R}_{A(b}{}^{A}{}_{c)}$	};

\draw (6*\x,-6) node [labels_down] { Level 6 \\ $\partial^3$, [$-2$]};
\draw (6*\x,-6) node (F3) [F,box_down] {\tiny $\! D_a\gamma^{A\!B}\!S_{+A\!B b}$ \\ $\approx 0$ by Bianchi};

\end{scope}

\draw[redarrows] (B1.east) -- (F1.west) node [midway,above] {$\epsilon_-$};
\draw[redarrows] (F1.east) -- (B2.west) node [midway,above] {$\epsilon_-$};
\draw[bluearrows] (B1down.north) -- (F1.south) node [pos=0.25,above] {$\epsilon_+$};
\draw[bluearrows] (F1down.north) -- (B2.south) node [pos=0.25,above] {$\epsilon_+$};
\draw[redarrows] (B1down.east) -- (F1down.west) node [midway,above] {$\epsilon_-$};
\draw[redarrows] (F1down.east) -- (B2down.west) node [midway,above] {$\epsilon_-$};
\draw[redarrows] (B2down.east) -- (F2.west) node [midway,above] {$\epsilon_-$};
\draw[redarrows](F2.east) -- (B3.west) node [midway,above] {$\epsilon_-$};
\draw[redarrows] (B3.east) -- (F3.west) node [midway,above] {$\epsilon_-$};
%


\end{tikzpicture} 

\caption{This diagram summarises the variation of constraints that appear when one begins with the initial geometric constraints $\hat T_{ab}{}^A = \hat T_a{}^{\{AB\}} = \hat f_{abcd} = \hat f_{Aabc} = 0$. As book-keeping notation, we note the number of derivatives $\partial$, while $[w]$ denotes the weight $w$ under the dilatation symmetry. 
The analysis of appendix \ref{tower_derivation} suggests that the tower of constraints terminates, using Bianchi identities and previous constraints, at Level 6.
Note that boosts act vertically mapping tensors in the bottom row into tensors in the top row.
}
\label{fig:diagram}
\end{figure}
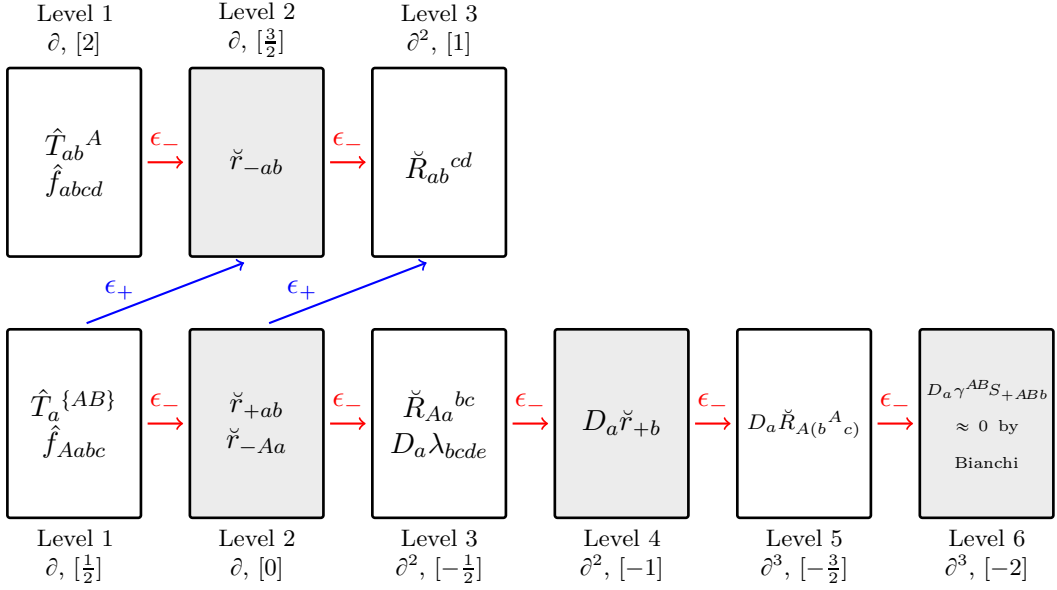

The tower of constraints terminates when varying $D_a \breve{R}_{A(b}{}^A{}_{c)}$.
At first glance, this leads to a new constraint involving a derivative of the superconformal curvature \eqref{SpABa}. However, as we explain in appendix \ref{tower_derivation} this can be argued to automatically vanish by a Bianchi identity, and so does not lead to an extra constraint.
The relationship between these constraints is depicted in figure \ref{fig:diagram}.

The constraints \eqref{allconstraints} are evidently very strong conditions.
For instance, it is easy to see that the equations of motion following from the action of the non-relativistic theory are all identically obeyed if these constraints are satisfied (for the bosons the equations of motion are \eqref{tauEom}, \eqref{EEom}, \eqref{EomChere} as well as $f^{(-)}_{abcd} = 0$; for the fermions they are \eqref{eom_psiplus} and \eqref{eom_psiminus}, noting that the gravitino curvatures $\breve{r}_\pm$ coincide with those denoted $\rl_{\pm}$).
In fact, imposing the constraints after \emph{any} variation of the action will give a vanishing variation up to total derivatives.
For the equations of motions of the vielbeins, this can be seen already from the variation \eqref{deltaStaue}.
We also note that the other super-Poisson equation \eqref{superfish2} is obeyed due to the constraint $\breve{r}_{-Aa} = 0$.

The bosonic Poisson equation $\breve{R}_{Aa}{}^{Aa}  + \tfrac{1}{4!} \lambda_{abcd} \lambda^{abcd} = 0$ must additionally be satisfied.
We also require the stronger conditions that $D_e\breve{R}_{Aa}{}^{Ab} = 0$ and $D_e \lambda_{abcd} = 0$, which are compatible with but do not imply the Poisson equation. 
This has important consequences for the interpretation of the theory with maximal supersymmetry as a Newton-Cartan supergravity with a Newton potential satisfying the usual Poisson equation.
We will discuss this in section \ref{woundM2}.

Despite the seemingly stringent nature of these constraints, they do allow for a family of supersymmetric solutions which includes a non-trivial limit of the M2 brane solution with a proposed holographic interpretation \cite{Lambert:2024uue}. We will demonstrate this in section \ref{LS}.

Let us also comment on the differences with respect to the set of constraints that emerges in the ten-dimensional $\mathcal{N}=1$ case of \cite{Bergshoeff:2021tfn}.
There the bosonic torsion constraints analogous to \eqref{geoconstraints} are themselves invariant under
supersymmetry, so that there are simply no fermionic constraints. For this to happen it is crucial that in this ten-dimensional stringy case, the longitudinal index decomposes as $A=(+,-)$ in terms of lightcone coordinates, and the constraints involve these projections. 
Then, due to some interplay with chirality the bosonic torsion constraints, are invariant under supersymmetry. 
In the eleven-dimensional MNC case, with a three-dimensional longitudinal index, this sort of structure is simply absent.

\subsection{Closure}
\label{fullclosure}

In this subsection we record some features of the closure of the algebra of symmetries.
Rather than be exhaustive, we illustrate the more interesting commutators. 
In particular, we focus on commutators involving boost transformations, which we denote by $\delta_G$, the fermionic shift or superconformal symmetries, which we denote by $\delta_S$, and the supersymmetry transformations, which we denote by $\delta_Q$.

\paragraph{Boost -- Superconformal}

Acting on the gravitini with boosts and superconformal transformations with parameters $\lambda^{Aa}$ and $\eta_-$, respectively, we have:
\begin{align}
    \big[\delta_G,\,\delta_S\big] = \delta_S\,,\qquad\mathrm{with}\qquad \rho'_{+A} = -\tfrac32\,\big(\eta_{AB} -\tfrac13\,\gamma_A\gamma_B\big)\lambda^{Bb}\gamma_b\eta_-\,,\quad \eta_-'=0 \,.
\end{align}

\paragraph{Boost -- Supersymmetry}

A simple calculation shows that 
\begin{align}
    &\big[\delta_G,\delta_Q\big] = \delta'_Q + \delta'_S& \qquad\mathrm{with}\qquad  &(\epsilon'_+,\epsilon'_-) = (\tfrac12\,\lambda^{Aa}\gamma_{Aa}\epsilon_-,0)\notag\\
    & &&(\eta'_-,\rho'_{+A}) = \big(0,-\tfrac18(\eta_{AB}-\tfrac13\,\gamma_A\gamma_B)\lambda^{Bb} \boldsymbol{\lambda} \gamma_b\epsilon_-\big)\,.
\end{align}

\paragraph{Superconformal -- Supersymmetry}

The commutator of supersymmetry and superconformal transformations can be verified to close in the following manner:
\begin{align}
    \big[\delta_S,\delta_Q\big] = \delta_D + \delta_{\mathsf{SO}(1,2)} + \delta_{\mathsf{SO}(8)} +\delta_G+\delta_{S'}\,,
\end{align}
where $\delta_D$ denotes the dilatation transformation, and with the following parameters:
\begin{subequations}
\begin{align}
\Lambda^A{}_B & = - \bar\epsilon_- \gamma^A{}_B \eta_- \,,\quad
\Lambda^a{}_b =  \tfrac12 \bar\epsilon_- \gamma^a{}_b \eta_- \,, \quad 
\lambda_A{}^a = \bar\epsilon_+ \gamma^a \gamma_A \eta_- + \bar \epsilon_- \gamma^a \rho_{A+}\,, \quad
\alpha = \bar\epsilon_- \eta_- \,,\\
\eta_-' &= \tfrac12 (\bar \eta_- \psi_{-}^A )\gamma_A\epsilon_-\,,\\
\rho'{}_+^A & = ( \bar \eta_- \psi_{-}^A ) \epsilon_+ - \tfrac12 (\bar \eta_- \psi_-^B ) \gamma^A{}_{B} \epsilon_+ 
- \tfrac12 ( \bar \rho_+^{(A} \gamma_a \psi_-^{B)} ) \gamma_B \gamma^a \epsilon_-
+ \tfrac16 ( \bar \rho_{+B} \gamma_a \psi_-^B ) \gamma^A \gamma^a \epsilon_-
 \,.
\end{align}
\end{subequations}

\paragraph{Supersymmetry -- Supersymmetry}

Finally we come to the commutator of two supersymmetry transformations.
This is, as expected, more involved.
The closure of the supersymmetry algebra involves all other symmetries of the theory, and takes the following form:
\begin{align}
    \big[\delta_{Q_1},\delta_{Q_2}\big] = \mathcal L_\xi + \delta_{\mathsf{SO}(1,2)} +\delta_{\mathsf{SO}(8)} + \delta_G + \delta_D + \delta_\theta+ \delta_{Q'}+ \delta_S + T\,.
 \label{QQ}
\end{align}
where $\mathcal L_\xi$ is a general coordinate transformation, $\delta_\theta$ denotes gauge transformations of the three-form, and $T$ indicates obstructions to the closure, proportional to intrinsic torsion components and to fermionic gravitino curvatures (including  $\gamma^a \breve{r}_{+a}$ which vanishes by the super-Poisson equation).

Verifying closure on $\tau^A{}_\mu$, $e^a{}_\mu$ and $c_{\mu\nu\rho}$ is relatively straightforward, and allows us to identify 
the following bosonic symmetry parameters appearing on the right-hand side of \eqref{QQ}:
\begin{subequations}\label{eq:QQparameters}%
\begin{align}
    &\xi^\mu = \tau^\mu{}_A\,\bar\epsilon_{2-}\gamma^A\epsilon_{1-} + e^\mu{}_a\big(\bar\epsilon_{2-}\gamma^a\epsilon_{1+} - \bar\epsilon_{1-}\gamma^a\epsilon_{2+}\big)\,,\\
    &\Lambda^A{}_{B} = -\xi^\mu\omega_\mu{}^A{}_{B}\,, \qquad 
    \Lambda^{ab} = -\xi^\mu\omega_\mu{}^{ab} + \tfrac{1}{4}\,\big(\bar\epsilon_{2-}\gamma^{[a}\boldsymbol{\lambda}\gamma^{b]}\epsilon_{1-}\big)\,,\\
    &\lambda^{Aa} = -\xi^\mu\omega_\mu{}^{Aa} +\tfrac{1}{12}\big(\bar\epsilon_{2-}\gamma^{Aa}\boldsymbol{\lambda}\epsilon_{1+} - \bar\epsilon_{1-}\gamma^{Aa}\boldsymbol{\lambda}\epsilon_{2+}\big)\,,\qquad 
    \alpha = -\xi^\mu b_\mu\,,\\
    %
    &\theta_{\mu\nu} =2\,\epsilon_{ABC}\,\bar\epsilon_{2+}\gamma^A\epsilon_{1+}\tau_\mu{}^B\tau_\nu{}^C + \bar\epsilon_{2-}\gamma_{ab}\epsilon_{1-}e_\mu{}^a e_\nu{}^b \notag\\
    &\qquad+ 2\,\big(\bar\epsilon_{2-}\gamma_{Aa}\epsilon_{1+}+\bar\epsilon_{2+}\gamma_{Aa}\epsilon_{1+}\big)\tau_{[\mu}{}^A e_{\nu]}{}^a - \xi^\rho c_{\rho\mu\nu}\,,
\end{align}
as well as the supersymmetry transformation parameter
\be
\epsilon'_\pm = -\xi^\mu\psi_{\pm\mu}\,.
\ee
\end{subequations} 
For the remaining fields, $\lambda_{abcd}$ and $\psi_{\pm \mu}$, we have verified closure at the linearised level. 
In particular, we find that the supersymmetry algebra closes on the gravitino at linearised level up to the above bosonic and supersymmetry transformation parameters appearing on the right-hand side of \eqref{QQ} together with the following superconformal transformations: 
\begin{subequations}\label{eq:QQparametersS}%
\begin{align}
    \eta_- & = \xi^B ( \tfrac{1}{4} \gamma_B{}^{CD} (\partial \psi)_{-CD} + \tfrac12 \gamma^C (\partial \psi)_{-CB} ) 
+ \tfrac13 \xi^a \gamma^B (\partial \psi)_{-Ba} 
+ 
\tfrac14 (  \bar \epsilon_{1-} \gamma^{bc} \epsilon_{2-}) \gamma_b \breve{r}_{+c}     
     \,,\\
\rho_{+A}  & = \xi^B \left(
\tfrac12  (\partial \psi)_{+BA} - \tfrac16 \gamma_B \gamma^C (\partial \psi)_{+CA} 
 - (\partial \psi)_{+BA}  
-  \varphi_{+(AB)} + \tfrac13 \gamma_B \gamma^C \varphi_{+(AC)} 
\right) \notag 
\\ & \quad
+ \xi^a \left( 
 (\partial \psi)_{+Aa}
+ \tfrac12 \gamma_a ( \gamma^B (\partial \psi)_{-AB} - \tfrac14 \gamma_A \gamma^{BC} (\partial \psi)_{-BC} ) 
-\tfrac13 \gamma_A \breve{r}_{+a} 
\right) 
 \notag\\
& \quad - \tfrac38 \left( \bar \epsilon_{1-} \gamma^B \gamma^a \epsilon_{2+} - \bar \epsilon_{2-} \gamma^B \gamma^a \epsilon_{1+}\right) (\eta_{AB} - \tfrac13 \gamma_A \gamma_B ) \breve{r}_{+a} \notag
\\ & \quad
+ \tfrac{1}{16} \left( \bar \epsilon_{1-} \gamma^B \gamma^{abc} \epsilon_{2+} - \bar \epsilon_{2-} \gamma^B \gamma^{abc} \epsilon_{1+} \right) ( \eta_{AB} - \tfrac{1}{3} \gamma_A \gamma_B) \gamma_{bc} \breve{r}_{+a}  \,,
\end{align}
where we use the shorthand $(\partial \psi)_{\pm \mu\nu} \equiv 2 \partial_{[\mu} \psi_{\pm \nu]}$.\footnote{Hence in \eqref{eq:QQparametersS} we have $\breve{r}_{+a} = \gamma^A(\partial \psi)_{+Aa} - \tfrac18 \gamma_a \gamma^{BC} (\partial \psi)_{-BC}$, while $\varphi_{+AB}$ is given by the linearisation of \eqref{varphiexplicit2}. Strictly speaking the latter is undetermined, but its gamma trace \eqref{gammatracevarphi} is not, and this is what is needed to verify here that $\gamma^A \rho_{+A} = 0$.}
\end{subequations} 

\section{Keeping only half-maximal supersymmetry}
\label{halfSUSYsection}

In this section, we analyse the alternative set of constraints, where we keep only 16 supersymmetry transformations parameters, $\epsilon_+ \neq 0$, $\epsilon_-=0$ and impose the restricted set of constraints \eqref{halfsusy}, i.e. just that $\hat T_{ab}{}^A = 0$.
We have seen in the previous section that having all 32 supersymmetry transformation parameters and imposing the initial constraints \eqref{geoconstraints} leads to a complicated sequence of constraints.
Inspecting figure \ref{fig:diagram}, it is clear that the bulk of the complexity of this sequence is due to repeated $\epsilon_-$ variations, in particular moving along the bottom row starting with the constraints $\hat{T}_a{}^{\{AB\}}, \hat{f}_{Aabc}$.
Requiring $\epsilon_-=0$ and imposing no constraints on $\hat{T}_a{}^{\{AB\}}, \hat{f}_{Aabc}$, the expectation would be that the situation will simplify drastically.

We will see below that this is indeed the case: variation leads only to the additional condition $\hat f_{abcd} = 0$. 
We will also check, via a linearised calculation, that the closure of supersymmetry transformations on the gravitinos does not require additional fermionic constraints.
This provides evidence that the non-relativistic theory with half-maximal supersymmetry is again self-consistent.
Compared to the maximal case, we will be rather brief in our discussion in this section, deferring a fuller treatment of this half-maximal theory to future work.

\subsection{Supersymmetry transformations}
\label{halfsusysummary}

If we impose the constraints $\epsilon_-=0$ and $\hat T_{ab}{}^A =0$, the supersymmetry transformations are:
\begin{subequations}
\label{halfsusy_transfs}
\begin{align}
\delta_Q \tau^A{}_\mu & = 0 \,,\\
\delta_Q e^a{}_\mu & =  \bar\epsilon_+ \gamma^a \psi_{-\mu}  \,,
\\
\delta_Q c_{\mu\nu\rho} & =
 6 \bar \epsilon_+   \epsilon_{ABC} \gamma^A \psi_{+[\mu} \tau^B{}_{\nu} \tau^C_{\rho]}
+6
\bar\epsilon_+ \gamma_{Aa} \psi_{-[\mu} \tau^A{}_\nu e^a{}_{\rho]}\,,
\\
 \delta_Q \psi_{-\mu} & =  \tau_{A\mu} \left(
\tfrac12 \hat T_{a}{}^{\{AB\}} \gamma_B \gamma^a
+\tfrac14 ( \eta^{AB} - \tfrac13 \gamma^A \gamma^B ) \boldsymbol{\hat{f}}_B \right) \epsilon_+
\notag
\\ & \qquad + e^a{}_{\mu} \left(
\tfrac{1}{24} \gamma_a \boldsymbol{\hat{f}}^{(-)} - \tfrac{1}{8} \boldsymbol{\hat f}^{(+)} \gamma_a
\right) \epsilon_+
\,,\\ \notag
\delta_Q \psi_{+\mu} & = (\partial_\mu  + \tfrac{1}{4}\hat \omega_\mu{}^{ab} \gamma_{ab}
+ \tfrac{1}{4} \hat \omega_\mu{}^{AB} \gamma_{AB} 	+ \hat b_\mu )\epsilon_+
\\ & \qquad \qquad  - \tfrac{1}{12}  \tau^A{}_\mu \gamma_A \boldsymbol{\lambda} \epsilon_+
+\tfrac{1}{12} \tfrac{1}{6} e^a{}_\mu \hat f_{A bcd} ( \gamma_a{}^{bcd} - 6 \delta_a^b \gamma^{cd} ) \gamma^A \epsilon_+
 \,.
\end{align}
\end{subequations}
We will see below that these transformations will require $\hat f_{abcd} = 0$, implying that the transverse terms in $\delta_Q \psi_{-\mu}$ can in fact be dropped.
The supersymmetry transformation of the
Lagrange multiplier following from the analysis of appendix \eqref{app_lambda} is:
\be
\delta_Q\lambda_{abcd} = 2 \bar \epsilon_{+} \gamma_{[abc}{}^e \breve{r}_{+d]e}  + \dots 
\ee
where the ellipsis denotes terms involving $\hat{f}_{Aabc}$ or $\hat{T}_a{}^{\{AB\}}$ that would not be captured by the (essentially linearised) derivation of that appendix.
We will only use this expression in a linearised calculation when checking closure of the supersymmetry algebra, which provides an initial consistency check on the half-maximally supersymmetric option.

\subsection{Tower of constraints}
\label{halftower}

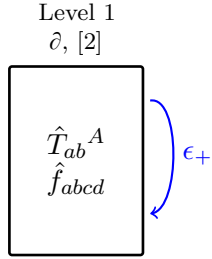
\begin{figure}[ht]

\centering
\begin{tikzpicture}

\draw (0,0) node [labels_up] {Level 1\\ $\partial$, [2] };
\draw (0,0) node (B1) [B,box_up] { $\hat T_{ab}{}^A$ \\ $\hat f_{abcd}$};

\draw[bluearrows] (0.8,-0.5) to [out=10,in=10] (0.8,-2);
\draw[blue] (1.6,-1.2) node {$\epsilon_+$};

\end{tikzpicture} 

\caption{Compared to figure \ref{fig:diagram}, the set of constraints does not grow under supersymmetry.}
\label{fig:diagramsilly}
\end{figure}

The only geometric constraint we impose at the beginning is
\be
\hat T_{ab}{}^A = 0\,.
\label{c1}
\ee
Varying this constraint using the supersymmetry transformations \eqref{halfsusy_transfs} leads to:
\be
\delta_Q \hat T_{ab}{}^A
=
2 \bar \epsilon_+ \gamma^c \psi_{-[a} \hat T_{b]c}{}^A
- 2 \bar \psi_{-[a} \gamma^A (
\tfrac{1}{24} \gamma_{b]} \boldsymbol{\hat{f}}^{(-)} - \tfrac{1}{8} \boldsymbol{\hat f}^{(+)} \gamma_{b]}
) \epsilon_+ \,,
\ee
which tells us that we also need to impose the constraint 
\be
\hat f_{abcd} = 0 \,.
\label{c2}
\ee
Varying this constraint produces:
\be
\begin{split}
\delta_Q \hat f_{abcd}
& = - 4 \bar \epsilon_+ \gamma^e \psi_{-[a} \hat f_{|e|bcd]}
+ 12 \bar \epsilon_+ \gamma_{A[d} \psi_{-b} \hat T_{ac]}{}^A
\\ & \quad
+ 12 \bar \epsilon_+ \gamma_{A[d} \psi_{-b} \bar \psi_{-a} \gamma^A \psi_{-c]}
- 12 \bar \epsilon_+ \gamma^e \psi_{-[b}  \bar \psi_{-a} \gamma_{|e|c} \psi_{-d]} \,.
\end{split}
\ee
Using Fierz identities (see appendix \ref{gammadetails}), the second line can be shown to equal $-1/2$ itself, and so it must vanish.
Then from the first line we again get the two constraints we have already imposed.
Hence the {\bf tower of constraints consists solely of \eqref{c1} and \eqref{c2}}, meaning that instead of the complicated structure captured by figure \ref{fig:diagram} in the maximal case, we now have the simple figure \ref{fig:diagramsilly}.

\subsection{Closure}
\label{halfclosure}

The closure of the supersymmetry algebra in the case of half-maximal supersymmetry is much simpler than in the maximally supersymmetric case. In particular, one finds that the commutator of two supersymmetry transformations takes the form
\begin{align}
    \big[\delta_{Q_1},\delta_{Q_2}\big] =  \delta_G  + \delta_\theta  + \delta_S + T\,. 
 \label{QQhalf}
\end{align}
where $\delta_\theta$ denotes gauge transformations of the three-form, and $T$ indicates obstructions to the closure, that are proportional to $\hat{T}_{ab}{}^A$ and $\hat{f}_{abcd}$, as well as fermionic equations of motion. These obstructions thus vanish upon imposing the constraints discussed in the previous subsection. 
Notably, the supersymmetry transformations do not close into diffeomorphisms.
Therefore this is a somewhat exotic realisation of local supersymmetry. 

Verifying closure on $\tau^A{}_\mu$, $e^a{}_\mu$ and $c_{\mu\nu\rho}$ shows that the bosonic symmetry parameters appearing on the right-hand side of \eqref{QQhalf} are given by:
\begin{align}
  \label{QQparhalf}
  \lambda^{Aa} &=  - \hat{T}^{a\{AB\}} \bar{\epsilon}_{2+} \gamma_B \epsilon_{1+} + \tfrac16 \hat{f}^{Aabc} \bar{\epsilon}_{2+} \gamma_{bc} \epsilon_{1+} + \tfrac{1}{6 \cdot 3!} \bar{\epsilon}_{2+} \gamma^{Aa}{}_{Bbcd} \epsilon_{1+} \hat{f}^{Bbcd} \,, \\
  \theta_{\mu\nu} &= 2\epsilon_{ABC} \left(\bar{\epsilon}_{2+} \gamma^A \epsilon_{1+}\right) \tau^B{}_{\mu} \tau^C{}_{\nu} \,.
\end{align}
For the Lagrange multiplier and the gravitinos, we have verified that the algebra \eqref{QQhalf} holds at the linearised level.
The Stuckelberg transformation on the right-hand side of \eqref{QQhalf} involves only a non-trivial $\rho_{+A}$ transformation.
Crucially, this closure can be seen to require fermionic equations of motion, specifically we need (the linearisations of) \eqref{eom_psiplus} and \eqref{eom_psiminus} as well as the super-Poisson equation \eqref{superfish2}. 
This shows in particular that no additional fermionic constraints are required, and the diagram \ref{fig:diagramsilly} is complete.

\section{Supersymmetric solutions}
\label{solutions}

\subsection{Killing spinor equations} 
\label{KSeqns} 

Given a bosonic solution satisfying the relevant  constraints, we can see if it is supersymmetric by solving the relevant Killing spinor equations.
Those of the theory with 32 supersymmetry parameters $\epsilon_\pm$ are:
\be
\begin{split}
0 & 
 =   (\partial_\mu + \tfrac14  \omega_{\mu}{}^{ab} \gamma_{ab} + \tfrac14  \omega_\mu{}^{AB}\gamma_{AB} - \tfrac12  b_\mu ) \epsilon_- 
 + \tau^A{}_\mu \gamma_A \eta_-   \,,
\\ 
0 & =
  (\partial_\mu  + \tfrac{1}{4} \omega_\mu{}^{ab} \gamma_{ab} 
+ \tfrac{1}{4}  \omega_\mu{}^{AB} \gamma_{AB} 	+  b_\mu )\epsilon_+
+ \tfrac12  \omega_\mu{}^{Aa} \gamma_{Aa} \epsilon_-
\\ & \qquad \qquad  - \tfrac{1}{12}  \tau^A{}_\mu \gamma_A \boldsymbol{\lambda} \epsilon_+
 - \tfrac18 e^a{}_\mu {\boldsymbol{\lambda}}  \gamma_a \epsilon_- 
  + \tau^A{}_\mu \rho_{+A}  - \tfrac12 e^a{}_\mu \gamma_a\eta_- 
\,.
\label{KSoriginal} 
\end{split} 
\ee
In the theory with the half-maximal condition $\epsilon_-=0$, the Killing spinor equations are:
\be
\begin{split} 
0& =  \tau_{A\mu} \left(
\tfrac12  T_{a}{}^{\{AB\}} \gamma_B \gamma^a 
+\tfrac14 ( \eta^{AB} - \tfrac13 \gamma^A \gamma^B ) \boldsymbol{{f}}_B \right) \epsilon_+
 + \tau^A{}_\mu \gamma_A \eta_- 
\,,\\ 
0 & = (\partial_\mu  + \tfrac{1}{4} \omega_\mu{}^{ab} \gamma_{ab} 
+ \tfrac{1}{4}  \omega_\mu{}^{AB} \gamma_{AB} 	+  b_\mu )\epsilon_+
\\ & \qquad \qquad  - \tfrac{1}{12}  \tau^A{}_\mu \gamma_A \boldsymbol{\lambda} \epsilon_+
+\tfrac{1}{12} \tfrac{1}{6} e^a{}_\mu  f_{A bcd} ( \gamma_a{}^{bcd} - 6 \delta_a^b \gamma^{cd} ) \gamma^A \epsilon_+  + \tau^A{}_\mu \rho_{+A}  - \tfrac12 e^a{}_\mu \gamma_a\eta_- 
 \,.
 \label{KShalf}
\end{split} 
\ee
The connections appearing are defined via equations \eqref{omegaBW} and \eqref{connections_nohats_limit}: we can always take the undetermined components $B^A$ and $W^{\{AB\} a}$ of \eqref{omegaBW} to vanish, as these can be absorbed into the Stuckelberg parameters $\eta_-$ and $\rho_{+A}$.
Note that the latter can in principle simply be solved for from the above equations in terms of $\epsilon_\pm$ and their derivatives.
Backsubstituting then gives an irreducible set of equations purely in terms of $\epsilon_\pm$ (and in which the undetermined connection components will automatically drop out).
However, it is convenient in practice to include these $\eta_-$ and $\rho_{+A}$ parameters, as the system of equations to solve is somewhat simpler before these are eliminated.
Note that this means we ultimately still have 32 (or 16) independent possible supersymmetry transformation parameters $\epsilon_\pm$.

\subsection{Flat Membrane Newton-Cartan background}
\label{flatBG}
Let's first consider the Killing spinor equations in a flat background, where $\tau^A{}_\mu = \delta^A_\mu$, $e^a{}_\mu = \delta^a{}_\mu$ and all other fields vanish.
The Killing spinor equations for the theory with maximal supersymmetry then become
\be
\partial_a \epsilon_- = 0 \,,\quad
\partial_A \epsilon_- + \gamma_A \eta_- = 0\,,
\quad
\partial_a \epsilon_+ - \tfrac12 \gamma_a \eta_- = 0 \,,\quad
\partial_A \epsilon_+ + \rho_{+A} = 0 \,.
\label{KSflat_orig} 
\ee
Eliminating the Stuckelberg parameters we obtain 
\be
\eta_- = - \tfrac13 \gamma^A \partial_A \epsilon_- = \tfrac14 \gamma^a \partial_a \epsilon_+\,,\quad
\rho_{+A} = - \partial_A \epsilon_+ + \tfrac13 \gamma_A \gamma^B \partial_B \epsilon_+ \,,
\ee
so that we are left needing to solve
\begin{align}
\partial_a \epsilon_- &  = 0 \,,\quad
\partial_A \epsilon_- - \tfrac13 \gamma_A \gamma^B \partial_B \epsilon_- = 0 \,,
\label{KS_flat_minus}
\\
\gamma^A \partial_A \epsilon_+  & = 0\,, \quad
\partial_a \epsilon_+ - \tfrac18 \gamma_a \gamma^b \partial_b \epsilon_+ = 0 \,,
\label{KS_flat_plus} 
\\
\gamma^A \partial_A \epsilon_- &= - \tfrac34 \gamma^a \partial_a \epsilon_+ \,.
\label{KS_flat_combined}
\end{align}
By taking second derivatives we learn that $\partial^A \partial_A \epsilon_- = 0 = \partial^a \partial_a \epsilon_+$.
However as $\partial_a \epsilon_- =0$ implies that $\partial_a \eta_- = 0$, from \eqref{KSflat_orig} we also have the stronger condition $\partial_a \partial_b \epsilon_+ = 0$.
We can solve for $\epsilon_+$ as 
\be
\epsilon_+ = \gamma_a \Theta_+ (x^A) x^a + \theta_{+}(x^A) \,,
\ee
with both $\Theta_{+}$ and $\theta_{+}$ obeying the massless longitudinal Dirac equation,
\be
\gamma^A \partial_A \Theta_{+} = 0 = \gamma^A \partial_A \theta_{+} \,.
\ee
Given this solution, we have $\eta_- = 2 \Theta_+$ and from the original equations \eqref{KSflat_orig} we obtain
\be
\partial_A \epsilon_- + 2 \gamma_A \Theta_+ = 0 \,,
\ee
or $\Theta_+ = - \tfrac16 \gamma^A \partial_A \epsilon_-$.
Evidently there is an infinite family of solutions with $\Theta_+ = 0$, namely
\be
\epsilon_- = \text{constant} \,,\quad
\epsilon_+ = \theta_{+}(x^A) \,,\quad
\eta_- = 0 \,,\quad
\rho_{+A} = - \partial_A \theta_{+}\,,
\ee
in terms of arbitrary solutions $\theta_{+}(x^A)$ of the longitudinal massless Dirac equation.
The truncation $\epsilon_- = 0$ gives a solution for the half-maximal theory.

The appearance of infinite dimensional Killing symmetries in non-relativistic backgrounds has been observed previously in stringy contexts, e.g. \cite{Batlle:2016iel} (for Killing vectors) and \cite{Blair:2020gng} (for Killing vectors and spinors). Therefore it is not wholly unexpected.
Indeed, the $\epsilon_+$ part of this solution is reminiscent of the Killing spinor solution found in \cite{Blair:2020gng} in the context of the putative double field theory description of the ten-dimensional $\mathcal{N}=2$ SNC limit.
It is not clear how this description is related to the dimensional reduction of the 11-dimensional theory we have obtained. For instance, it naively lacks both Stuckelberg symmetries and constraints.
One can observe that the Killing spinor equations (2.53) and (2.54) of \cite{Blair:2020gng} include the longitudinal dimensional reduction of the conditions $\partial_a \epsilon_\pm=0$ and $\gamma^A \partial_A \epsilon_\pm = 0$.

\subsection{Wound M2 background and Newton potential}
\label{woundM2}

We next analyse a particularly simple solution of the non-relativistic supergravity.
This is the bosonic uplift to eleven dimensions of the wound F1 solution of the 10-dimensional SNC supergravity, which was analyzed in section 4.1 of \cite{Bergshoeff:2022pzk} and shown to be a half-supersymmetric solution of the $\mathcal{N}=1$ SNC supergravity.
It can also be derived as the U-dual (on two spatial and one null direction) of the 11-dimensional pp-wave solution.
Both these origins indicate that it can be viewed as `wound M2' solution describing a non-relativistic M2 brane wrapping the longitudinal directions of the MNC geometry. 
This background has trivial vielbeins and only the longitudinal component of the three-form is non-zero.
Let the 11-dimensional coordinates be $x^\mu= (x^A, x^a)$ with $A=0,1,2$ and $a=3,\dots,10$.
Then we specify the geometry by:
\be
\tau^A{} = \dd x^A \,,\quad
e_a  = \partial_a \,,\quad
c_{012} = \Phi(x) \,,\quad
\lambda_{abcd} = 0 \,.
\label{woundM2bg}
\ee
The solution coming from U-duality will have $\Phi(x) = \Phi(x^a)$ depending on transverse directions only, however for the moment we take \eqref{woundM2bg} as an ansatz inspired by duality or uplift, and remain agnostic about whether we can also have longitudinal coordinate dependence.
The expectation is that $\Phi(x)$ will serve as a Newtonian potential, as we previously discussed and following expectations from \cite{Bergshoeff:2022pzk}.

All the intrinsic torsion constraints vanish, and the only non-trival components of a spin connection are
\be
\omega_\mu{}^{Aa} = \tfrac13 \delta^A{}_\mu \partial^a \Phi \,,
\ee
so that the only non-trivial curvature is:
\be
 R_{\mu\nu}{}^{Aa} = \tfrac23 \delta^A_{[\nu} \partial_{\mu]} \partial^a \Phi
\Rightarrow
\breve{R}_{Ab}{}^{Aa} = -  \partial_b \partial^a \Phi \,.
\ee
Taking the trace of $\breve{R}_{Ab}{}^{Aa}$ gives the Poisson equation.
This is the only non-trivial equation of motion and dictates that the function $\Phi$ should be harmonic in the transverse space:
\be
\partial^a \partial_a \Phi = 0 \,.
\ee
This is the expected behaviour for a Newtonian gravitational potential, and is consistent with the ten-dimensional uplift as well as the U-duality to known relativistic solutions.
Note that the equations of motion do not constrain the longitudinal dependence of $\Phi$, so in principle we may allow $\Phi$ to have arbitrary dependence on the coordinates $x^A$. 

In the theory with half-maximal supersymmetry, we can immediately solve the Killing spinor equations \eqref{KShalf}. 
They reduce for the background \eqref{woundM2bg} to the Killing spinor equations for flat spacetime considered in the previous subsection, with the imposition of $\epsilon_- = 0$.
Accordingly, there is again an infinite family of solutions
\be
\epsilon_+ = \theta_{+}(x^A) \,,\qquad
\eta_- = 0 \,,\qquad
\rho_{+A} = - \partial_A \epsilon_{+}\,,
\ee
where $\gamma^A \partial_A \theta_{+}(x^B) = 0$.

Now let us consider the theory with maximal supersymmetry.
We now have to take into account the curvature constraints \eqref{allconstraints_bos2}, in particular that $D_a \breve{R}_{A(b}{}^A{}_{c)} = 0$, which requires the stronger condition
\be
\partial_a \partial_b \partial_c \Phi = 0 \,,
\ee
on the function $\Phi$.
We write the solution in terms of arbitrary functions of the longitudinal coordinates as
\be
\Phi = M_{ab}(x^A) x^a x^b + N_a(x^A) x^a + C(x^A) \,,
\ee
where $M_{ab}$ is symmetric and also, in order to satisfy the Poisson equation, traceless i.e. $\delta^{ab} M_{ab} = 0$,
The Killing spinor equations \eqref{KSoriginal} are:
\be
\partial_a \epsilon_- = 0 \,,\quad
\partial_A \epsilon_- + \gamma_A \eta_- = 0 \,,\quad
\partial_a \epsilon_+ - \tfrac12 \gamma_a \eta_- = 0 \,,\quad
\partial_A \epsilon_+ + \rho_{+A} + \tfrac16 \partial_a \Phi \gamma_A \gamma^a \epsilon_- = 0 \,,
\ee
the first three of which appeared in the case of the flat background.
Accordingly we can write
\be
\epsilon_+ = \gamma_a x^a \Theta_+(x^A) + \theta_+(x^A) \,,
\ee
where now from the gamma trace of the final equation we obtain
\be
\gamma^A \partial_A  (\gamma_a x^a \Theta_+ + \theta_+)
+  M_{ab} x^a \gamma^b \epsilon_- + \tfrac12 N_a \gamma^a \epsilon_- = 0 \,.
\ee
Taking $\partial_a$ of this equation we get $
\gamma^A \gamma_a \partial_A \Theta_+ + M_{ab} \gamma^b \epsilon_- = 0$, 
and contracting with $\gamma^a$ we find 
$\gamma^A \partial_A \Theta_{+} = 0$,
using that $M_{ab}$ is symmetric and traceless.
This in turn implies
\be
M_{ab} \gamma^b \epsilon_-  =0 \,.
\ee
Given that $M_{ab}$ is symmetric, we can diagonalise it using an orthogonal matrix, which will not affect the Clifford algebra obeyed by the $\gamma^a$. Then as long as $M_{ab}$ is not identically zero we find $\gamma^a \epsilon_- = 0$ for some $a$, which means that $\epsilon_- = 0$.
The latter implies $\eta_- = 0$, hence $\partial_a \epsilon_+ = 0$, and so $\Theta_+ = 0$.
In this case we find again the flat background solutions
\be
M_{ab} \neq 0 : \quad 
\epsilon_-=0 \,,\quad
\epsilon_+ = \theta_+(x^A) \,,\quad
\eta_- = 0 \,,\quad
\rho_{+A} = - \partial_A \theta_+ \,,\quad
\gamma^A \partial_A \theta_+ = 0 \,.
\ee 
This solution has $16$ independent Killing spinors. 

If on the other hand, $\Phi$ is only linear in the transverse coordinates, so $M_{ab} = 0$ identically, then we still need to solve
\be
\gamma^A \partial_A \theta_+ + \tfrac12 N_a (x^A) \gamma^a \epsilon_- = 0\,.
\ee
If $N_a$ is in fact constant and not an arbitrary set of functions of the longitudinal coordinates, one solution is
\be
M_{ab} = 0 : \quad
\epsilon_- = \text{constant} \,\quad
\epsilon_+ = -  \tfrac16 \gamma_A x^A N_a \gamma^a \epsilon_- + \epsilon_+'\,,\quad
\eta_- =0 \,,\quad
\rho_{+A} = 0\,,
\ee
with $\epsilon_+'$ constant. 
This solution has 32 independent Killing spinors.

In the maximally supersymmetric non-relativistic supergravity, the solution for the Newton potential $\Phi$ satisfies stronger constraints than just the Poisson equation and consequently does not constitute a supersymmetrisation of standard Newton-Cartan gravity. Surprisingly, it thus seems that there is no maximally supersymmetric version of 11-dimensional Newton-Cartan gravity. Note that examples of supersymmetrisations of Newton-Cartan gravity in lower dimensions and with less supersymmetry have been found in 3 and 10 dimensions (see \cite{Bergshoeff:2022iyb} for a recent overview).

\subsection{Lambert-Smith holographic M2 background}
\label{LS} 

An interesting solution to the bosonic part of the non-relativistic supergravity theory was obtained recently in \cite{Lambert:2024uue}.
In this paper, the authors take an MNC limit of an M2 background solution, where the longitudinal directions of the limit only overlap with the time direction of the longitudinal directions of the solution. This orientation can be illustrated as follows:

\begin{center}
\begin{tabular}{c|c|c|c|c|c|c|c|c|c|c|c|}
 & $0$ &  $1$ & $2$ & $3$ & $4$ & $5$ & $6$ & $7$ & $8$ & $9$ & ${10}$ \\
MNC limit & $\times$ & $\times$&  $\times$ & --& --& --& --& --&-- &-- & -- \\
M2 solution & $\times$  &-- &-- &$\times$ & $\times$ &-- &-- &-- &-- & -- &--\\
\end{tabular}
\end{center}

\noindent This limit involved rescaling the coordinates $(x^0,x^1,x^2)$ longitudinally (i.e. by a factor of $c$) and the coordinates $x^3,\dots, x^{10}$ transversally (by a factor of $c^{-1/2}$). The harmonic function of the M2 brane solution was $\cH = 1 + R^6/r^6$, with $r^2 = (x^1)^2 + (x^2)^2 + (x^5)^2 + \dots + (x^{10})^2$ and in order to obtain a non-trivial solution the constant $R$ was rescaled as $R \rightarrow c R$.

The latter rescaling is not part of the definition of the MNC limit that we have used to obtain the limit of the theory itself.
With just the rescaling of the coordinates, the limit of the M2 solution gives a trivial background.
The additional rescaling of $R$ (which could potentially be viewed as sending the number of M2 branes sourcing the geometry to infinity) ensures that the limit produces a background geometry with a non-trivial profile.
This additional rescaling can be viewed as a manipulation required to engineer the relativistic M2 solution into a form from which the MNC limit can be taken. 
It is not however guaranteed a priori that this manipulation guarantees that the limit of the relativistic solution is a solution of the limit of the relativistic theory, which must be checked independently.

It was argued in \cite{Lambert:2024uue} that the resulting MNC geometry in its near horizon limit is holographically dual to a non-relativistic limit of the theory on multiple membranes. As such it should be viewed as analogous to (and indeed a limit of) the maximally supersymmetric $\text{AdS}_4 \times S^7$ solution of the relativistic 11-dimensional supergravity. 
Similar rescalings were used for other brane solutions in other non-relativistic limits in \cite{Fontanella:2024rvn,Lambert:2024yjk}, again with holographic interpretations: it would be interesting to further explore the physical meaning of this mechanism (and compare to the discussion in \cite{Guijosa:2023qym,Avila:2023aey}).

Here we simply accept the solution of \cite{Lambert:2024uue} as a geometry of potential interest for the 11-dimensional non-relativistic supergravity.
In fact, we will generalise the background of \cite{Lambert:2024uue} and obtain an infinite family of supersymmetric backgrounds. 
Let us introduce some (flat) indices adapted to the M2/MNC limit configuration:
\be
A = (0,\alpha) \,,\quad \alpha=1,2\,,\quad
a=(m,I) \,,\quad m=3,4 \,,\quad I = 5,\dots,10 \,.
\ee
Below we will denote the corresponding curved indices with a hat. (Coordinates $x^\mu$ and derivatives $\partial_\mu$ will be written `naturally' with unhatted indices; however, this does \emph{not} denote contraction with a vielbein.)
We take the bosonic fields to be:
\be
\begin{split} 
\tau^0 & = \cH^{-1/3} \dd x^0 \,,\quad
\tau^\alpha = \cH^{1/6} \dd x^\alpha \,,\quad
e^{m} = \cH^{-1/3} \dd x^m \,,\quad
e^I = \cH^{1/6} \dd x^I \,,\\
c_{(3)} &= \cH^{-1} \dd x^0 \wedge \dd x^3 \wedge \dd x^4\,,\quad
\lambda_{abcd} = 0 \,.
\end{split}
\label{LSansatz}
\ee
The powers of the function $\cH=\cH(x^1,x^2)$ are determined from the relativistic M2 solution.
We make the obvious choice for the inverse vielbein pair. To compare with the results of \cite{Lambert:2024uue}, we note that the coordinates used in  \cite{Lambert:2024uue} are related to the ones chosen here via $(u,\bar u) \leftrightarrow (x^1,x^2)$, $(z,\bar z) \leftrightarrow (x^3,x^4)$ and $v \leftrightarrow (x^5,\dots, x^{10})$. Moreover, the solution of \cite{Lambert:2024uue} is given by \eqref{LSansatz} with $\cH$ given by $\cH =1 + R^6/( (x^1)^2+(x^2)^2)^3$. The near horizon limit of this solution is then obtained simply by dropping the constant term $1$ in $\cH$. 

The field strength is
\be
f_{(4)} = -\partial_{\alpha} \cH^{-1} \dd x^0 \wedge \dd x^{\alpha} \wedge \dd x^3 \wedge \dd x^4 \,.
\ee
It follows that the only non-zero field strength components are those of $f_{ABab}$, and  the only non-zero torsion components are those of $T_{AB}{}^C$, with:
\be
f_{0\alpha mn} = - \cH^{5/6} \partial_\alpha \cH^{-1}\epsilon_{mn} \,,\quad
T_{0 \alpha}{}^0 = \tfrac13 \cH^{-7/6} \partial_{ \alpha} \cH\,,\quad
T_{\beta \gamma }{}^\alpha =  \tfrac16 2 \cH^{-7/6} \delta^\gamma_{[\alpha} \partial_{\beta]} \cH\,.
\ee
Thus this background satisfies all intrinsic torsion constraints \eqref{geoconstraints} imposed in the theory with maximal supersymmetry.
The non-zero components of the spin connections are:
\begin{subequations} 
\begin{align}
\omega_\mu{}^{AB} :\quad &
\omega_{\hat 0}{}^{0 \alpha} = - \tfrac13 \cH^{-3/2} \partial^\alpha \cH \,,\quad
\omega_{\hat \alpha}{}^{\beta \gamma}= \tfrac16 \cH^{-1} 2 \delta^{[\beta}_{\alpha} \partial^{\gamma]} \cH \,,
\\ 
\omega_\mu{}^{ab}: \quad &
\omega_{\hat \alpha}{}^{34} = \tfrac12 \cH^{-1} \epsilon_{\alpha \beta} \partial^\beta \cH\,,
\label{LS_omega_ab}
\\
\omega_\mu{}^{Aa}: \quad &
\omega_{\hat m}{}^{\alpha n} =
\cH^{-3/2} ( 
\tfrac13 \delta^n_m \partial^\alpha \cH 
+\tfrac12 \epsilon^{\alpha \beta} \epsilon_m{}^n \partial_\beta \cH 
)
\,,\quad
\omega_{\hat I}{}^{\alpha J} = - \tfrac16 \cH^{-1} \delta_I^J \partial^\alpha \cH \,,
\end{align}
\end{subequations} 
with $b_\mu$ vanishing. 
Using these expressions, we compute the curvatures appearing in the constraints of the non-relativistic supergravity with maximal supersymmetry.
The only non-zero well-defined curvatures are $\breve{R}_{AB}{}^{ab}$ and $\breve{R}_{Aa}{}^{Ab}$, with:
\be
\breve{R}_{\alpha \beta}{}^{mn} = -\tfrac12 \epsilon_{\alpha \beta} \epsilon^{mn} \cH^{-1/3} \partial_\gamma \partial^\gamma \log \cH \,,\quad
\breve{R}_{A(a}{}^A{}_{b)} =   \tfrac38 \cH^{-1/3}\partial_\alpha \partial^\alpha \log \cH 
\begin{pmatrix}
	1_2 & 0 \\
	0 & -\tfrac13 1_6
\end{pmatrix}\,.
\label{LSR}
\ee
Note that $\breve{R}_{Aa}{}^{Aa} = 0$, which means the Poisson equation is trivially satisfied.
The vanishing of the intrinsic torsions as well as all other curvature conditions means that all the equations of motion are satisfied \emph{with no restrictions on the function $\cH(x^1,x^2)$}.
This means that the background \eqref{LSansatz} is a solution of the bosonic theory for $\cH$ an arbitrary function of the longitudinal spatial coordinates.
It is then also automatically a solution of the theory with half-maximal supersymmetry.

In order for it to be a solution of the theory with maximal supersymmetry, from the constraints \eqref{allconstraints_bos2}, we need to impose that $D_c \breve{R}_{A(a}{}^{A}{}_{b)}=0$.
It follows from the form of the $\mathrm{SO}(8)$ spin connection \eqref{LS_omega_ab} and the curvature in \eqref{LSR} that this is the case.

Let us make a remark about the possibility of imposing stronger conditions on this geometry.
At the end of appendix \ref{tower_derivation}, we mention that instead of imposing the constraints that the transverse covariant derivatives of $\lambda_{abcd}$, $\breve{r}_{+a}$ and $\breve{R}_{A(b}{}^A{}_{c)}$ must vanish, we could actually require that these tensors vanish themselves, which from the variation of $\breve{r}_{+a}$ would also require $\breve{R}_{AB}{}^{ab} = 0$ (when covariant derivatives are taken one obtains a possible constraint $D_c \breve{R}_{AB}{}^{ab}=0$ which is equivalent to a Bianchi identity by other constraints, and satisfied by the curvature in \eqref{LSR}).
Requiring both curvatures in \eqref{LSR} to vanish identically amounts to requiring that
 $\log \cH$ must be harmonic in two dimensions, i.e. $\log \cH \propto \log ((x^1)^2 + (x^2)^2)$, or
\be
\cH(x^1,x^2) = A ((x^1)^2 + (x^2)^2)^\beta 
\label{Hbeta} 
\ee
for arbitrary power $\beta$ and some (dimensionful) constant $A$.
The solution of \cite{Lambert:2024uue} in the near horizon limit has $\beta=-3$ and $A = R^6$.


We now seek to solve the Killing spinor equations of the maximal supersymmetric theory for this background.
We restrict to `near horizon' solutions with $\cH$ given by \eqref{Hbeta}, and set $A=1$ for simplicity.
The Killing spinor equations are given in equation \eqref{KSoriginal}.
The Killing spinor equations \eqref{KShalf} of the half-maximally supersymmetric theory are for this background simply the $\epsilon_-=0$ truncation.
For the ansatz we have made, we find that \eqref{KSoriginal} in components leads to:
\begin{subequations}
\begin{align}
\label{plusKSa}
0 & = \partial_0 \epsilon_+ -\tfrac16 \cH^{-1/2} \partial_\alpha \log \cH \gamma_0 \gamma^\alpha \epsilon_+ + \cH^{-1/3} \rho_{+0}\,,
\\
\label{plusKSb}
0 & = \partial_\alpha \epsilon_+ +  \epsilon_{\alpha \beta} \partial^\beta \log \cH ( \tfrac{1}{12} \gamma_{12} + \tfrac14 \gamma_{34} ) \epsilon_+  + \cH^{1/6} \rho_{+\alpha}\,,
\\
\label{plusKSc}
0 & = \partial_m \epsilon_+ 
+ \cH^{-1/2} \tfrac12 ( \tfrac13  \partial^\alpha\log \cH \gamma_\alpha \gamma_m 
+ \tfrac12 \epsilon^{\alpha \beta} \epsilon_{mn} \partial_\beta \log \cH \gamma_\alpha \gamma^n ) \epsilon_- - \tfrac12 \cH^{-1/3} \gamma_m \eta_- \,,\\
0 & = \partial_I \epsilon_+ - \tfrac{1}{12} \partial^\alpha \log \cH \gamma_\alpha \gamma_I \epsilon_-
- \tfrac12 \cH^{1/6} \gamma_I \eta_- \,,
\label{plusKSd}
\\
\label{minusKSa}
0 & = \partial_0 \epsilon_- -\tfrac16 \cH^{-1/2} \partial_\alpha \log \cH \gamma_0 \gamma^\alpha \epsilon_- + \cH^{-1/3} \gamma_0 \eta_- \,,
\\
\label{minusKSb}
0 & = \partial_\alpha \epsilon_- +  \epsilon_{\alpha \beta} \partial^\beta \log \cH ( \tfrac{1}{12} \gamma_{12} + \tfrac14 \gamma_{34} ) \epsilon_-  + \cH^{1/6} \gamma_\alpha \eta_- \,,
\\ 
\label{minusKSc}
0 & = \partial_m \epsilon_-\,,
\quad
0  = \partial_I \epsilon_-\,.
\end{align}
\end{subequations} 
It's convenient to define polar coordinates $x^1 = r\cos \theta$, $x^2 = r\sin \theta$ and take $\cH = r^{2\beta}$.
Then the Killing spinor equations become
\begin{subequations}
\begin{align}
\label{KSpolara}
0 & = \partial_0 \epsilon_+ -\tfrac{\beta}{3} r^{-\beta-1}  \gamma_0 \gamma_1 \exp(- \gamma_0 \theta)  \epsilon_+ + r^{-2\beta/3} \rho_{+0}\,,\\\label{KSpolarb}
0 & = \partial_r \epsilon_+ + r^{\beta/3}(  \cos \theta \rho_{+1} + \sin \theta \rho_{+2} )\,, \\\label{KSpolarc}
0 & = \partial_\theta \epsilon_+ - 2 \beta ( \tfrac{1}{12} \gamma_{12} + \tfrac14 \gamma_{34} ) \epsilon_+  +r^{\beta/3+1} ( \cos \theta \rho_{+2}  - \sin \theta \rho_{+1}) \,,\\\label{KSpolard}
0 & = \partial_m \epsilon_+ 
- r^{-\beta-1} \beta \left( \tfrac{1}{3}   \gamma_m  \gamma_1  
- \tfrac{1}{2} \epsilon_{mn} \gamma^n  \gamma_2    \right) \exp ( \gamma_0 \theta) \epsilon_-
- \tfrac12 r^{-2\beta/3} \gamma_m \eta_- \,,\\\label{KSpolare}
0 & = \partial_I \epsilon_+ + \tfrac12 r^{\beta/3} \gamma_I \left( \tfrac{\beta}{3} r^{-1-\beta/3} \gamma_1 \exp(\gamma_0\theta)  \epsilon_-
- \eta_-\right) \,,
\\\label{KSpolarf}
0 & = \partial_0 \epsilon_- - r^{-2\beta/3} \gamma_0 \left( \tfrac{\beta}{3} r^{-1-\beta/3}  \gamma_1 \exp ( \gamma_0 \theta) \epsilon_- - \eta_-\right) \,,\\\label{KSpolarg}
0 & = \partial_r \epsilon_- + r^{\beta/3}\gamma_1 \exp(\gamma_0 \theta) \eta_-\,, \\\label{KSpolarh}
0 & = \partial_\theta \epsilon_- - 2 \beta ( \tfrac{1}{12} \gamma_{12} + \tfrac14 \gamma_{34} ) \epsilon_-  +r^{\beta/3+1}\gamma_2 \exp( \gamma_0 \theta) \eta_- \,,\\\label{KSpolari}
0 & = \partial_m \epsilon_-\,,\quad 0  = \partial_I \epsilon_-\,.
\end{align}
\end{subequations}
Here we used extensively the fact that the spinors have definite chirality under the longitudinal projectors $\Pi_\pm=\tfrac12 (1\pm \gamma_{012})$, so in particular $\gamma_1 \gamma_2 \epsilon_\pm = \mp \gamma_0 \epsilon_\pm$.

Now, suppose we attempted to find a solution of these equations with the Stuckelberg transformation parameters set to zero, $\rho_{+A} = 0 = \eta_-$.
Then we would have $\partial_r \epsilon_+ = 0$, $\partial_r \epsilon_-=0$.
Differentiating the $\mu=0$ equations, \eqref{KSpolara} and \eqref{KSpolarf} would then imply that $\epsilon_\pm = 0$ if $\beta \neq -1$.
If $\beta = -1$ the solutions of the $\mu=\theta$ equations, \eqref{KSpolarc} and \eqref{KSpolard} would give $\epsilon_\pm = \exp(\pm \tfrac16 \gamma_0 \theta) \exp(-\tfrac12 \gamma_{34} \theta)\tilde \epsilon(x^0,x^m,x^I)$.
However from the $\mu=0$ equations we would find $\partial_0 \tilde \epsilon_\pm = -\tfrac13 \gamma_0\gamma_1 \exp(\mp \tfrac23 \gamma_0 \theta) \tilde \epsilon_\pm$, where the left-hand side is independent of $\theta$ but the right-hand side is not.
So we cannot find a solution in this case.

We therefore have to allow for non-vanishing Stuckelberg parameters.
Inspecting the above equations, a simplifying possibility is to assume $\partial_0 \epsilon_- = 0$ or equivalently $\partial_I \epsilon_+ = 0$, which determines $\eta_-$ from \eqref{KSpolare} or \eqref{KSpolarf}.
Substituting this value of $\eta_-$ into the other equations, it is straightforward to obtain the following solution:
\begin{subequations}
\begin{align}
\epsilon_+ & = \epsilon_+' + \epsilon_+'' \,,\\
\epsilon_+' & =  r^{\beta/3} \exp ( - \tfrac{2\beta}{3} \gamma_0 \theta) \tilde \epsilon_{++}
+ r^{\beta/3} \exp (  \tfrac{\beta}{3} \gamma_0 \theta) \tilde \epsilon_{+-}\,, \\ 
\epsilon_+'' & =  \beta  r^{-1-4\beta/3}  \exp ( [1+\beta] \gamma_0 \theta) x^m \gamma_m \gamma_1 \tilde \epsilon_{--}\,,
\\ 
\epsilon_- & = r^{-\beta/3}\tilde \epsilon_{-+}
+r^{-\beta/3} \exp( \beta \gamma_0 \theta ) \tilde \epsilon_{--} \,,
 \\
\rho_{+0} & =\gamma_0 \gamma_1 \tfrac{\beta}{3} r^{-1-\beta/3} \exp(-\gamma_0 \theta) (\epsilon_+'+\epsilon_+'')\,,
\\ 
 \rho_{+1} & =r^{-1-\beta/3}  \left( - \tfrac{\beta}{3} \cos \theta \epsilon_+'
+ \tfrac{\beta}{3} \sin \theta \gamma_0 \epsilon_+'' 
+(1+\tfrac{4\beta}{3}) \exp(\gamma_0 \theta) \epsilon_+''
\right)\,,
\\
\rho_{+2} & = r^{-1-\beta/3} 
\left(
- \tfrac{\beta}{3} \sin \theta \epsilon'_+
- \tfrac{\beta}{3} \cos \theta \gamma_0 \epsilon_+'' 
-(1+\tfrac{4\beta}{3}) \gamma_0 \exp(\gamma_0 \theta) \epsilon_+''
\right)\,,
\\ 
\eta_- & = \tfrac{\beta}{3} r^{-1-\beta/3} \gamma_1 \exp(\theta \gamma_0) \epsilon_- \,,
\end{align} 
\end{subequations}
where $\tilde \epsilon_{+\pm}$ and $\tilde \epsilon_{-\pm}$ are constant and obey $\gamma_{034} \tilde \epsilon_{+\pm} = \pm \tilde \epsilon_{+\pm}$, $\gamma_{034} \tilde \epsilon_{-\pm} = \pm \tilde \epsilon_{-\pm}$. 
Therefore this is a solution involving 32 independent Killing spinors.
We have not demonstrated that this is the unique solution to the Killing spinor equations.
Nonetheless, its existence demonstrates that we can view the family of solutions given by \eqref{LSansatz} and \eqref{Hbeta} as supersymmetric solutions to the non-relativistic eleven-dimensional supergravity, when we impose the constraints compatible with either maximal or half-maximal supersymmetry (in the latter case, the Killing spinor solution is simply the truncation $\tilde \epsilon_{-\pm} = 0$).

It would be interesting to further explore the geometry of this family of solutions.
In particular, requiring the above Killing spinor solution to be globally defined may restrict the allowed values of $\beta$.
It would also be of interest to see if the solutions other than the case of \cite{Lambert:2024uue} have an origin as a limit of a relativistic background, with or without a holographic interpretation.  
We leave this as interesting future work.

\section{Discussion}
\label{discussion}

In this paper, we have explored the membrane Newton-Cartan non-relativistic limit of 11-dimensional supergravity.
This limit has thrown up various surprises.  

We have found that we have to impose constraints, compatible either with maximal or half-maximal supersymmetry.
In the maximal case, this leads to a surprisingly complicated series of constraints -- which remarkably do not trivialise the theory, but admit an infinite family of supersymmetric solutions including the putative holographic M2 near horizon non-relativistic geometry of \cite{Lambert:2024uue}.

In the case with half-maximal supersymmetry, we found an unusual realisation of supersymmetry.
The attractive feature of this scenario is that it imposes fewer constraints on the theory -- in particular no constraints on bosonic curvatures are needed. This avoids the surprising tension between the maximal supersymmetry constraints and the possibility of having a Newton potential of standard form in the solution of section \ref{woundM2}.
However, the supersymmetry algebra \eqref{QQhalf} with parameters \eqref{QQparhalf} no longer closes into diffeomorphisms. 
Therefore it is not clear to what extent this should be viewed as supergravity in the standard sense.

In the case with maximal supersymmetry, our results 
show at the linearised level that there is a consistent multiplet  of constraints defining a consistent supermultiplet realizing a representation of an underlying non-relativistic superalgebra. In order to complete this multiplet at the non-linear level it would be beneficial to
reformulate the linearised results of this paper  in terms of constrained superfields on  a suitable non-relativistic superspace. For earlier discussions of such a non-relativistic superspace and superfields in lower dimensions, see e.g. \cite{deAzcarraga:1991fa,Nakayama:2009ku,Bergman:1995zr,Clark:1983ne,Auzzi:2019kdd}.
Furthermore, it would be interesting to achieve an accurate counting of the number of bosonic and fermionic non-gauge field components by gauge fixing along the lines of \cite{Andringa:2013mma}. In the present context this is more involved due to the presence of more constraints whose geometric meaning is not as clear-cut as in the three-dimensional case. We hope to return to this question in the future.

It is interesting to note that the underlying geometry of the theories considered in this paper, constitutes a generalization of a Newton-Cartan geometry with three new features. Firstly, it includes a three-form (instead of a one-form) and a tangent space splitting in 3 longitudinal and 8 transversal directions (instead of a time and spatial directions). Secondly, it has local conformal symmetry built in and finally, it also incorporates supersymmetry. This geometry exhibits several features that are already present in $2$-brane Galilean geometry (that does not include a three-form, nor a type of superconformal symmetry), such as the presence of intrinsic torsion and undetermined connection components \cite{Bergshoeff:2023rkk}. In the 2-brane Galilean case, these can both be studied and classified via Spencer cohomology. It would be interesting to investigate whether these techniques can be extended to the present situation. 

We have not shown that the non-relativistic limit that we defined in this paper is the unique membrane non-relativistic limit.
For instance, the ansatz for the gravitino expansion \eqref{ExpFer} is not the most general one that is covariant with respect to SO(1,2) and SO(8) transformations.
Modifying the ansatz would alter the structure of the divergences that appear in the supersymmetry transformations, and hence potentially lead to limits with different (ideally simpler) sets of constraints.
We explored some possible expansions along these lines by decomposing the gravitino further (e.g. introducing different $c$--rescalings of the gamma trace and gamma traceless parts of the longitudinal and transverse projections)
but came to the preliminary conclusion that this did not lead to significant simplifications in the structure of the theory.
We nevertheless cannot exclude that there exist other field redefinitions that lead to different  consistent limits as well (for earlier work on algebraic aspects of a classification of such limits, see \cite{Barducci:2019jhj}).

The standard dimensional reduction procedure should lead to a non-relativistic supergravity with $\mathcal{N}=2$ supersymmetry.
We can dimensionally reduce either along a direction longitudinal to the membrane Newton-Cartan limit or transverse to the limit. This should lead to two distinct non-relativistic supersymmetric IIA supergravity theories, one with an underlying string Newton-Cartan geometry and the other with a D2 brane Newton-Cartan geometry, extending the bosonic sectors obtained by these reductions in  \cite{Blair:2021waq}. The latter has recently appeared in \cite{Blair:2023noj} and been termed M$2$T geometry there. T-duality of the SNC type IIA would then connect to the IIB case, whose bosonic sector was obtained in \cite{Bergshoeff:2023ogz}.

In addition, the naive expectation would be that the longitudinal reduction to type IIA SNC supergravity could be truncated to $\mathcal{N}=1$ in order to match the $\mathcal{N}=1$ SNC supergravity of \cite{Bergshoeff:2021tfn}.
However, this is not the case: reducing and truncating, we find a different $\mathcal{N}=1$ expansion and set of constraints.
This can be demonstrated through a representation theory argument, which traces the rearrangement and truncation of the $\mathbf{32}\otimes \mathbf{11}$ fermionic gravitino components in both cases. The truncation from eleven-dimensional supergravity to minimal supergravity in ten dimensions follows a specific process: first, reduce to type IIA in ten dimensions, then truncate the right-handed gravitino ($\mathbf{16}^v_R$) and the left-handed dilatino ($\mathbf{16}_L^s$). We apply this same basic procedure in the non-relativistic setting, based on the observation that both should result from a limit. However, we will demonstrate that these procedures do not commute.
When first truncating to $\mathcal{N}=1$ and taking the limit as described in \cite{Bergshoeff:2021tfn}, we obtain:
\begin{align}
    \mathbf{32}^v\quad\to\quad \mathbf{16}_L^v \oplus \mathbf{16}_R^s\quad\to\quad \big(\mathbf{8}_{L+}^v\oplus\mathbf{8}_{R+}^s\big)\oleft \big(\mathbf{8}_{L-}^v\oplus\mathbf{8}_{R-}^s\big)\,.
\end{align}
Here, $v/s$ denote gravitino/dilatino, $L/R$ indicate eigenvalues under $\gamma_2$, and $+/-$ represent eigenvalues under $\gamma_{01}$. Importantly, $+/-$ also signify behavior under boosts. We use the notation $X_+\oleft X_-$ to indicate that under boosts, $X_+\to X_-\to 0$.
Conversely, if we begin with the decomposition introduced in this paper, we arrive at:
\begin{align}
    \mathbf{32}^v\quad\to\quad \mathbf{16}_+^v\oleft\mathbf{16}_-^v\quad\to\quad \big(\mathbf{8}_{L+}^v\oplus\mathbf{8}_{R-}^s\big)\oleft \big(\mathbf{8}_{L-}^v\oplus\mathbf{8}_{R+}^s\big)\,.
\end{align}
Note that the sublabels on $\mathbf{16}_\pm^v$ indicate eigenvalues of $\gamma_{012}$, whereas the $+/-$ on $\mathbf{8}_{L\pm}^{v}/\mathbf{8}_{R\mp}^s$ indicates eigenvalues of $\gamma_{01}$. This result does not align with the representation structure of the $\mathcal{N}=1$ limit, as the $\gamma_{01}$-chiralities of the dilatini $\mathbf{8}^s_{R\mp}$ do not align with the boost structure as they did above. This discrepancy can be traced back to the observation that decomposing the spinors into $\gamma_{012}=\gamma_{01}\gamma_2$ chiralities effectively couples the $\gamma_2$ and $\gamma_{01}$ chiralities in ten dimensions.
We have attempted to find an alternative, twisted truncation procedure but have been unsuccessful thus far. The intriguing question of how to relate maximal and half-maximal multiplets remains open for future investigation.
Potentially the known ten-dimensional $\mathcal{N}=1$ theory simply does not embed directly into type IIA but should be viewed as a truncation of heterotic supergravity (the bosonic non-relativistic limit of which has been studied in \cite{Bergshoeff:2023fcf,Lescano:2024url}), which may require some Ho\v{r}ava-Witten dimensional reduction on an interval to obtain from the 11-dimensional theory.



In addition to the membrane Newton-Cartan limit considered here, the eleven-dimensional supergravity should also admit a non-relativistic limit associated with the M5 brane, with six longitudinal and five transverse directions.
This would lead to a further non-uniqueness of non-relativistic eleven-dimensional supergravity.

All these possible non-relativistic limits, and their ten-dimensional reductions, should be related by dualities, e.g. the membrane and five-brane decoupling limits are U-dual to each other.
The membrane Newton-Cartan limit considered in this paper is U-dual to the null reduction or DLCQ of M-theory, which leads to the Matrix theory description. See \cite{Blair:2023noj} for a modern overview of the non-relativistic duality web. 
It would be interesting to explore further the implications of these observations for our understanding of M-theory and its decoupling limits in curved supergravity backgrounds. In the same vein, it would be interesting to study Carrollian decoupling limits of eleven dimensional supergravity.

Finally, apart from constructing non-relativistic supergravity theories we also considered solutions. In particular, we found that the  recent solution found by  Lambert and Smith in \cite{Lambert:2024uue} solves the eleven-dimensional non-relativistic supergravity theory constructed in this paper.  This solution, in its near horizon limit, is  a particular non-relativistic limit of AdS${}_4 \times S^7$ that constitutes a holographic dual to a decoupling limit of ABJM theory which describes dynamics on the Hitchin moduli space.  As such it was conjectured that this solution would be maximally supersymmetric which indeed we confirmed. Surprisingly, we found that the Lambert-Smith solution is a particular member of a larger class of solutions. It would be interesting to explore the consequences of this larger set of solutions. 
It was further suggested in \cite{Lambert:2024yjk} that some of the supersymmetries of the supergravity background dual to the non-relativistic limit of ABJM may be non-physical. 
Our Killing spinor solution indicates that this background does admit 32 supersymmetries, at least locally. It would be interesting to further study this and other brane solutions of the non-relativistic 11-dimensional supergravity, in search of other surprises.

\section*{Acknowledgements}
The work of CB is supported through the grants CEX2020-001007-S and PID2021-123017NB-I00, funded by MCIN/AEI/10.13039/501100011033 and by ERDF A way of making Europe. 
The work of JR is supported by the Croatian Science
Foundation project IP-2022-10-5980 “Non-relativistic supergravity and applications”. Nordita is supported in part by NordForsk.
We would like to thank Martin Cederwall, Neil Lambert, Rishi Mouland, Niels Obers, Luca Romano, Ingmar Saberi, Joseph Smith, and Ziqi Yan for useful discussions. 
Without the M.O.N.S.T.E.R., this paper would not have been finished.
The authors have incurred debts of hospitality to numerous institutes where parts of this work have been done: this includes 
KU Leuven (EB, CB, JL, JR, not all at the same time); 
the Erwin Schr\"odinger Institute in Vienna during the program ``ESI workshop on Carrollian physics and holography'' (EB, JL, JR);
the Rijksuniversiteit Groningen (JL, JR); 
the Rudjer Bo\v{s}kovi\'c Institute, Zagreb during the Mini-symposium on ``Physics and Geometry'' 2022 (EB) and 2024 (EB, CB); 
Nordita (EB, CB, JR); 
the Instituto de Física Teórica UAM/CSIC (EB);
the Mainz Institute for Theoretical Physics during the MITP scientific program ``Higher structures, gravity, and fields'' (EB);
and the Vrije Universiteit Brussel (CB).

\appendix

\addtocontents{toc}{\protect\setcounter{tocdepth}{2}}

\section{Technical details}
\label{details}
\subsection{Gamma matrix identities}
\label{gammadetails}

\paragraph{Useful identities}
The product of all gamma matrices (with upper indices) is $\gamma^{012 3 \dots 10} = + 1$.
Hence
\be
\gamma^{\hat a_1 \dots \hat a_p} =
\tfrac{(-1)^{p(p-1)/2} }{(11-p)!} \epsilon^{\hat a_1 \dots \hat a_p \hat b_1 \dots \hat b_{11-p}} \gamma_{\hat b_1 \dots \hat b_{11-p}} \,.
\ee
We split $\hat a= (A,a)$.
The epsilon tensors with flat indices are $\epsilon_{ABC}$ with $\epsilon_{012}=+1$ and $\epsilon^{ABC}$ with $\epsilon^{012} = -1$, as well as $\epsilon^{a_1\dots a_8}$, and $\epsilon_{a_1 \dots a_8}$ which are numerically identical.
The following relationships are then useful: 
\be
\gamma_{ABC} =  \epsilon_{ABC} \gamma_{012} \,,\quad
\gamma_{AB} =  \epsilon_{ABC} \gamma^C \gamma_{012} \,,
\label{usefullong}
\ee
\be
\gamma^{a_1 \dots a_p} = \tfrac{1}{8-p!} (-1) (-1)^{\tfrac12p(p-1)} \epsilon^{a_1 \dots a_p b_1 \dots b_{8-p}} \gamma_{b_1 \dots b_{8-p}} \gamma_{012} \,.
\label{usefultransfs}
\ee
The longitudinal projectors $\Pi_{\pm} = \tfrac12 ( 1 \pm \gamma_{012})$ are such that
\be
\gamma_A \Pi_\pm  = \Pi_\pm \gamma_A \,,\quad
\gamma_a \Pi_{\pm}  = \Pi_{\mp} \gamma_a \,,
\label{gammaPi}
\ee
If $\chi_\pm \equiv \Pi_\pm\chi$, then $\bar \chi_\pm =  \bar \chi \Pi_\pm$ so conjugates are projected in the same way.
Accordingly one has
\be
\bar\chi \gamma_{A_1 \dots A_q a_1 \dots a_p} \psi
=
\begin{cases}
\bar\chi_+ \gamma_{A_1 \dots A_q a_1 \dots a_p} \psi_+
+ \bar\chi_- \gamma_{A_1 \dots A_q a_1 \dots a_p} \psi_-\,,
 & p\, \text{even}\,, \\
\bar\chi_+ \gamma_{A_1 \dots A_q a_1 \dots a_p} \psi_-
+ \bar\chi_- \gamma_{A_1 \dots A_q a_1 \dots a_p} \psi_+\,,
 & p\,	 \text{odd}\,. \end{cases}
\label{usefulbilinears}
\ee
A useful identity relating the spinorial projectors and the transverse tangent space self-duality projectors is:
\be
\label{proj_mixes}
\gamma^{abcd} \Pi_\pm = \tfrac{1}{4!} P^{(\mp)}{}^{abcd}{}_{efgh} \gamma^{efgh} \,.
\ee

\paragraph{Fierz identities}
From the relativistic 11-dimensional Fierz identities \cite{Freedman:2012zz} we can derive the following Fierz identities for spinors in the non-relativistic theory, according to their chiralities under the projectors $\Pi_\pm$.
For $\lambda_1$ and $\lambda_3$ with the same chirality the Fierz identity is:
\be
\begin{split}
(\bar \lambda_1 \lambda_2) \lambda_3
 & =
 - \tfrac{1}{32} \Big[
 2 (\bar \lambda_1 \lambda_3) 	 \lambda_2
 +2 (\bar \lambda_1 \gamma^A \lambda_3) \gamma_A	 \lambda_2
\\ & \qquad \qquad
+ 2 \tfrac12 (\bar \lambda_1 \gamma^{ab} \lambda_3) \gamma_{ba} 	 \lambda_2
+ 2 \tfrac12 (\bar \lambda_1 \gamma^A \gamma^{ab} \lambda_3) \gamma_A\gamma_{ba} 	 \lambda_2
\\ & \qquad \qquad
 + \tfrac{1}{4!}   (\bar \lambda_1 \gamma^{abcd} \lambda_3) \gamma_{dcba} \lambda_2
 + \tfrac{1}{4!}   (\bar \lambda_1 \gamma^{Aabcd} \lambda_3) \gamma_{dcbaA} \lambda_2
 \Big] \,.
 \end{split}
\label{FierzSameChirality}
\ee
For $\lambda_1$ and $\lambda_3$ with the opposite chirality the Fierz identity is:
\be
\begin{split}
(\bar \lambda_1 \lambda_2) \lambda_3
 & =
 - \tfrac{1}{16} \Big[
  (\bar \lambda_1\gamma^a \lambda_3) 	\gamma_a \lambda_2
 + (\bar \lambda_1 \gamma^{Aa} \lambda_3) \gamma_{aA}	 \lambda_2
 \\ & \qquad \qquad
  + \tfrac{1}{3!}   (\bar \lambda_1 \gamma^{abc} \lambda_3) \gamma_{cba} \lambda_2
+  \tfrac{1}{3!} (\bar \lambda_1 \gamma^{Aabc} \lambda_3) \gamma_{cbaA} 	 \lambda_2
 \Big] \,.
\end{split}
\label{FierzOppositeChirality}
\ee


\subsection{Expansion of supersymmetry transformations of the fermions} 
\label{gorydetails}

Starting from the relativistic supersymmetry transformations \eqref{susytransforig}, making the expansions \eqref{ExpBos} and \eqref{ExpFer} of the bosonic and fermionic fields, one finds that the supersymmetry transformations can be expanded as $\delta \Psi_\mu = \sum_{n=-3}^{2} c^{3n/2} (X_{3n/2})_\mu \epsilon$ in terms of particular operators $(X_m)_\mu$.
Those appearing with integer powers $m=(3,0,-3)$ of $c$ map between spinors of the same chirality, while those with half-integer powers $m=(3/2,-3/2,-9/2)$ change chirality.
After using the equation of motion symmetry to remove divergent terms from the variation of $\psi_{+\mu}$ acccording to \eqref{deltaextra} and hence \eqref{deltaprimepsi0}, a straightforward but tedious calculation shows that the finite part of the supersymmetry variations of the fermions take the form:
\be
\begin{split}
\delta'_0 \psi_{+\mu} & = (X_0)_\mu \epsilon_+ + (X_{-3/2}){}_{\mu} \epsilon_-  - \tfrac{1}{12} \tau^A{}_\mu \gamma_A {\boldsymbol{\lambdaone}} \epsilon_+
- \tfrac{1}{8} e^a{}_\mu {\boldsymbol{\lambdaone}} \gamma_a \epsilon_- \,,\\
\delta'_0 \psi_{-\mu} & = (X_0)_\mu\epsilon_- + (X_{3/2})_\mu \epsilon_+ \,,
\end{split}
\label{deltapsiAfterEomSymm}
\ee
where the explicit forms of the operators that appear are as follows:
\begin{subequations}
\begin{align}
(X_0)_\mu& = \nonumber
\mathbb{I} \, \partial_\mu + \tfrac14 \hat\omega(\tau)_{\mu}{}^{AB} \gamma_{AB}
\\ & \qquad \nonumber
+ \tfrac14 (\hat\omega(e)_\mu{}_{ab} - \tfrac{1}{12}\tau_{A \mu} \epsilon^{ABC} \hat f'_{BCab} \gamma_{012} + \tfrac{1}{3} \tau^A{}_\mu \hat f'_{ABab}\gamma^B
- \tfrac13 e_a{}_\mu \hat{T}_{b A}{}^A \gamma_{012}
) \gamma^{ab}
\\ & \qquad
+ \tfrac16 e^a{}_\mu \hat T_{aA}{}^A \gamma_{012}
- \tfrac{1}{12} \tau^A{}_\mu \gamma_{A} \boldsymbol{Z}
+  \tfrac{1}{12} \tfrac{1}{3!} e^a{}_\mu \hat f_{Abcd} ( \gamma_{a}{}^{bcd} - 6 \delta^b_a \gamma^{cd} ) \gamma^A \,,\\
(X_{3/2})_\mu \epsilon_+ & =\nonumber
\tau_{A \mu} \Big( \frac12 \gamma_B ( {\hat T}_{a}{}^{(AB)}  + \tfrac13 \eta^{AB}  {\hat T}{}_{aC}{}^C )\gamma^a
+ \tfrac14 ( \eta^{AB} - \tfrac13 \gamma^A \gamma^B ) \boldsymbol{\hat f}_B
\Big) \epsilon_+ \\
& \qquad
+ e^a{}_\mu \Big(
- \tfrac12 \gamma_a [ - \tfrac{1}{12} \boldsymbol{\hat f}^{(-)}  + \tfrac{1}{6} \gamma_a {\boldsymbol{\hat T}^A} \gamma_A  ]
+ \tfrac12 \hat T_{ab}{}^A \gamma^b \gamma_A
- \tfrac18 \boldsymbol{\hat f}^{(+)} \gamma_a
\Big)\epsilon_+
\,,\\
(X_{-3/2})_\mu  \epsilon_-
& =\nonumber
\Big(
\tfrac12 \hat R_{\mu}{}^{Aa} \gamma_{Aa}
+ e^a{}_\mu \tfrac{1}{4} \hat f'_{bcAB} ( \tfrac{1}{24} \gamma_a \gamma^{bc} - \tfrac{1}{8} \gamma^{bc} \gamma_a) \gamma^{AB}
+ e^a{}_\mu \tfrac{1}{24} ( \gamma_a \boldsymbol{Z} - 3 \boldsymbol{Z} \gamma_a )
 \\ & \qquad
+ \tau^A{}_\mu( 	\tfrac16 \hat f'_{012a} \gamma^a \gamma_A  + \tfrac16 \boldsymbol{Z}_A - \tfrac{1}{12} \boldsymbol{Z}_B \gamma_A{}^B)
\Big)\epsilon_- \,.
\end{align}
\end{subequations}
Here boldface symbols denote contraction with transverse gamma matrices, and
\begin{subequations}
\be
\begin{split}
\hat\omega(\tau)_{\mu}{}^{AB}  & = \hat T_{\mu}{}^{[AB]} - \tfrac12 \tau^C{}_\mu \hat T^{AB}{}_C \,,
\,\\
\hat\omega(e)_{\mu}{}^{ab} & = \hat \Omega_\mu{}^{[ab]} - \tfrac12 e_{c\mu} \hat \Omega^{abc} + \tfrac12 \tau^A{}_\mu \bar\psi_{+}{}^a \gamma_A \psi_+{}^b
\,,\\
\hat R_\mu{}^{Aa} & =
\tfrac12 \hat \Omega_\mu{}^{Aa} - \tfrac12 e_{\mu b} \hat \Omega^{Aab}
+ \tfrac12 \tau^B{}_\mu ( - \bar\psi_+{}^a \gamma^A \psi_{+B}  + \bar\psi_+{}^A \gamma_B \psi_{+}{}^a )
- \tfrac12 \bar\psi_+{}^a \gamma^A \psi_{+b} e^b{}_\mu
\end{split}
\ee
while the field strength combinations are:
\be
\begin{split}
\hat f_{abcd} & \equiv f_{abcd} - 6 \bar\psi_{-[a} \gamma_{bc} \psi_{-d]}
\,,\\
\hat f_{Aabc} & \equiv f_{Aabc}
-3 \bar \psi_{-A} \gamma_{[ab} \psi_{-c]}
+ 6 \bar\psi_{+ [a} \gamma_{|A|b} \psi_{-c]}
\,,\\
\hat f'_{ABab} & \equiv f_{ABab}
- 4 (\bar\psi_{+[A} \gamma_{B] [a} \psi_{-b]} + \bar\psi_{-[A} \gamma_{B] [a} \psi_{+b]})
- \bar\psi_{+a} \gamma_{AB} \psi_{+b}
- \bar\psi_{-A} \gamma_{ab} \psi_{-B}
\,,\\
\hat f'_{ABCa} & \equiv  f_{ABCa}
+ 6  \bar\psi_{+[A} \gamma_{B|a|} \psi_{-C]}
- 3 \bar\psi_{+[A} \gamma_{BC]} \psi_{+a}
\,,\\
\end{split}
\ee
and also some purely fermionic objects appear:
\be
Z_{abcd} \equiv - 6  \bar\psi_{+[a} \gamma_{bc} \psi_{+d]}\,,\quad
Z_{Aabc} \equiv - 3  \bar \psi_{+A} \gamma_{[ab} \psi_{+c]} \,,\quad
Z_{ABab} \equiv - \bar\psi_{+A} \gamma_{ab} \psi_{+B}\,.
\ee
\end{subequations}
To simplify the above expressions, we make field-dependent Stuckelberg transformations with the following parameters:
\be
\begin{split}
\rho_{A+} & =
\tfrac{1}{12} (
\epsilon_{ABC} \hat f'^{BC}{}_{ab} - \tfrac{1}{3} \hat f'_{ABab} \gamma^B) \gamma^{ab} \epsilon_+
- \tfrac{1}{6} ( \boldsymbol{Z}_A  - \tfrac12 \boldsymbol{Z}_B \gamma_A{}^B )  \epsilon_-
\\ & \qquad - \tfrac12  \left(
\bar\psi_+{}_{(A} \gamma_{B)} \psi_{+}^a
-\tfrac13  \bar\psi_{+C}  \gamma^C \psi_+^a\eta_{AB} \right)\gamma^B \gamma_a \epsilon_-
 \,,\\
\eta_- &
= -\tfrac13 \hat T_{aA}{}^A \gamma^a \epsilon_+
+ \tfrac{1}{12} \boldsymbol{Z} \epsilon_-
- \tfrac{1}{24} \gamma^{BC} \hat f'_{BCab} \gamma^{ab} \epsilon_- \,.
\end{split}
\label{StuckBeautifiers}
\ee
This then leads to the supersymmetry transformations in the form \eqref{finiteSUSY} stated and used in the main text.
Note that the purely fermionic second line of the parameter $\rho_{A+}$ in \eqref{StuckBeautifiers} ensures that it is indeed the supercovariant combination $\hat \omega_{\mu}{}^{Aa}$ that appear in the final expressions \eqref{finiteSUSY}, while $\hat \omega_{\mu}{}^{AB}$ and $\hat \omega_{\mu}{}^{ab}$ come out automatically.
In particular, this means that the following correct supercovariant field strengths appear (in place of the quantities $\hat f'$ which come out originally from the expansion):
\begin{subequations}
\begin{align}
 \hat{f}_{ABab} & \equiv f_{ABab}
- 4 (\bar\psi_{+[A} \gamma_{B] [a} \psi_{-b]} + \bar\psi_{-[A} \gamma_{B] [a} \psi_{+b]})
- \bar\psi_{-A} \gamma_{ab} \psi_{-B}
- 2 \bar\psi_{+a} \gamma_{AB} \psi_{+b}
\label{defhathatfABab}
\,,\\
\hat{f}_{ABCa} & \equiv  f_{ABCa}
+ 6 \bar\psi_{+[A} \gamma_{B|a|} \psi_{-C]}
- 6 \bar\psi_{+[A} \gamma_{BC]} \psi_{+a}
\,.
\label{defhathatfABCa}
\end{align}
\end{subequations}

\subsection{Gravitino kinetic terms at subleading order and super-Poisson equations} 
\label{RS_subleading}

In this appendix, we show how to extract useful information from the limit of the gravitino kinetic term, treating the background non-relativistic geometry (described by $\tau^A{}_\mu$ and $e^a{}_\mu$) as being fixed.
With this assumption, we consider the kinetic term:
\be
\begin{split}
\mathcal{L}_{\text{RS}} & =  
- 2 \sqrt{|g|} \bar \Psi_\mu \gamma^{\mu\nu\rho} \partial_\nu \Psi_\rho 
\\ & =
- 2 \Omega \bar \Psi_\mu \big(
c^{1/2} e^\mu{}_a e^\nu{}_b e^\rho{}_c \gamma^{abc} 
+ 3 c^{-1} e^{[\mu}{}_a e^\nu{}_b \tau^{\rho]}{}_A \gamma^{abA} 
\\ & \qquad\qquad
+ 3 c^{-5/2} \tau^{[\mu}{}_A \tau^\nu{}_B e^{\rho]}{}_a \gamma^{ABa} 
+ c^{-4} \tau^\mu{}_A \tau^\nu{}_B \tau^\rho{}_C \gamma^{ABC} 
\big)  \partial_\nu \Psi_\rho \,.
\end{split}
\label{RSexp2}
\ee
Using the gravitino expansion \eqref{ExpFer} and integrating by parts 
\be
\begin{split} 
\Omega^{-1} \mathcal{L}_{\text{RS}} & = 
-4 \bar \psi_{+a} \gamma^{abc} \partial_b \psi_{-a} 
-4 \bar \psi_{-A} \gamma^{abA} \partial_a \psi_{-b} +2 \bar \psi_{-a} \gamma^{abA} (\partial_A \psi_{-b} -\partial_b \psi_{-A})
\\ & 
	\quad + c^{-3}\Big[
-4 \bar \psi_{+ A} \gamma^{abA} \partial_a \psi_{+b} +2 \bar \psi_{+a} \gamma^{abA} \partial_A \psi_{+b} 
\\ & \qquad\qquad
-4 \bar \psi_{+A} \gamma^{ABa} ( \partial_B \psi_{-a} - \partial_a\psi_{-B}) - 4 \bar \psi_{+a} \gamma^{ABa} \partial_A \psi_{-B} 
\\& \qquad\qquad
-2 \bar\psi_{-A} \gamma^{ABC} \partial_B \psi_{-C}  
\Big] + \mathcal{O}(c^{-6})\,.
\end{split} 
\ee
Varying according to \eqref{deltaSgrav}, we find the following expansion for the equations of motion:
\be
\begin{split} 
\mathcal{E}(\psi_+)^a & =- 4  \gamma^{abc} \partial_b  \psi_{-c}- 4 c^{-3} \left( 
\gamma^{ABa} \partial_A  \psi_{-B}  +\gamma^{abA}  ( \partial_b  \psi_{+A} - \partial_{A}  \psi_{+b} ) 
\right) \,,\\
\mathcal{E}(\psi_+)^A & =- 4 c^{-3} \left( \gamma^{abA} \partial_a  \psi_{+b}  +  \gamma^{ABa}  ( \partial_B  \psi_{-a} - \partial_a  \psi_{-B}) \right) \,,\\
\mathcal{E}(\psi_-)^a & =  -4 \left(  \gamma^{abc} \partial_b  \psi_{+c} + \gamma^{abA} ( \partial_b  \psi_{-A} - \partial_A  \psi_{-b} )  \right) + \mathcal{O}(c^{-3}) \,,\\
\mathcal{E}(\psi_-)^A & = - 4 \gamma^{abA}  \partial_a  \psi_{-b} 
+ c^{-3} \left(
4 \gamma^{ABa} (  \partial_a  \psi_{+B} - \partial_B \psi_{+a} )  - 4 \gamma^{ABC} \partial_B  \psi_{-C} 
\right)
\,.
\end{split} 
\label{lineompsi}
\ee
The subleading parts cannot in general be expressed in terms of the covariant determined gravitino curvatures \eqref{gravitinocurvatures}.
However, for the dynamics we are only interested in extracting the equations of motion conjugate to the Stuckelberg variation, which gives equations that start at order $c^{-3}$.
Firstly, using $\delta \psi_{-A} = \gamma_A \eta_-$, $\delta \psi_{+a} = - \tfrac12 \gamma_a \eta_-$ we find the combination:
\be
( - \gamma_A \mathcal{E}(\psi_-)^A  + \tfrac12 \gamma_a \mathcal{E}(\psi_+)^a ) \big|_{c^{-3}}
= 6 \left( \gamma^a \gamma^A ( \partial_A \psi_{+a} - \partial_{+a} \psi_{+A} ) - 2 \gamma^{AB} \partial_B \psi_{-C} \right) \,,
\ee
from which we infer, referring to the definitions \eqref{gravitinocurvatures}, that we are obtaining here the linearisation of the covariant super-Poisson equation \eqref{superfish1}.
For the other  Stuckelberg symmetry, we have $\delta \psi_{+A} = \rho_{+A}$, and we have
\be
( \mathcal{E}(\psi_+)^A - \tfrac13 \gamma^A \gamma_B  \mathcal{E}(\psi_+)^B) \big|_{c^{-3}}= 4 \gamma^a ( \partial^A \psi_{-a} - \partial_a \psi_{-}^A - \tfrac13 \gamma^A \gamma^B ( \partial_B \psi_{-a} - \partial_a \psi_{-B} ) \,,
\ee
from which we infer that we are obtaining here the linearisation of the covariant super-Poisson equation \eqref{superfish2}.

\subsection{Supersymmetry transformation of Lagrange multiplier} 
\label{app_lambda}

In this appendix, we discuss the supersymmetry transformation of $\lambda_{abcd}$, which was determined in \eqref{deltalambdadesired} to have the form:
\be
\begin{split} 
\delta_Q \lambda_{abcd} & = 
 c^3 \left( \delta'_\epsilon \hat f^{(-)}_{abcd}
- \tfrac{1}{24} \bar \epsilon_+  \gamma_{abcd} \gamma_A\,{\mathcal{E}}(\psi_+)^A
- \tfrac{1}{16} \bar \epsilon_- \gamma_e \gamma_{abcd} \, {\mathcal{E}}(\psi_+)^e 
\right)
\\ & \quad
- \tfrac12 P^-{}_{abcd}{}^{efgh} \bar\psi_{+e} \gamma_{fg} ( \delta'_\epsilon \psi_{+h})\,,
\end{split}
\label{deltalambdadesired_app}
\ee
where $\mathcal{E}(\psi_+)^\mu$ denotes the equation of motion of $\psi_{+\mu}$.
The supersymmetry transformation $\delta'_\epsilon \psi_{+ \mu}$ is defined in \eqref{deltaprimepsi3} and \eqref{deltaprimepsi0}.
The supersymmetry transformation $\delta'_\epsilon \psi_{-\mu}$ is defined in \eqref{deltaprimepsi3} at order $c^3$ and at order $c^0$ is given in \eqref{deltapsiAfterEomSymm}.
Acting on the bosons, $\delta'_\epsilon$ coincides at order $c^0$ with the transformations $\delta_Q$ given in \eqref{deltaQbos}; below we will also use the subleading transformation acting on $c_{\mu\nu\rho}$, given by
\be
\delta'_{-3} c_{\mu\nu\rho} = 3 \bar \epsilon_+ \gamma_{ab} \psi_{+[\mu} e^a{}_\nu e^b{}_{\rho]} \,.
\label{deltaQ_c_subleading}
\ee
Let's first consider the terms at order $c^3$ in \eqref{deltalambdadesired_app}.
This starts with the finite variation of $\hat f^{(-)}_{abcd}$, which can be obtained by projecting
\be
\delta'_\epsilon \hat f_{abcd} \big|_{c^0} = 6 \bar \epsilon_- \gamma_{[ab} \breve{r}_{-cd]} +  (\text{constraints}) \,,
\label{deltalambda_div_1}
\ee
as follows from a short calculation, 
where we omit writing terms which vanish either when imposing the initial geometric constraints needed for maximal or half-maximal supersymmetry.
The gravitino curvature $\breve{r}_{-ab}$ is defined in \eqref{gravitinocurvatures} and coincides with that $\rl_{-ab}$ obtained in the equations of motion derived from the limit of the action.
We similarly know that the variation $\delta'_\epsilon \psi_{+h}$ at order $c^3$ has the form \eqref{deltaprimepsi3} and so also vanishes when we impose these constraints.
We then need to consider the terms involving the equations of motion $\mathcal{E}(\psi_+)^\mu$ of $\psi_{+\mu}$ at order $c^0$, which were obtained directly by varying the non-relativistic action and are recorded in \eqref{eom_psiplus}.
We easily see that all terms contributing at order $c^3$ from these equations of motion and involving the bosonic intrinsic torsions vanish when the constraints are applied, such that
\be
\left(
- \tfrac{1}{24} \bar \epsilon_+  \gamma_{abcd} \gamma_A\,{\mathcal{E}}(\psi_+)^A
- \tfrac{1}{16} \bar \epsilon_- \gamma_e \gamma_{abcd} \, {\mathcal{E}}(\psi_+)^e 
\right)\big|_{c^0}
=  \tfrac18 \bar \epsilon_- \gamma_e \gamma_{abcd} \gamma^{efg} \breve{r}_{-fg} +  (\text{constraints}) \,.
\label{deltalambda_div_2}
\ee
The sum of the gravitino curvature terms in \eqref{deltalambda_div_1} and \eqref{deltalambda_div_2} then turns out to vanish after some gamma matrix algebra.
We conclude that the supersymmetry transformation of $\lambda_{abcd}$ is guaranteed to be finite simply by the initial constraints \eqref{geoconstraints} or \eqref{halfsusy} (with the further implied constraint $f_{abcd} = 0$) imposed for maximal or half-maximal supersymmetry.

Now we have to work out the finite part of the transformation.
To simplify matters, we again assume that the vielbein are fixed in order that we can focus solely on terms involving derivatives of $\psi_{\pm \mu}$.
We then invoke covariance to upgrade these to covariant expressions.
The cost of doing so is that we may miss contributions involving the intrinsic torsions \eqref{intrinsictorsions}. These are all set to zero in the maximal supersymmetry case, so this procedure should be at least unambiguous there.

Proceeding in this manner, using \eqref{deltaQ_c_subleading} and the subleading equations of motion \eqref{lineompsi}, we find
\begin{align} 
\delta'_\epsilon \hat f_{abcd} \big|_{c^{-3}} & \sim 6 \bar \epsilon_+ \gamma_{[ab} r_{+cd]} + 12 \partial_{[c} \bar \epsilon_+  \gamma_{ab} \psi_{+d]} \,,\\
- \tfrac{1}{24} \bar \epsilon_+  \gamma_{abcd} \gamma_A\,{\mathcal{E}}(\psi_+)^A \big|_{c^{-3}} & \sim \tfrac16 \bar \epsilon_+ \gamma_{abcd} \left( \tfrac32 \gamma^{ef} r_{+ef} + 2 \gamma^A \gamma^e r_{-Ae} \right)\,,\\
- \tfrac{1}{16} \bar \epsilon_- \gamma_e \gamma_{abcd} \, {\mathcal{E}}(\psi_+)^e\big|_{c^{-3}} & \sim \tfrac14 \bar \epsilon_- \gamma^e \gamma_{abcd} \gamma^A r_{+Ae} \,,
\end{align} 
where at the order we are working at, $r_{\pm \mu\nu} \sim 2 \partial_{[\mu} \psi_{\pm \nu]}$.
The term involving a derivative of $\epsilon_+$ cancels against the finite variation of $\delta'_\epsilon \psi_{+h}$ coming from the final term of \eqref{deltalambdadesired_app}: this is the explicit realisation of the statement made in the main text that it is $\lambda_{abcd}$ defined via \eqref{deflambda} by adding this term which is supercovariant.
Inserting the above expressions into \eqref{deltalambdadesired_app} and using a slew of gamma matrix identities we ultimately find a transformation consistent with a covariant transformation of the form:
\be
\delta_Q \lambda_{abcd} = \tfrac14 \bar \epsilon_- \gamma^e \gamma_{abcd} \breve{r}_{+e} + 2 \bar \epsilon_+ \gamma_{[abc}{}^e \breve{r}_{+d]e} + \dots\,,
\ee
where the ellipsis denotes the possibility that there are modifications to this expression missed by the above (essentially linear) analysis, involving bosonic intrinsic torsion tensors.

\subsection{The dependent fermionic connections}
\label{app_varphi}

In this appendix, we explain how to solve for the fermionic gauge fields $\varphi_{-\mu}$ and $\varphi_{+A\mu}$ introduced in \eqref{hatDpsi}.
To do so, let's rewrite the gravitino curvatures \eqref{fermioniccurvatures} as 
\begin{equation}
\begin{split}
	r_{-\mu\nu} = 2\,D_{[\mu}\psi_{-\nu]} + 2\,\gamma_A\tau^A{}_{[\mu}\varphi_{-\nu]}\,,\quad
	r_{+\mu\nu} = 2\,\tilde D_{[\mu}\psi_{+\nu]} - \gamma_a e^a{}_{[\mu}\varphi_{-\nu]} + 2\,\tau^A{}_{[\mu}\varphi_{+A\nu]}\,,
\end{split}
\end{equation}
where we define the following bosonic covariant derivative
\be
\tilde D_{[\mu} \psi_{+ \nu]} = 2\,D_{[\mu} \psi_{+ \nu]} - \tfrac{1}{6} \boldsymbol{\lambda} \tau^A{}_{[\mu} \gamma_A  \psi_{+\nu]}- \tfrac14 \boldsymbol{\lambda}e^e{}_{[\mu}  \gamma_e \psi_{-\nu]} \,,
\ee
Below we use the shorthand notation $(D\psi)_{-\mu\nu} \equiv 2 D_{[\mu} \psi_{-\nu]}$ and $(\tilde D \psi)_{+\mu\nu} \equiv 2 \tilde D_{[\mu} \psi_{+\nu]}$.
Before we start, note that since by definition $\gamma^A \varphi_{+A \mu} = 0$ we have
\be
\gamma^A \varphi_{+[AB]} + \gamma^A \varphi_{+(AB)} = 0 \,,\quad
\gamma^{AB} \varphi_{+AB} = \eta^{AB} \varphi_{+AB} \,.
\ee
To solve for $\varphi_{-\mu}$ we use the following conventional constraints:
\be
r_{-AB} = 0 \,,\quad
\gamma^A r_{-Aa} = 0 \,.
\ee
Here we have $3+8=11$ constraints and $11$ unknowns (not counting spinor degrees of freedom).
There is a unique result 
\be
 \varphi_{-\mu} = \tau^A{}_\mu\,\big(\gamma^B (D\psi)_{-AB} - \tfrac14\,\gamma_A\gamma^{BC}(D\psi)_{-BC}\big) - \frac13\,e^a{}_\mu\gamma^A(D\psi)_{-Aa}\,.
\ee
To solve for $\varphi_{+A \mu}$ we use the following conventional constraints:
\be
r_{+Aa} - \tfrac13 \gamma_A \gamma^B r_{+Ba} = 0 \,,\quad
r_{+AB} = 0 \,.
\ee
Counting $\gamma^A \varphi_{+A\mu} = 0$ as a constraint, we have $3\times 11=33$ unknowns and $(3-1)\times 8 + 3 + 11 = 30$ constraints.
Hence we obtain a non-unique answer: 
\be
\begin{split}
\varphi_{+A\mu} &= -\tfrac12\,\tau^B{}_\mu (\tilde D\psi)_{+AB} + \tau^B{}_\mu\varphi_{+(AB)} 
\\ & \quad
- e^a{}_\mu\big(( \tilde D \psi)_{+Aa} +\tfrac12\,\gamma_a\varphi_{-A} - \tfrac13 \gamma_A  \gamma^B ( ( \tilde D \psi)_{+Ba} +\tfrac12\,\gamma_a\varphi_{-B} )\big)\,.
\end{split}
\label{varphiexplicit2}
\ee
The symmetric part $\varphi_{+(AB)}$ naively gives 6 undetermined components, but the gamma trace condition amounts to three conditions on this quantity:
\be
\gamma^A \varphi_{+(AB)} = \tfrac12 \gamma^A (\tilde D \psi)_{+AB} \,.
\label{gammatracevarphi}
\ee
So in the end there are three unknowns left over, as expected.
Following from \eqref{gammatracevarphi}, we also have
\be
\eta^{AB} \varphi_{+(AB)} = -\tfrac12 \gamma^{A B} (\tilde D\psi)_{+AB}
\,,\quad
\gamma^B \varphi_{+AB} = \gamma^B (\tilde D \psi)_{+BA} \,.
\ee
Note that the three curvature components not used in the above are $\gamma^a r_{+Aa}$ which can be evaluated as
\be
\gamma^a r_{+Aa} = \gamma_A \gamma^B( \tfrac13 \gamma^a (\tilde D\psi)	_{+Ba} - \tfrac12 \varphi_{-B} ) \,,
\label{whatdoesthismean}
\ee
and so do not help to solve for the undetermined components of $\varphi_{+A\mu}$.

\section{Bianchi identities} 
\label{appBI}

\subsection{Bosonic Bianchi identities} 
\label{appBI_bos}

In this appendix, we work out the bosonic Bianchi identities for an MNC geometry obeying the following conventional constraints: 
\begin{align}
    \mathcal{T}^A &= \rmd\tau^A + \omega^A{}_B\tau^B - b\tau^A = \frac12\,e^a e^b\,T_{ab}{}^A + e^a\tau_B\,T_a{}^{\{AB\}}\,,\\
    \mathcal{T}^a &= \rmd e^a + \omega^a{}_b e^b +\frac12\, b e^a - \omega_A{}^a\tau^A = 0\,,\\
    F_4 &= \rmd c_3 + \tfrac12\,\epsilon_{ABC}\omega^A{}_a e^a\tau^B\tau^C = \frac{1}{4!}\,e^a e^b e^c e^d \,f_{abcd} + \frac{1}{3!}\,\tau^A e^a e^b e^c\,f_{Aabc}\,,
\end{align}
where in particular we do not assume that the right-hand sides vanish.
For convenience we have adopted differential form notation, with wedge products implicit. 

\paragraph{Algebraic BIs}

These constraints can be easily seen to imply the following algebraic Bianchi identities:
\begin{align}
    R^A{}_B\tau^B - R\,\tau^A &= \frac12\,e^a e^b\,D T_{ab}{}^A + e^a\tau_B\,D T_a{}^{\{AB\}} - e^a\mathcal{T}_B T_a{}^{\{AB\}}\,,\\
    R^a{}_b e^b + \frac12\, R\,e^a - R_A{}^a\tau^A &= 0\,,\\
    \frac12\,\epsilon_{ABC}R^A{}_a e^a\tau^B\tau^C &= \frac{1}{4!}\,e^a e^b e^c e^d\,D f_{abcd} \notag\\
    &\quad+ \frac{1}{3!}\,\tau^A e^a e^b e^c\,D f_{Aabc} + \frac{1}{3!}\,\mathcal{T}^A e^a e^b e^c\, f_{Aabc}\,.
\end{align}
Note that the curvatures here contain explicit torsion contributions that are in principle not straightforward to figure out; see \eqref{modRab} for an example where we worked out the explicit contributions. Since the starting point was covariant it is guaranteed that everything re-collects into covariant expressions appropriately.
Expanding the implicit differential forms in terms of $\tau^A$ and $e^a$, we read component forms of the Bianchi identities as follows.

\noindent {Completely transversal:} 
\vspace{-1em}
\begin{align}
    0 &= 3\,D_{[a}T_{bc]}{}^A -3\,T_{[abB}T_{c]}{}^{\{AB\}}\,,\\
    3\,R_{[ab}{}^d{}_{c]} + \frac32\,R_{[ab}\delta^d{}_{c]} &= 0\,,\label{eq:0longe}\\
    0 &= 5\,D_{[a}f_{bcde]} + 10\,T_{[ab}{}^A f_{Acde]}\,.
\end{align}
\noindent {One longitudinal:} 
\vspace{-1em}
\begin{align}
    R_{ab}{}^A{}_B - R_{ab}\delta^A{}_B &= D_B T_{ab}{}^A + 2\,D_{[a}T_{b]}{}^{\{AB\}} + 2\,T_{[a\{BC\}}T_{b]}{}^{\{AC\}}\,,\label{eq:1longtau}\\
    2\,R_{A[a}{}^b{}_{c]} + R_{A[a}\delta^b{}_{c]} -R_{acA}{}^b&= 0\,,\\
    0 &= D_A f_{abcd}+ 4\,D_{[a}f_{Abcd]} - 4\,T_{[a\{AB\}}f^B{}_{bcd]}\,.
\end{align}
\noindent {Two longitudinal:} 
\vspace{-1em}
\begin{align}
    2\,R_{a[A}{}^B{}_{C]} - 2\,R_{a[A}\delta^B{}_{C]} &= 2\,\eta_{D[A}D_{C]}T_a{}^{\{BD\}}\,,\\
    R_{AB}{}^a{}_b +\frac12\,R_{AB}\delta^a{}_b - 2\,R_{b[AB]}{}^a &= 0\,,\\
    3\,R_{[ab}{}^A{}_{c]} &= -\epsilon^{ABC}D_B f_{Cabc}\,.
\end{align}
\noindent {Three longitudinal:} 
\vspace{-1em}
\begin{align}
    3\,R_{[AB}{}^C{}_{D]} - 3\,R_{[AB}\delta^C{}_{D]} &= 0\,,\\
    3\,R_{[ABC]}{}^a &= 0\,,\\
    R_{A[a}{}^A{}_{b]} &= 0\,.
\end{align}
These can be further simplified into irreps by contracting indices and combining equations appropriately. This gives rise to a number of useful identities 
\begin{align}
    R_{ab} &= -\frac13\,D_A T_{ab}{}^A\,,\\
    \epsilon^A{}_{BC}R_{ab}{}^{BC} &= -\epsilon^{AB}{}_C\big(D_B T_{ab}{}^C + 2\,T_{[a\{BD\}}T_{b]}{}^{\{CD\}}\big)\,,\\
    R_{aABC} + 2\,\eta_{A[B}R_{aC]} &= 2\,\delta_{[B}{}^DD_{C]}T_{a\{AD\}}\,.
\end{align}

\paragraph{Differential BIs}

Given the definitions of the curvature two-forms we can also derive associated differential Bianchi identities. These are given by:
\begin{align}
    &D(T)R=0\,,&&D(T)R^{AB}=0\,, && D(T)R^{ab}=0\,,&& D(T)R^{Aa}=0\,.
\end{align}
The covariant derivative used here contains connections according to the index structure. The expression of the boost connection is more complicated $D(T)R^{Aa} = \rmd R^{Aa} + \omega^A{}_B R^{Bb} + \tfrac32\,b R^{Aa} + \omega^a{}_b R^{Ab}-\omega^{Ba}R^A{}_B - \omega^{Ab}R^a{}_b - \tfrac32\,\omega^{Aa}R+\Delta(T)$.
We have also absorbed potential torsion terms $\Delta(T)$ in the definition of $D(T)$ without being explicit, since we effectively only need the case where torsion is set to zero. As above, we can project the Bianchi identities along directions longitudinal and transverse to the foliation. In practice, we only use
\begin{align}\label{eq:diffBianchi}
    D_a \breve{R}_{AB}{}^{bc} + 2\,D_{[A}\breve{R}_{B]a}{}^{bc}=0\,.
\end{align}

\subsection{Fermionic Bianchi identities}
\label{appBI_fer}

We summarize the conventional constraints for the superconformal gauge fields, in form notation, as:
\begin{align}
    r_- &
    = \tfrac12\,e^a e^b\,\breve{r}_{-ab} + \tau^A e^a\,\breve{r}_{-Aa}\,,\\
    r_+ &
    = \tfrac12\,e^a e^b\,\breve{r}_{+ab} + \tfrac13\,\tau^A e^a\,\gamma_A\breve{r}_{+a}\,,
\end{align}
where $r_\pm$ are defined in \eqref{fermioniccurvatures}.
These equations are invariant under the following shift of the superconformal gauge field:
\begin{align}
    \varphi_{+A\mu} \to \varphi_{+A\mu} + \tau^B{}_\mu\tilde{f}_{+(AB)}\,,
\end{align}
where $\tilde{f}_{+(AB)}$ is defined to manifestly satisfy $\gamma^B\tilde{f}_{+(AB)}=0$ and thus also $\eta^{AB}\tilde{f}_{+(AB)}=0$. 
This shift ambiguity captures the fact that we are three conventional constraints short of the number of connection components. This is in complete analogy to the bosonic shift ambiguity \eqref{connectionshifts}.
The superconformal curvature two-forms $S_{-\mu\nu}$ and $S_{+A\mu\nu}$, defined in \eqref{superconformal}, naturally transform under this fermionic shift 
as $S_{+A}\to S_{+A} - \tau^B\,D\tilde{f}_{+(AB)}$. This means that not all components of the superconformal curvature components $S_{+A\mu\nu}$ are unambiguous under the fermionic shift. In order to further analyse these, we decompose the \textbf{48} components $S_{+ABa}$, each carrying sixteen fermionic components themselves, into the irreducible representations $S_{+[AB]a}$, $S_{+\{AB\}a}$ and $\eta^{AB}S_{+ABa}$. The gamma-tracelessness condition $\gamma^A S_{+ABa}=0$ turns into the statement $\gamma^BS_{+[AB]a} = \gamma^BS_{+\{AB\}a} + \tfrac13\gamma_A\eta^{BC}S_{+BCa}$ which effectively subtracts $\mathbf{24}$ components. Note, however, that as a consequence we also have $\gamma^{AB}S_{+ABa}=\eta^{AB}S_{+ABa}$---meaning we should read the gamma traceless condition as $\mathbf{16}\oplus\mathbf{8}$ conditions. We subtract the $\mathbf{16}$ from the $S_{+\{AB\}a}$ meaning we count it effectively as $\mathbf{24}$ components and use the $\mathbf{8}$ to read the trace as a dependent expression. This is useful because the remaining components $S_{+\{AB\}a}$ are then the ones that are ambiguously defined with respect to the shift $\tilde{f}_{+(AB)}$. A similar analysis can be done for the completely longitudinal components $S_{+ABC}$. In summary, we can separate the $\mathbf{110}$ components into 
\begin{align}
    &\mathrm{well-defined}~(\mathbf{80}): && S_{+Aab}~(\mathbf{56})\,, &&  S_{+[AB]a}~(\mathbf{24})\,,\notag\\
    & &&  \epsilon^{ABC}S_{+ABC}~(\mathbf{1})\,, \notag\\
    &\mathrm{ill-defined}~(\mathbf{29}):  && S_{+(AB)C}~(\mathbf{5})\,, && S_{+\{AB\}a}~(\mathbf{24})\,.
\end{align}

\paragraph{Algebraic BIs}

The algebraic Bianchi identities are:
\begin{subequations}
    \begin{align}
        -\tau^A\gamma_A S_- &= \tfrac12\,e^a e^b\, \widehat D\breve{r}_{-ab} + \tau^A e^a\,\widehat D\breve{r}_{-Aa}+ \mathcal{T}^A e^a\breve{r}_{-Aa}\,,\\
        \tfrac12\,e^a\gamma_a S_- - \tau^A S_{+A} &= \tfrac12\,e^a e^b\, \widehat{D}\breve{r}_{+ab} + \tfrac13\,\tau^A e^a\,\gamma_A \widehat{D}\breve{r}_{+a}+ \tfrac13\,\mathcal{T}^A e^a\,\gamma_A\breve{r}_{+a}\notag\\
        &\quad {-\tfrac{1}{12}\,\tau^A\boldsymbol{\lambda}\gamma_A r_+ - \tfrac18\,e^a\boldsymbol{\lambda}\gamma_a r_-}\,,
    \end{align}
\end{subequations}
where we've omitted $\mathcal{O}(\lambda^2)$ terms for simplicity. The covariant
derivatives are defined as $\widehat{D} r_- = D r_- - \mathcal{T}^A\gamma_A\varphi_-$ and $\widehat{D} r_+ = D r_+ - \mathcal{T}^A\varphi_{+A}$. Projecting this gives rise to the following constraints in terms of components.

\noindent {Completely transversal:}\vspace{-1em}
\begin{subequations}
    \begin{align}
        0 &= 3\,\widehat D_{[a}\breve{r}_{-bc]} +3\,\breve{r}_{-A[a}T_{bc]}{}^A\,,\\
        3\,\gamma_{[a}S_{-bc]} &= 6\,\widehat D_{[a}\breve{r}_{+bc]} + 2\,T_{[ab}{}^A\gamma_A\breve{r}_{+c]} {-\tfrac38\boldsymbol{\lambda}\gamma_{[a}\breve{r}_{-bc]}}\,.
    \end{align}
\end{subequations}
\noindent {One longitudinal:}\vspace{-1em}
\begin{subequations}
    \begin{align}
        S_{-ab} &= -\tfrac13\,\gamma_A\big(\widehat D^A\breve{r}_{-ab} -2\,T_{[a}{}^{\{AB\}}\breve{r}_{-Bb]}\big)\,,\\
        S_{+Aab} + \gamma_{[a}S_{-Ab]} &= -\widehat D_A\breve{r}_{+ab} + \tfrac23\,\gamma_A \widehat D_{[a}\breve{r}_{+b]} \notag\\
        &\quad+ \tfrac23\,T_{[a\{AB\}}\gamma^B\breve{r}_{+b]} {-\tfrac{1}{24}\,\gamma_A\boldsymbol{\lambda}\breve{r}_{+ab}}\,.
    \end{align}
\end{subequations}
\noindent {Two longitudinal:}\vspace{-1em}
\begin{subequations}
    \begin{align}
        2\,\gamma_{[A}S_{-B]b} &= -2\,\widehat D_{[A}\breve{r}_{-B]b}\,,\\
        2\,S_{+[AB]a} - \tfrac12\,\gamma_a S_{-AB} &= \tfrac23\,\gamma_{[A}\widehat D_{B]}\breve{r}_{+a} {-\tfrac{1}{18}\,\gamma_{AB}\boldsymbol{\lambda}\breve{r}_{+a}}\,.\label{SpABaBI}
    \end{align}
\end{subequations}
\noindent {Three longitudinal:}\vspace{-1em}
\begin{subequations}
    \begin{align}
        3\,\gamma_{[A}S_{-BC]} &= 0\,,\\
        S_{+[ABC]} &= 0\,.
    \end{align}
\end{subequations}
All the components appearing in these expressions are unambiguously defined with respect to both the bosonic and the fermionic shift ambiguities. Upon imposing level two constraints $\breve{r}_{-ab} = 0 = \breve{r}_{+ab} = \breve{r}_{-Aa}$ we can further simplify these expressions to find
\begin{align}\label{eq:superconformalbianchitorsionless}
    &S_{-ab} = 0\,, &&S_{-Aa}=0\,,&& \gamma^{AB}S_{-AB}=0\,,\notag\\
    &S_{+Aab}=0\,, &&S_{+[ABC]}=0\,, &&2\,D_{[a}\breve{r}_{+b]}=0\,,
\end{align}
which further implies 
\begin{align}\label{eq:thelastSCBianchi}
  \gamma^{AB}S_{+ABa} + \tfrac23\,\gamma^A D_A\breve{r}_{+a} - {\tfrac16\,\boldsymbol{\lambda}\breve{r}_{+a}= 0}\,.
\end{align}
\paragraph{Differential BIs}
The superconformal curvatures \eqref{superconformal} also satisfy differential Bianchi identities. These follow simply from the definition and can be written as three-form constraints
\begin{align}
  &DS_-=0\,, && D S_{+A}=0\,,
\end{align}
where the definition of the derivatives includes second order bosonic curvatures. These can be projected to give rise to the following constraints on the components of the superconformal curvatures that are not set to zero by the algebraic Bianchi identities \eqref{eq:superconformalbianchitorsionless}:
\begin{align}
  &D_a S_{-AB}=0\,, &&D_{[a}S_{+[AB]b]}=0\,.
\end{align}

\section{The tower of constraints for maximal supersymmetry}
\label{tower_derivation} 

In this appendix, we discuss the derivation of the tower of constraints present when we impose the bosonic constraints \eqref{geoconstraints} compatible with maximal supersymmetry.
The full set of constraints must be invariant under all symmetries of the theory.
This means in particular that we have to vary \eqref{geoconstraints} under supersymmetry (and superconformal transformations) and require that the variation itself vanishes, using both \eqref{geoconstraints} and new fermionic constraints.
Then we have to repeat this process with the new constraints that appear, until we have determined all possible constraints. 

We start with the supersymmetry transformations of the constraints \eqref{geoconstraints}.
Let us refer to these as the {\bf Level 1} constraints.
We find:
\begin{subequations}\label{leveloneconstraints}%
  \begin{align}
    \delta_Q \hat T_{ab}{}^A & \approx \bar\epsilon_-\gamma^A \breve r_{-ab}\,, \\
    \delta_Q \hat T_a{}^{\{AB\}} & \approx -\bar\epsilon_-\gamma^{\{A}\breve{r}_-{}^{B\}}{}_a\,,\\
    \delta_Q \hat f_{abcd} & \approx 6\,\bar\epsilon_-\gamma_{[ab}\breve r_{-cd]}\,, \\
    \delta_Q \hat f_{Aabc} &\approx + 3\,\bar\epsilon_+\gamma_{A[a}\breve r_{-bc]} - 3\,\bar\epsilon_-\gamma_{A[a}\breve r_{+bc]} - 3\,\bar\epsilon_-\gamma_{[ab}\breve r_{-c]A}\,.
\end{align}
\end{subequations}
Here we have written $\approx$ to mean that these equations are true up to the imposition of the constraints \eqref{geoconstraints}.
For instance, what we explicitly find is that:
\be
\begin{split}
\delta_Q \hat T_{ab}{}^A 
& = \bar \epsilon_- \gamma^A \breve{r}_{-ab} + 2\, \hat T_{c[a}{}^A ( \bar \epsilon_+ \gamma^c \psi_{-b]} + \bar \epsilon_- \gamma^c \psi_{+b]}) 
 + 2 \bar \epsilon_- \gamma_B \psi_{-[a} \hat T_{b]}{}^{\{AB\}}
 \\ &
 \qquad -\tfrac{1}{12} ( \bar \epsilon_- \gamma^A \gamma_{[a} \boldsymbol{\hat f}^{(-)} \psi_{+b]} 
 + \bar \psi_{-[a} \gamma^A \gamma_{b]} \boldsymbol{\hat f}^{(-)} \epsilon_+ )
\approx \bar \epsilon_- \gamma^A \breve{r}_{-ab} \,.
\end{split} 
\label{varyhatT}
\ee
We remark that this shows again that $\hat f^{(-)}{}_{abcd}$, which did not appear in the original divergent terms in the supersymmetry transformation, should also be included as a constraint in \eqref{geoconstraints}.

We thus find from the variations \eqref{leveloneconstraints} that we should impose the following {\bf Level 2} constraints as superpartners of the bosonic constraints \eqref{geoconstraints}:
\be
\label{fermconstraints}
    \breve{r}_{-ab} = 0\,, \quad \breve{r}_{-Aa} = 0\,,\quad \breve{r}_{+ab} = 0\,.
\ee
Note that not all curvatures appearing in \eqref{gravitinocurvatures} are set to zero.
Next, we must vary \eqref{fermconstraints}.
On general grounds we expect to find the well-defined bosonic curvatures $\breve{R}$ appearing as new constraints. 

A tempting shortcut at this point is to resort to a linearised calculation.
This allows the curvature combinations that must appear in the variation to be easily identified from their leading terms, using covariance to conclude that the full variation must follow. 
\begin{subequations}
This way one gets:
\begin{align}
\delta_Q \breve{r}_{-ab} &\approx \tfrac14 \breve R_{ab}{}^{cd} \gamma_{cd} \epsilon_-\,,\quad
\delta_Q \breve{r}_{-Aa}  \approx \frac14 \breve R_{Ba}{}^{bc} ( \delta_A^B-\tfrac13 \gamma_A \gamma^B ) \gamma_{bc} \epsilon_- \,,\\
\delta_Q \breve{r}_{+ab} & \approx \tfrac14\breve R_{ab}{}^{cd} \gamma_{cd} \epsilon_+ 
+ \left( 
-\breve{R}_{A[ab]c}\gamma^c - \tfrac{1}{12} \breve{R}_{A[a}{}^{cd} \gamma_{b]} \gamma_{cd}\right)  \gamma^A  \epsilon_- 
- \tfrac{1}{4} D_{[a} {\boldsymbol \lambda} \gamma_{b]} \epsilon_-
\,,
\end{align} 
\label{deltalevel2}%
\end{subequations} 
after using the Bianchi identities of appendix \ref{appBI_bos}.

The issue with the linearised calculation is that it may miss terms.
For instance, suppose we started only with the knowledge that $\hat T_{ab}{}^A$ was a constraint.
Then a linearised calculation would show the result \eqref{leveloneconstraints} from which we would conclude that $\hat r_{-ab}$ appears as a new constraint.
However, the explicit variation \eqref{varyhatT} also contains, at the non-linear level, other bosonic constraints, namely $\hat T_{a}{}^{\{AB\}}$ and $\hat f^{(-)}_{abcd}$. 
This means that in principle we could find $\breve r_{+a}$ appearing in the variation of the other gravitino curvatures, which would imply that we have to set this curvature to zero too and add it to the level 2 fermionic constraints \eqref{fermconstraints}.

An involved calculation of the non-linear variation\footnote{Albeit omitting higher order terms involving $\lambda_{abcd}$.} of $\breve{r}_{-ab}$, $\breve{r}_{-Aa}$ and $\breve{r}_{+ab}$, taking into account the supersymmetry transformation of the bosonic connections and the higher order supercovariant completions of the bosonic curvatures, shows that in fact their variation takes exactly the form \eqref{deltalevel2}, modulo imposing the level 1 and 2 constraints \eqref{geoconstraints} and \eqref{fermconstraints} -- up to terms involving the gamma trace $\gamma^a \breve{r}_{+a}$, which appears in term bilinear in fermions.
We therefore also require 
\be
\gamma^a \breve{r}_{+a} \approx 0 \,,
\ee
which is in fact one of the super-Poisson equations. 
Let us discuss the variation of this equation.
A linearised calculation gives
\begin{align}
    \delta_Q \breve{r}_{+a} \approx \tfrac14\,D_a\boldsymbol{\lambda}\epsilon_+
    +\tfrac14\,\gamma^{AB}\breve{R}_{AB}{}^{bc}\big(\delta_{ab}\gamma_c - \tfrac18\gamma_a\gamma_{bc}\big)\epsilon_-
     + \tfrac32\,\breve{R}_{A(a}{}^A{}_{b)}\gamma^b\epsilon_-\,,
\label{deltara}
\end{align}
from which we infer
\be
\delta_Q (\gamma^a \breve{r}_{+a} ) \approx\tfrac14\,\gamma^a D_a\boldsymbol{\lambda}\epsilon_+
 + \tfrac32 \breve{R}_{Aa}{}^{Aa} \epsilon_- \,.
 \label{deltagammabrever}
\ee
We note in passing that the bosonic Poisson equation is $\breve{R}_{Aa}{}^{Aa} + \tfrac{1}{4!} \lambda_{abcd} \lambda^{abcd} = 0$, and so this must be what appears in the $\epsilon_-$ term in the non-linear completion of \eqref{deltagammabrever}.
The remaining term in this variation implies that $D_{[a} \lambda_{bcde]} = 0$ (or equivalently using anti-self-duality $D_a \lambda^{abcd} = 0$). This corresponds, modulo the level 1 constraints, to the equation of motion \eqref{EomChere} of the three-form.
We also note that under Stuckelberg transformations, $\delta_S \breve{r}_{+a} = - \tfrac14\,\boldsymbol{\lambda}\gamma_a\eta_-$, so that $\delta_S \gamma^a \breve{r}_{+a} = 0$.

Returning to \eqref{deltalevel2}, we then pick out the following {\bf Level 3} constraints:
\be
\breve{R}_{ab}{}^{cd} = 0\,,\quad
\breve{R}_{Aa}{}^{bc} = 0 \,,\quad 
D_a\lambda_{bcde} = 0
\,,
\label{level3}
\ee
the last of which follows from combining $D_{[a} \lambda_{bcde]} = 0$ with $D_{[a} \lambda_{b]cde} = 0$, coming from equation \eqref{deltalevel2}.
Next, a linearised variation of the level 3 constraints then produces:
\begin{subequations}
\label{deltalevel3}%
\begin{align}
\delta_Q \breve{R}_{ab}{}^{cd} \approx 0 \,,\quad
\delta_Q\breve{R}_{Aa}{}^{bc}\approx -\bar\epsilon_-\gamma_A\gamma^{[b}D^{c]}\breve{r}_{+a}\,, \quad
\delta_Q D_a\lambda_{b_1\cdots b_4} \approx \frac14\,\bar\epsilon_-\gamma^c\gamma_{b_1\cdots b_4}D_a\breve{r}_{+c}\,,
\end{align}
\end{subequations} 
on using the Bianchi identities $S_{-ab}\approx 0\approx S_{-Aa}\approx S_{+Aab}\approx D_{[a}\breve{r}_{+b]}$. 
This leads us to the solitary new {\bf Level 4} constraint:
\be
D_a\breve{r}_{+b}\approx 0\,.
\label{level4}
\ee
The linearised variation of this, using \eqref{deltara}, leads to the following new {\bf Level 5} constraint:
\be
\label{level5}
D_a \breve{R}_{A(b}{}^A{}_{c)} = 0
\,.
\ee
(We also appear to find $D_a \breve{R}_{AB}{}^{bc} = 0$, however, this is zero identically due to the differential Bianchi identity \eqref{eq:diffBianchi}.)
Varying \eqref{level5} at the linearised level, we find
\begin{align}
    \delta_Q \breve{R}_{A(a}{}^A{}_{b)} &\approx \bar\epsilon_-\gamma^{AB}\gamma_{(a} S_{+ABb)} + \frac13\,\bar\epsilon_-\gamma^A\gamma_{(a}D_A\breve{r}_{+b)} \,.
\end{align}
after using Bianchi identities. 
We can then infer that 
\be
\delta_Q D_a \breve{R}_{A(b}{}^A{}_{c)}
\approx \tfrac32 \bar \epsilon_- D_c \gamma_{(a} \gamma^{AB} S_{+ABb)}\,,
\ee
using the Bianchi identity \eqref{eq:thelastSCBianchi}.
This suggests a possible further constraint:
\be
D_a \gamma^{AB} S_{+ABb} = 0 \,.
\label{putativelevel6}
\ee
However, this is identically satisfied at the linearised level by using the derivative of the same Bianchi identity \eqref{eq:thelastSCBianchi} together with the level 3 and 4 constraints. We also have that at the non-linearised level that if we hit \eqref{eq:thelastSCBianchi} with $D_a$, write $D_a D_A = [D_a,D_A] + D_A D_a$, rewrite the commutator in terms of bosonic curvatures and apply constraints, we find:
\be
 D_a \gamma^{AB} S_{+ABb} + \tfrac23 R_{aA} \gamma^A \breve{r}_{+b} \approx 0 \,.
\label{DSalmost}
\ee
Here the curvature components, $R_{Aa}$ can be seen to act as a gauge field for the special conformal transformations acting as shifts of $b_A$, as noted in footnote \ref{fn:SCTgauge}, such that  the whole expression \eqref{DSalmost} is independent of shifts of the $b_A$ and thus a well defined quantity.
We expect that working at non-linear order the `full' definition of the covariant derivative $D_a$ appearing in \eqref{putativelevel6} absorbs these $R_{Aa}$ terms: in that case \eqref{putativelevel6} is automatically satisfied as a consequence of the Bianchi identity \eqref{eq:thelastSCBianchi} and previous constraints.
At the linearised level, where we can simply ignore the $R_{aA}$ term in \eqref{DSalmost}, we can conclude however that the tower of constraints terminates at level 5.

Let us summarise our findings.
The variation of the constraints \eqref{geoconstraints} required to have a maximally supersymmetric non-relativistic 11-dimensional supergravity leads to a complicated tower of constraints, given in \eqref{fermconstraints}, \eqref{level3}, \eqref{level4} and \eqref{level5}.
The variation of \eqref{level5} leads to a possible constraint \eqref{putativelevel6}, however at least at linear level this appears to be zero by Bianchi.
We are fairly confident that this signals that we have a finite set of constraints.

The presence of a constraint involving a covariant derivative of a curvature (therefore a third derivative bosonic equation) may seem unusual.
Instead of imposing $D_e \lambda_{abcd}$ at level 3, we could directly require the stronger constraint $\lambda_{abcd}$.
This then directly leads to the following set of constraints, which amount to lifting the covariant derivatives at level 4 and 5:
\be
\lambda_{abcd} = 0 = \breve{r}_{+a} = \breve{R}_{A(b}{}^A{}_{c)} = \breve{R}_{AB}{}^{bc} \,.
\label{strongerConstraints} 
\ee
Using the Bianchi identity \eqref{eq:thelastSCBianchi} we infer that $\gamma^{AB} S_{+ABa} \approx 0$ directly and hence the variation of $\breve{R}_{A(a}{}^A{}_{b)}$ vanishes.
However, we also have to compute the variation of $\breve{R}_{AB}{}^{bc}$.
This gives:
\begin{align}
\delta \breve{R}_{AB}{}^{bc} &\approx 
+\tfrac12\,\bar\epsilon_-\gamma^{bc}S_{-AB}
\approx \tfrac14 \bar\epsilon_-\gamma^{bc} \gamma^a S_{+[AB]a} \,,
\end{align}
using a Bianchi identity \eqref{SpABaBI} at the second step.
Therefore we need to add $S_{-AB} = 0$ (or equivalently $\gamma^a S_{+[AB]a}=0$) to the set \eqref{strongerConstraints}.
We would then need to vary this additional fermionic constraint. On general grounds it should lead to a derivative $D \breve{R}$ of a bosonic curvature.
However, all such curvatures have already been set to zero, so any such constraint will automatically be satisifed.
The possibility \eqref{strongerConstraints} is therefore a valid set of sufficient constraints.

\bibliographystyle{JHEP}
\bibliography{references}

\end{document}